\documentclass[
    reprint,
    superscriptaddress,
    nofootinbib,   
    amsmath,amssymb,
    aps,
    prc,floatfix
]{revtex4-2}
\usepackage{style}

\begin{document}

\preprint{APS/123-QED}

\title{\texorpdfstring{\ce{^{117m}Sn} and \ce{^{119m}Te}}{117mSn and 119mTe} Production via Proton Bombardment on Natural Antimony and Implications for Modeling Charged Particle Reactions}

\newcommand{\UCBAffil}{Department of Nuclear Engineering, University of California, Berkeley, Berkeley, California 94720, USA}
\newcommand{\LANLAffil}{Los Alamos National Laboratory, Los Alamos, New Mexico 87544, USA}
\newcommand{\BNLAffil}{Brookhaven National Laboratory, Upton, New York 11973, USA}

\author{Catherine E. Apgar}
\email{ceapgar@lanl.gov}
\thanks{Current Affiliation: \LANLAffil}
 \affiliation{\UCBAffil}

\author{Andrew S. Voyles}
\email{andrew.voyles@berkeley.edu}
 \affiliation{\UCBAffil}

\author{Jon C. Batchelder}
 \affiliation{\UCBAffil}

\author{Cathy S. Cutler}
 \affiliation{\BNLAffil}
 
 \author{Morgan B. Fox}
 \thanks{Current Affiliation: Terrestrial Energy, Oakville, Ontario L6H 0C3, CA}
 \affiliation{\UCBAffil}

\author{Yun-Hsuan Lee}
 \affiliation{\UCBAffil}

 \author{Dmitri G. Medvedev}
 \affiliation{\BNLAffil}

 \author{Jonathan T. Morrell}
 \thanks{Current Affiliation: \BNLAffil}
 \affiliation{\LANLAffil}

  \author{Ellen M. O'Brien}
 \affiliation{\LANLAffil}

 \author{Michael Skulski}
  \thanks{Current Affiliation: NorthStar Medical Radioisotopes, LLC, Beloit, Wisconsin 53511, USA}
 \affiliation{\BNLAffil}
 
 \author{C. Etienne Vermeulen}
 \affiliation{\LANLAffil}

\author{Lee A. Bernstein}
 \affiliation{\UCBAffil}
 \affiliation{Lawrence Berkeley National Laboratory, Berkeley, California 94720, USA}

\date{\today}

\begin{abstract}
    \ce{^{117m}Sn} and \ce{^{119}}Sb, the latter of which is produced via a \ce{^{119m}Te} generator, are promising radionuclides for the targeted treatment of both osteoarthritis and small mass tumors via Auger therapy.
Experiments were conducted at Lawrence Berkeley National Laboratory, Los Alamos National Laboratory, and Brookhaven National Laboratory to measure the
\ce{^{nat}Sb}(p,x)\ce{^{117m}Sn} and \ce{^{nat}Sb}(p,x)\ce{^{119m}Te} cross sections for incident proton energies up to 200\,MeV. 
Additional measurements for co-produced isotopes are included as well.
In addition to this dataset, this paper investigates improvements for proton-induced reaction modeling capabilities through comparison of these experimental dataset against theoretical models in TALYS 1.95. 
Parameter adjustments affecting level density, optical model potential, and pre-equilibrium emission were explored, with a goodness-of-fit metric established by the largest independent cross section channels and cross-validated with remaining channels.

\end{abstract}

\maketitle

    \section{\label{sec:Introduction}Introduction\protect}
        Radionuclides have been used for medical applications since the early $\mathrm{20^{th}}$ century \cite{obaldo_early_2021}.
Theranostics, a portmanteau of \enquote{therapy} and \enquote{diagnostic}, are pharmaceutical radionuclides that selectively target and bind to treatment areas and emit radiation to facilitate either imaging and monitoring the drug pathway or deposition of a targeted dose to treat the underlying condition \cite{kelkar_theranostics_2011}.

Ideal candidates for theranostic use have a sufficient half-life ($\mathrm{t_{1/2}}$) for administration, precise radiotoxicity to target sites with minimal damage to healthy tissue, ease of radiolabeling to a targeting vector, and a feasible production pathway \cite{engle_recommended_2019}.
Tin-117m (\ce{^{117m}Sn}) is of particular interest due to its emission of low energy Auger electrons, conversion electrons (CE) ranging from 126.82\,(3)\,keV to 155.88\,(3)\,keV, and a characteristic decay gamma at 158.56\,(2)\,keV \cite{blachot_nuclear_2002}. 
The emitted CE from \ce{^{117m}Sn} have a sub-mm range in tissue \cite{berger_estar_2005} and the decay gamma has comparable imaging capabilities as the 140.511\,(1)\,keV gamma from Technicium-99m (\ce{^{99m}Tc}) \cite{browne_nuclear_2017,krishnamurthy_tin-117m4dtpa_1997}.
These unique characteristics position \ce{^{117m}Sn} for dual use as a therapeutic and diagnostic radioisotope.

Similarly, Antimony-119 (\ce{^{119}Sb}) produces CE and Auger electrons in the range of 19.405\,(8)\,keV to 22.986\,(8)\,keV \cite{symochko_nuclear_2009}, providing high precision dose deposition with a range of less than 10\,\mmicro m. 
Due to the short half-life of \ce{^{119}Sb}, practical clinical applications require an isotope generator. 
Tellurium-119m (\ce{^{119m}Te}), the parent product, can be used as a generator since it will decay to \ce{^{119}Sb}.

This paper presents cross section measurements for these isotopes via proton bombardment on natural antimony (\ce{^{nat}Sb}) from stacked target experiments at Lawrence Berkeley National Laboratory (LBNL), Los Alamos National Laboratory (LANL), and Brookhaven National Laboratory (BNL), covering 43 energy points for incident proton energies up to 200\,MeV. 
Along with \ce{^{117m}Sn} and \ce{^{119m}Te}, an additional 22 \ce{^{nat}Sb}(p,x) residual product cross sections are included.
Residual product cross sections are also provided for proton-induced reactions on natural copper, titanium, and niobium (\ce{^{nat}Cu} and \ce{^{nat}Ti}, and \ce{^{nat}Nb} respectively).

In addition, following the efforts of Fox\,\etal\ and Morrell\,\etal \cite{fox_investigating_2021, fox_measurement_2021,morrell_measurement_2024}, this experimental dataset was used as a comparator against theoretical modeling capabilities, with systematic and physics-informed adjustments made to level density, optical model potential, and pre-equilibrium emission parameters. 
This optimization was validated by a goodness-of-fit comparison with the largest independent cross sections based on qualitative and quantitative checks and confirmed against smaller independent and cumulative cross sections. 
    
    \section{\label{sec:ExpMethod}Experimental Method\protect}
        The experiments detailed in this work were carried out at the LBNL 88-Inch Cyclotron (88") for incident proton energies ($\mathrm{E_p}$)\,$\mathrm{<55\,MeV}$, the LANL Isotope Production Facility (IPF) for $\mathrm{E_p<100\,MeV}$ and the BNL Brookhaven Linac Isotope Producer (BLIP) for $\mathrm{100\,MeV<E_p<200\,MeV}$.
The experiments in each facility utilized a stacked target measurement technique \cite{graves_nuclear_2016, voyles_excitation_2018, morrell_measurement_2020,voyles_proton-induced_2021, fox_investigating_2021, fox_measurement_2021}, in which a stack of thin target foils is placed sequentially along the incident particle beam.
This technique offers the advantage of measuring cross section data across a range of energies during a single experiment.
Each target foil is accompanied by "monitor" foils with robust, evaluated cross section measurements - used to calculate the current and energy for the target foil. 
Each set of target and monitor foils is separated by a material that reduces incident particle energy, here referred to as "degraders".

The target material was \ce{^{nat}Sb}, accompanied by \ce{^{nat}Cu}, \ce{^{nat}Ti}, and \ce{^{nat}Nb} monitor foils to characterize proton-induced reactions based on recommended cross section data provided by the IAEA \cite{hermanne_reference_2018}. 
Full details on the composition of each target stack are provided in Appendix \ref{app:FoilCharacterization}.
At the time of analysis, \ce{^{nat}Nb} monitor reactions had not been published \cite{tarkanyi_extension_2024}.
The following sections detail the setup and irradiation at each experimental facility. 
\subsection{\label{sec:StackedTarget}Stacked Target Design and Irradiations}
        \subsubsection{\label{LBNLIrrad2020}Lawrence Berkeley National Laboratory}
            Two measurements were conducted in 2020 at LBNL's 88" Cyclotron \cite{kireeff_covo_88-inch_2018} with incident proton energies of 55\,MeV and 35\,MeV.
            The two stacks were comprised of 10 of each foil: 25\,\mmicro m\,\ce{^{nat}Cu} (99.95\%, Lot \#300711914, Goodfellow Cambridge Ltd., Huntingdon PE29 6WR United Kingdom), 25\,\mmicro m\,\ce{^{nat}Ti} (99.6+\%, Lot \#300711931, Goodfellow Cambridge Ltd.), 25\,\mmicro m\,\ce{^{nat}Nb} (99.8\%, Lot \# T23A035, Alfa Aesar, Ward Hill, MA 01835 USA), and 25\,\mmicro m\,\ce{^{nat}Sb} on a 25\,\mmicro m\,Permanent Polyester backing with 25\,\mmicro m\,Bostick 1430 adhesive layer (95+\%, Lot \#300858965, Goodfellow Cambridge Ltd.).

\begin{figure}[htbp]
                \centering
                \subfloat[Mounted \ce{^{nat}Sb} foil.]{
                    \includegraphics[width=.45\linewidth]{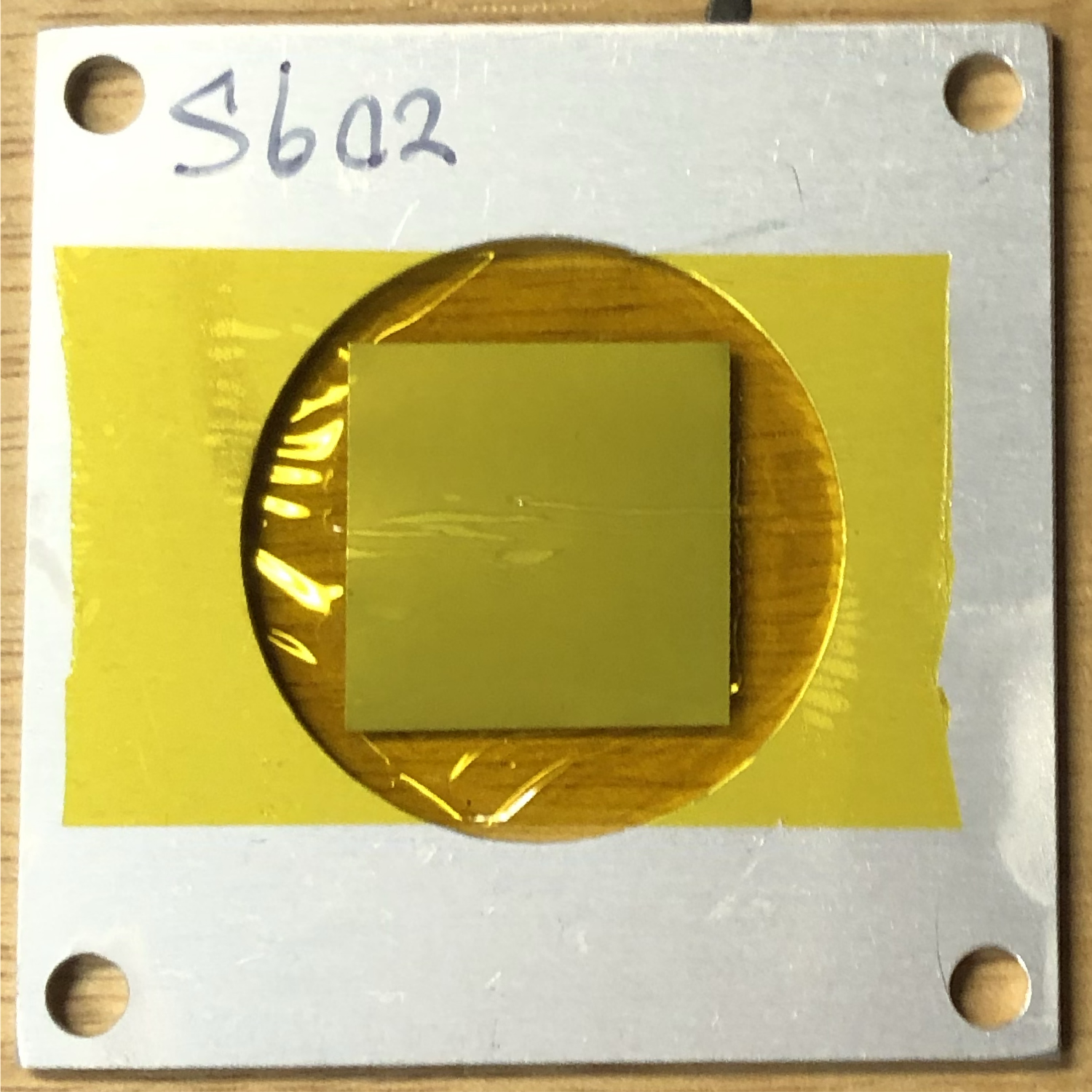}
                    \label{fig:Sb_foil}
                }
                \subfloat[Mounted \ce{^{nat}Nb} foil.]{
                    \includegraphics[width=.45\linewidth]{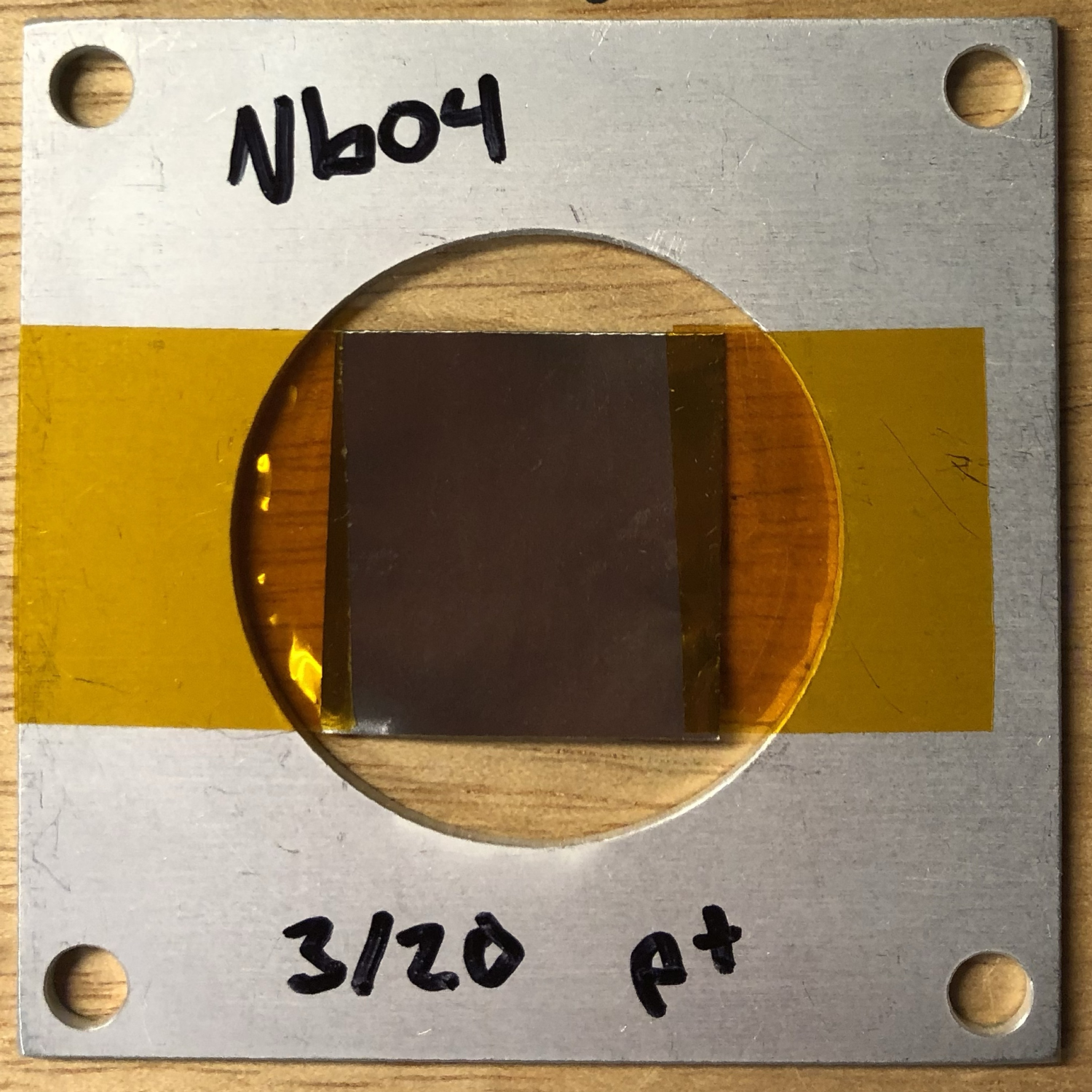}
                    \label{fig:Nb_foil}
                }
                \caption{Representative mounted foils prior to irradiation.}
                \label{fig:foil_mounts}
            \end{figure}

            The adopted value of 95\% Sb purity was based on the Goodfellow XRF results using a Fischerscope X-RAY XDV-SDD spectrometer. 
            Foils were cut to approximately 2.54\,cm\,x\,2.54\,cm, cleaned with isopropyl alcohol, and characterized by taking 4 measurements each of length and width with digital calipers (Mitutoyo America Corp.), thickness using a digital micrometer (Mitutoyo America Corp.), and mass to 0.1\,mg precision on an analytical balance (Mettler Toledo AL204).
            \ce{^{nat}Sb} foils were sealed with kapton tape and mounted on 1.5\,mm thick acrylic or aluminum frames.
            \ce{^{nat}Nb}, \ce{^{nat}Cu}, and \ce{^{nat}Ti} foils were mounted to the frames using kapton outside the beam strike area.
            Representative examples of this mounting procedure are shown in \autoref{fig:foil_mounts}.
            The 10 sets of foils were separated by \ce{^{nat}Al} degraders.
            Foils and degraders were loaded into the beam box as seen in \autoref{img:LBNL_Target_Box}.
            In addition, measured and characterized stainless steel plates were placed in the front and back of the target stack.
            These plates were used to assess the beam profile using radiochromic film (Gafchromic EBT3).
            The current was not steady across the irradiation periods, requiring the use of reaction rate in the subsequent analysis.
            The 55\,MeV irradiation was carried out for 8961\,s, with 4670\,s of the beam on target.
            The 35\,MeV irradiation ran without interruption for 3595\,s.
            
        \subsubsection{\label{LANLIrrad}Los Alamos National Laboratory}
            The LANL target stack consisted of 10 of each: 25\,\mmicro m\,\ce{^{nat}Cu}, 25\,\mmicro m\,\ce{^{nat}Ti}, 25\,\mmicro m\,\ce{^{nat}Nb} foils, and 25\,\mmicro m\,\ce{^{nat}Sb} foils adhered to polyester backings, sourced from the same lots and prepared as described in \autoref{LBNLIrrad2020}.
            \ce{^{nat}Al} degraders were used to separate the sets of foils, with stainless steel plates at the front and back of the stack to monitor the beam profile. 
            Detailed characterization of degraders \cite{fox_investigating_2021}, beam box schematics, and upstream beamline components \cite{obrien_novel_2020} was incorporated into the calculations of incident particle energy loss.
            Foils were mounted on 1\,mm thick acrylic frames and frame sets were bound with baling wire to facilitate removal via remote manipulators in the IPF hot cell.
            The beam box with foil packs is shown in \autoref{img:LANL_Target_Box}.
            The stack was irradiated with a 100\,MeV incident proton beam for 3594\,s with a steady 200\,nA nominal current measured via an inductive pickup.
            
        \subsubsection{\label{sec:BNLIrrad}Brookhaven National Laboratory}
\begin{figure}[!!!htbp]
                \centering
                \subfloat[LBNL]{
                    \includegraphics[width=0.29\linewidth]{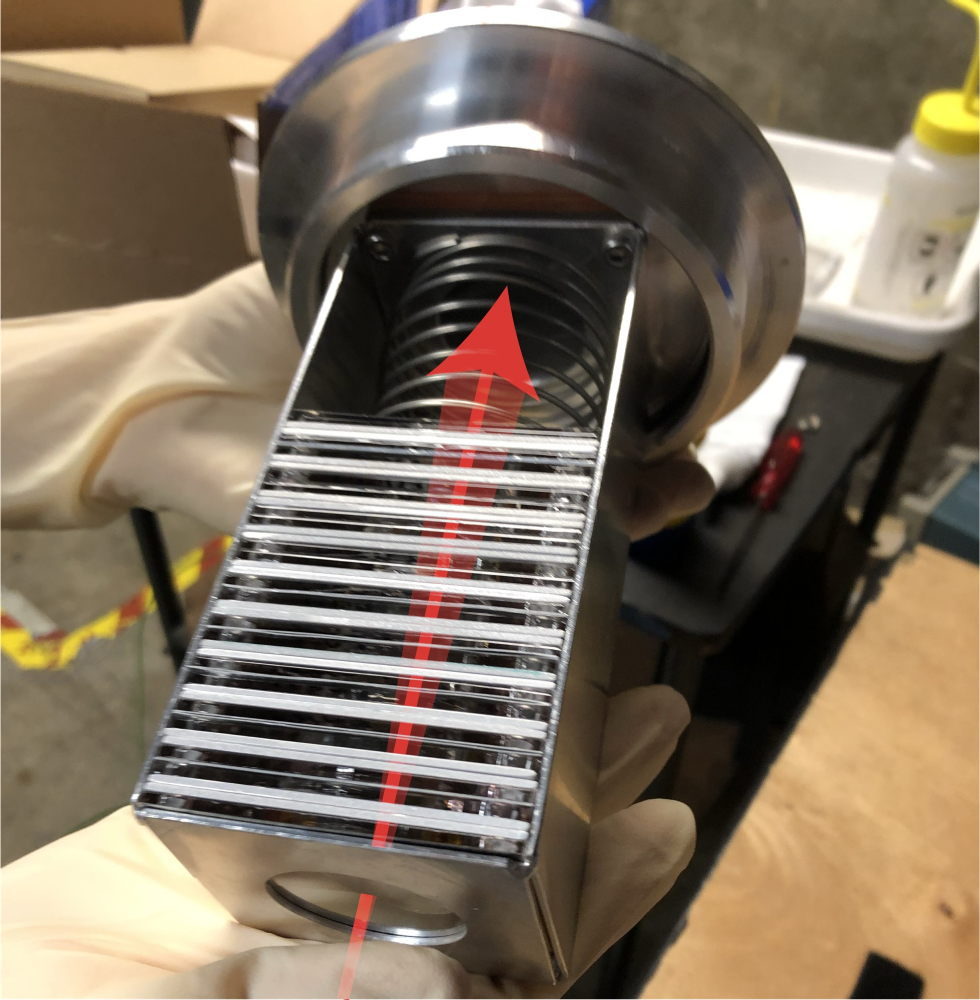}
                    \label{img:LBNL_Target_Box}
                }
                \subfloat[LANL]{
                    \includegraphics[width=0.29\linewidth]{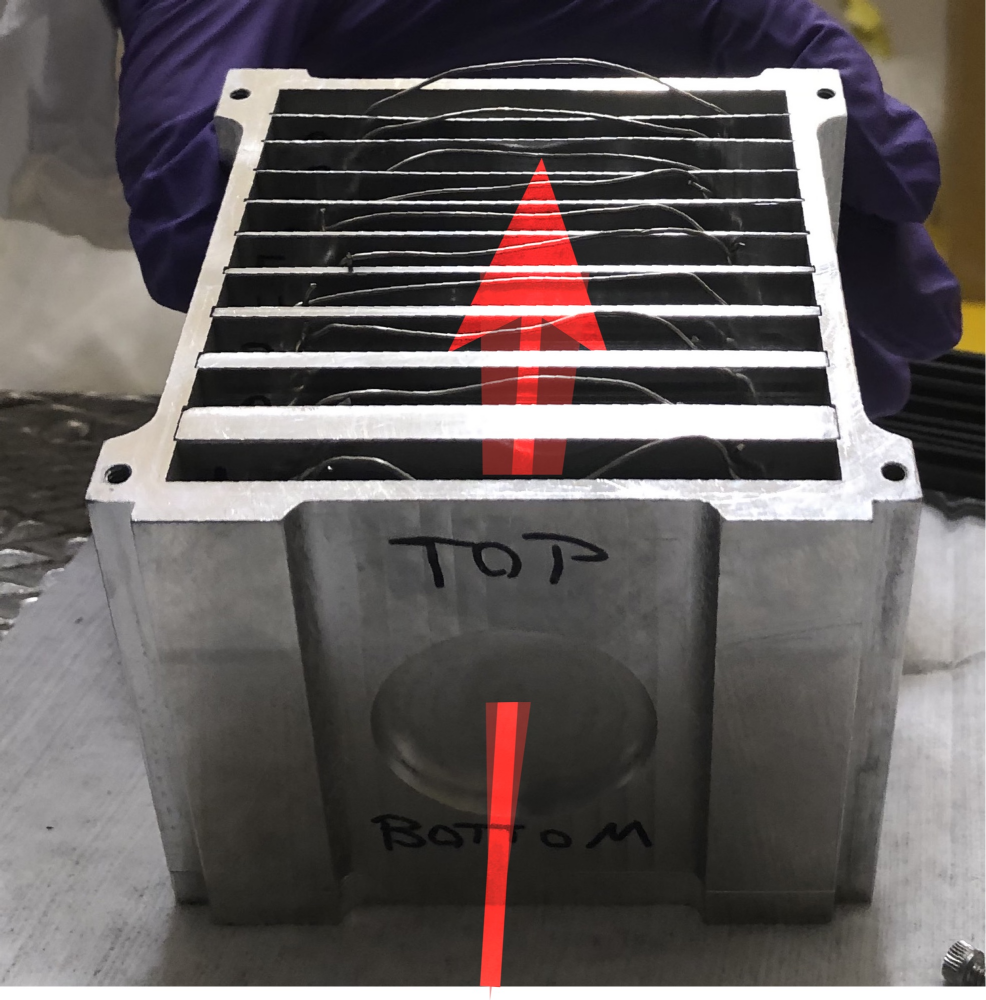}
                    \label{img:LANL_Target_Box}
                }
                \subfloat[BNL]{
                    \includegraphics[width=0.29\linewidth]{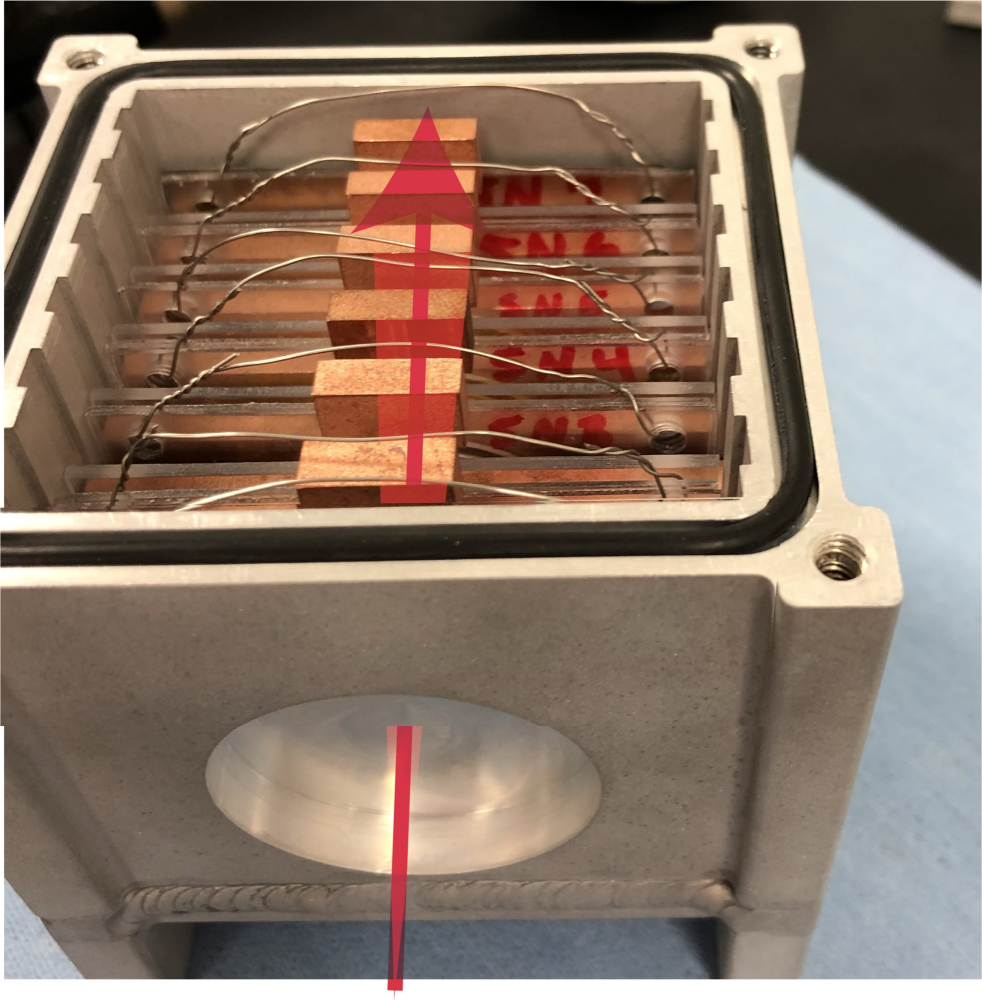}
                    \label{img:BNL_Target_Box}
                }
                \caption{Target boxes for the irradiations. Red arrows indicate beam direction.}
                \label{fig:beam_boxes}
            \end{figure}

            The BNL stack consisted of 7 of each: 25\,\mmicro m\,\ce{^{nat}Cu} and 25\,\mmicro m\,\ce{^{nat}Nb} foils, and  25\,\mmicro m\,\ce{^{nat}Sb} targets adhered to polyester backings, originating from the same batches as \autoref{LBNLIrrad2020} and characterized and prepared in the same manner.
            The stack included six \ce{^{nat}Cu} degraders and stainless steel plates at the front and back of the stack to assess the beam profile.
            Foils were mounted on 0.5\,mm thick acrylic frames.
            The individual target and monitor foil frame sets were held together with baling wire to facilitate removal after irradiation in the BLIP hot cell.
            The beam box is shown in \autoref{img:BNL_Target_Box}.
            The stack was irradiated for 7362\,s with an incident 200\,MeV proton beam with a steady 150\,nA nominal current, measured using toroidal beam transformers.
        \subsubsection{\label{LBNLIrrad2022}Lawrence Berkeley National Laboratory (2022)}
            With a number of residual products decaying through the 158.562\,keV gamma shared by \ce{^{117m}Sn} \cite{blachot_nuclear_2002} and a previously unresolvable 156.02\,keV decay gamma for \ce{^{117m}Sn}, an additional irradiation was performed at LBNL to be assayed using a Low-Energy Photon Spectrometer (LEPS).
            The potential activation products with decay gammas in this energy region are provided in \autoref{tab:158.6_keV}.
            This subsequent run validated the characterization of \ce{^{117m}Sn} production around the reaction threshold of its primary contaminant,\ce{^{113}Sn}.
            \begingroup
            \squeezetable
            \begin{longtable}{ccccc}
                \caption{Isotopes With Direct $\gamma$ Decays Around 158.6\,keV.} \\
                \label{tab:158.6_keV} \\
                \hline\hline
                Parent & Half life & Decay mode & E$_{\gamma}$\,(keV) & I$_{\gamma}$\,(\%) \Tstrut\Bstrut\\
                \hline
                \endfirsthead
                \hline\hline
                Parent & Half life & Decay mode & E$_{\gamma}$\,(keV) & I$_{\gamma}$\,(\%) \Tstrut\Bstrut\\
                \hline
                \endhead
                \ce{^{119m}Te} \cite{symochko_nuclear_2009} & 4.70\,(4)\,d & EC$+\beta^+$ & 153.59\,(3) & 66\,(3)\% \Tstrut\\
                \ce{^{117m}Sn} \cite{blachot_nuclear_2002} & 13.60\,(4)\,d & IT & 156.02\,(3) & 2.113\,(6)\% \\
                \ce{^{116}Te} \cite{blachot_nuclear_2010} & 2.49\,(4)\,h & EC+$\beta^+$ & 157.14\,(9) & 0.435\,(21)\% \\
                \ce{^{117m}In} \cite{blachot_nuclear_2002} & 116.2\,(3)\,m & IT+$\beta^-$ & 158.562\,(12) & 15.9\,(16)\% \\
                \ce{^{117}Sb} \cite{blachot_nuclear_1998} & 2.80\,(1)\,h & EC+$\beta^+$ & 158.562\,(12) & 85.9\,(4)\% \\
                \ce{^{117m}Sn} \cite{blachot_nuclear_2002} & 13.60\,(4)\,d & IT & 158.562\,(12) & 86.4\,(4)\% \\
                \ce{^{117}In} \cite{blachot_nuclear_2002} & 43.2\,(3)\,m & $\beta^-$ & 158.562\,(12) & 87\,(9)\% \\
                \ce{^{123m}Te} \cite{chen_nuclear_2021} & 119.7\,(1)\,d & IT & 158.97\,(5) & 84.3\,(3)\%  \Bstrut\\
                \hline\hline
            \end{longtable}
            \endgroup
            The 2022 LBNL target stack consisted of 6 of each: 25\,\mmicro m\,\ce{^{nat}Cu} (99.95\%, Lot \#300958705, Goodfellow Cambridge Ltd.),
            25\,\mmicro m\,\ce{^{nat}}Ti (99.6+\%, Lot \#300942464, Goodfellow Cambridge Ltd.), and 25\,\mmicro m\,\ce{^{nat}Sb} adhered to polyester backings, from the batch listed in \autoref{LBNLIrrad2020}.
            The stack included \ce{^{nat}Al} degraders and stainless steel plates to assess the beam profile and was prepared as described in \autoref{LBNLIrrad2020}.
            Irradiation lasted for 3511\,s with an incident 55\,MeV proton beam with a stable 150\,nA nominal current.
    \subsection{\label{sec:GammaSpec}Decay Gamma Spectroscopy and Detector Calibration Sources}
        Post-irradiation, activated foils were assayed on location-specific HPGe detector setups.
        The initial two irradiations at LBNL utilized four ORTEC IDM-200-V and one ORTEC GMX-50220-S detectors to assay foils.
        Counting began roughly 25 minutes after the end of bombardment (EoB) and continued over the following 40 days.

        Decay spectroscopy at LANL began approximately 3\,hours following EoB using an ORTEC IDM-200-V and an ORTEC GEM20P-PLUS detector.
        Activated foils were counted for 3\,days and then relocated to the LANL TA-48 Countroom, where measurements continued for an additional 14\,days on 3 HPGe detectors, numbered 26, 43, and 48.
        The foils were then moved to four ORTEC IDM-200-V and one ORTEC GMX-50220-S detector for an additional 21 days, with a final measurement done on two ORTEC IDM-200-V detectors approximately 10 months later.
        
        At BNL, foil assay began roughly 2 hours after EoB using an ORTEC GMX-13180 and an ORTEC IDM-200-V detector.
        After several hours, foils were moved to an ORTEC GEM20P4-70-PL, an ORTEC IDM-200-V, and an ORTEC tran-SPEC-N detector, where they were measured for approximately four weeks.
        The foils were then relocated to four ORTEC IDM-200-V and one ORTEC GMX-50220-S detector for an additional 2 weeks of counting.
        
        Assay on the foils from the 2022 experiment at LBNL began less than 15\,minutes after EoB.
        These measurements were performed on an ORTEC IDM-200-V, an ORTEC GMX-50220-S, and an ORTEC GLP-36360/13-P-S LEPS.
        The ORTEC GLP-36360/13-P-S LEPS was later replaced with an ORTEC GLP-32340/13P4-SMN LEPS.
        Foils were counted for approximately 2 months, with additional measurements taken approximate one year later on the ORTEC GLP-32340/13P4-SMN LEPS.
             
    Detector energy and efficiency calibrations were calculated using 1\,\mmicro Ci sealed source calibration standards from Eckert \& Ziegler (Eckert \& Ziegler, Valencia, CA 91355 USA) at all locations excluding the LANL Countroom, where calibration was provided by staff. Calibrations were performed at distances ranging from 10\,cm to 80\,cm. The sources used included \ce{^{22}Na}, \ce{^{57}Co}, \ce{^{60}Co},\ce{^{109}Cd}, \ce{^{133}Ba}, \ce{^{137}Cs}, \ce{^{152}Eu}, \ce{^{210}Po}, \ce{^{226}Ra}, and \ce{^{241}Am}.
        
    \section{\label{sec:Analysis}Data Analysis\protect}
        Residual product cross sections were derived from gamma-ray spectra collected during the experiments, leveraging the CURIE python toolkit \cite{morrell_curie_2023} for detector calibration, peak fitting, and reaction rate calculations.
Proton energy and current was characterized for each target foil using the recommended monitor reactions from the 2017 IAEA CRP evaluation \cite{hermanne_reference_2018}. 
A variance minimization technique, employed in several previous stacked target experiments \cite{graves_nuclear_2016, voyles_excitation_2018,voyles_proton-induced_2021, fox_measurement_2021, fox_investigating_2021,morrell_measurement_2020, morrell_measurement_2024}, reduced the uncertainty in the current and energy profile caused by proton range straggling.
The data were then used to compute the flux-weighted average cross sections $\tilde{\sigma}$ for the residual products in each foil.
The flux-weighted average cross section is an integral cross section that accounts for incident proton energy broadening, seen in \autoref{eq:weightedxs}. 
       
        \begin{equation}
            \tilde{\sigma_i}=\dfrac{\int{\sigma_i(E)\phi(E)dE}}{\int{\phi(E)dE}}
            \label{eq:weightedxs}
        \end{equation}
        
        where $\phi(E)$ is the proton flux in a given foil, calculated using the Anderson and Ziegler methodology \cite{ziegler_srim_2010,morrell_curie_2023}.

Assuming a negligible loss of current and energy within a given foil, this calculation simplifies to a thin-target activation equation, seen in \autoref{eq:xs_current_r}.
        \begin{equation}
            \tilde{\sigma_i}=\dfrac{R_i}{I_{p}(\rho_N\Delta r)}
            \label{eq:xs_current_r}
        \end{equation}
Where $R_i$ is the calculated reaction rate for a residual product $i$, $I_p$ is the proton beam current, and $\rho_N\Delta r$ is the areal density of the foil.

\subsection{\label{sec:PeakfittingCalib}Calibration and Peak Fits}
        Detector energy, efficiency, and resolution calibrations were calculated in CURIE using spectra obtained from sealed sources for all locations with the exception of the LANL TA-48 Countroom, where detector calibration was provided by the staff.
        CURIE employs a quadratic fit for energy calibration, a built-in routine for peak fitting for resolution, and a semi-empirical formula based on the work of Vidmar\,\etal for efficiency calibration \cite{vidmar_semi-empirical_2001}. 
        
        CURIE's peak fitting function was further used to fit the peaks from residual products in the activated foils. 
        Despite the superior resolution of HPGe detectors, the large number of residual products and their decay gamma-rays led to several partial or fully overlapping peaks in the spectra.
        A primary isotope of interest, \ce{^{117m}Sn}, shares its 158.6\,keV peak with several other residual products in the A=117 drip line as well as other products with decay gammas within a few keV. 
        To address this, the decay gammas and their intensities, as well as the half-lives of all potential products, were systematically reviewed as outlined below to reduce potential uncertainties from overlapping peaks in the spectra.
        In general, decay gammas were divided into three categories for analysis - fully included, fully excluded, or included only for spectra taken after a specific time dictated by the competing products' half-lives.
        This was done manually, since several factors influenced the downselection of gammas, including the number of observable decay gammas for a residual product, gamma intensities, and the option for observation via progeny in secular equilibrium.
        
\begin{figure*}[htbp]
            \centering
            \includegraphics[width=.9\linewidth]{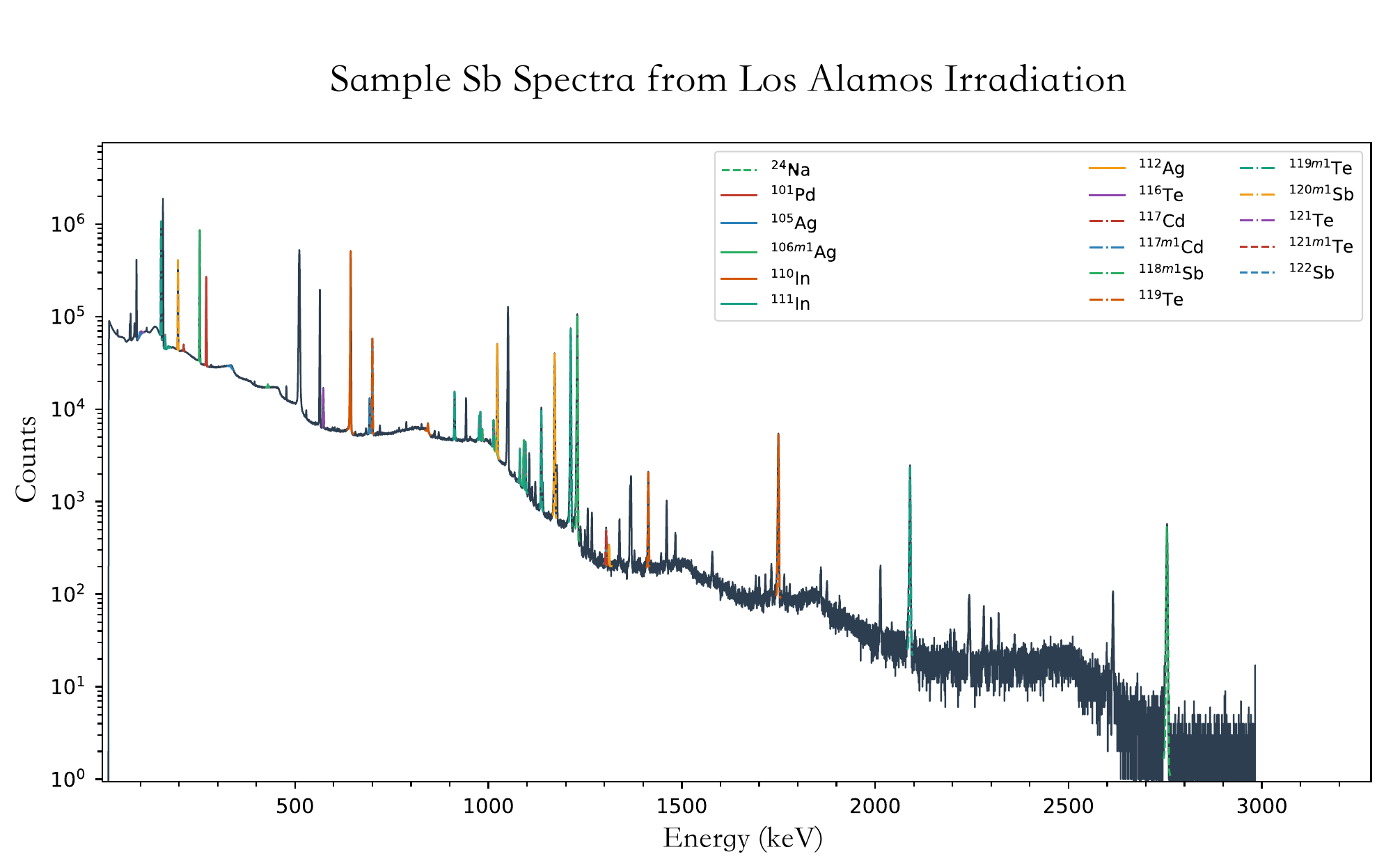}
            \caption{\ce{^{nat}Sb} spectrum taken on an ORTEC GEM20P-PLUS HPGe detector at LANL. While more products were fit than are provided in the results, their cross sections could not be confidently calculated.}
            \label{fig:LANLSbSpectra}
        \end{figure*}

        The characteristic decay gammas used to measure residual products from \ce{^{nat}Sb}(p,x) reactions are listed in Appendix \ref{app:Sb_product_data} and \ce{^{nat}Ti}, \ce{^{nat}Cu}, and \ce{^{nat}Nb} residual product gammas are provided in Appendix \ref{app:monitor_product_data}.
        A sample gamma spectrum from an activated \ce{^{nat}Sb} foil is shown in \autoref{fig:LANLSbSpectra}.
        \subsection{\label{sec:production}Reaction Rate Calculation}
\begin{figure}[ht]
              \centering
              \includegraphics[width=\linewidth]{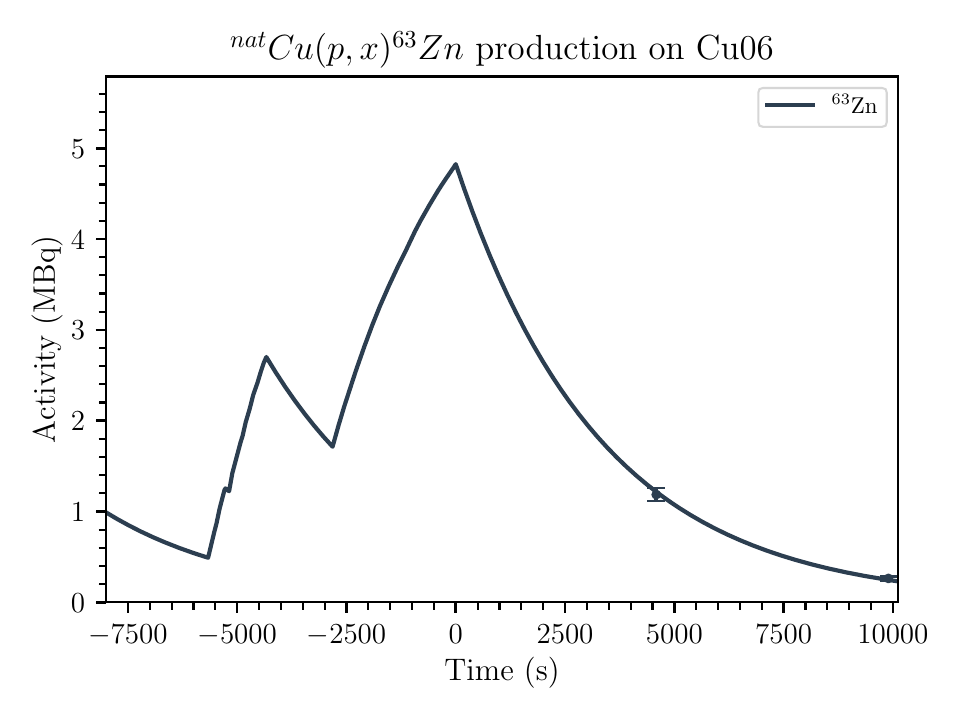}
              \caption{\ce{^{63}Zn} production and decay, calculated based on beam current and the measured decay gammas.}
              \label{fig:63ZnProdDecay}
          \end{figure}

        This analysis leveraged the nominal reaction rate ($R$), also referred to as the production rate, instead of activity at EoB ($A_0$) to accommodate fluctuations in the incident beam current.
          These fluctuations were most significant during the LBNL 55\,MeV irradiation in 2020, discussed in \autoref{LBNLIrrad2020}. 
          Calculated activities at EoB for shorter-lived isotopes were greatly impacted by beam loss, as demonstrated in the plot of \ce{^{nat}Cu}(p,x)\ce{^{63}Zn} production during the 55\,MeV run at LBNL in \autoref{fig:63ZnProdDecay}. 
          Here, t=0 is EoB.

        Decays of a residual product at time $t$, ($D(t)$) were calculated using \autoref{eq:Activityt1}.
        \begin{equation}
        D(t)=\dfrac{N_t F_{att}}{I_\gamma \epsilon_D (t_L/t_R)}
        \label{eq:Activityt1}
        \end{equation}
        where $N_t$ is the net counts in a given peak in the spectrum, $F_{att}$ is a correction factor for photon attenuation, $I_\gamma$ is gamma intensity, $\epsilon_D$ is detector efficiency as a function of energy and distance, and $t_L$/$t_R$ is the ratio of live time ($t_L$) to real time ($t_R$), accounting for detector dead time.
        Photon attenuation was based on data from the XCOM-Photon Cross Section Database \cite{seltzer_xcom-photon_1987}.
        
        The reaction rate $R_i$ for a residual product $i$ is calculated using \autoref{eq:reactionrate}.
        \begin{equation}
        R_i = \frac{D_i(t) e^{\lambda t}}{t_{irr} (1 - e^{-\lambda t_{irr})}}
        \label{eq:reactionrate}
        \end{equation}
        where $\lambda$ is the decay constant, $t$ is the time since EoB, and $t_{irr}$ is the irradiation time.
        
        If a residual product was fed by the decay of other residual nuclides, the calculated reaction rate during irradiation could be determined utilizing a higher order Bateman equation. 
        With multiple measurements of each residual product, a more precise $R$ could be confidently determined through regression analysis in CURIE.
        When possible, an independent $R$ was calculated; for complex decay chains, or for isotopes with a lack of measurable gammas in the decay chain, a cumulative value was calculated.
        
        \subsection{\label{sec:BeamCurrent}Characterizing Beam Current and Incident Particle Energy}
        Energy and range straggling of protons traversing the experimental stacks introduce uncertainties in the calculated beam energy and current.
        The empirical Anderson \& Ziegler formula for stopping power provides a predicted energy and current in the target stack materials, but a more accurate assessment can be made using monitor foils and their evaluated reactions.
        The monitor reactions used in these experiments were based on the 2017 IAEA CRP evaluation \cite{hermanne_reference_2018}, with the following reactions used at each site:
    
        \begin{itemize}
            \item{LBNL: \\
            \indent \ce{^{nat}Cu}(p,x): \ce{^{56}Co}, \ce{^{58}Co}, \ce{^{62}Zn}, \ce{^{63}Zn}, \ce{^{65}Zn}\\
            \indent \ce{^{nat}Ti}(p,x): \ce{^{46}Sc}, \ce{^{48}V}}
            \item{LANL: \\
            \indent \ce{^{nat}Cu}(p,x): \ce{^{56}Co}, \ce{^{58}Co}, \ce{^{62}Zn}, \ce{^{65}Zn}\\
            \indent \ce{^{nat}Ti}(p,x): \ce{^{46}Sc}, \ce{^{48}V}}
            \item{BNL: \\
            \indent \ce{^{nat}Cu}(p,x): \ce{^{58}Co}}
        \end{itemize}
        The scandium-46 (\ce{^{46}Sc}) experimental data deviated from IAEA evaluated values for energies approaching 100\,MeV, as seen in Appendix \ref{app:monitorXstable}. 
        
         With known cross sections $\tilde{\sigma_i}$, reaction rates $R_i$, and foil areal densities $(\rho_N\Delta r)$, \autoref{eq:xs_current_r} can be rearranged to calculate the proton beam current ($I_{p,i}$) for the $i^{th}$ monitor reaction as follows in \autoref{eq:currentR}.
    
        \begin{equation}
            I_{p,i}=\dfrac{R_i}{(\rho_N\Delta r)\tilde{\sigma_i}}
            \label{eq:currentR}
        \end{equation}
        
\begin{figure}[ht]
    \centering
    \subfloat[]{
    \includegraphics[width=\linewidth]{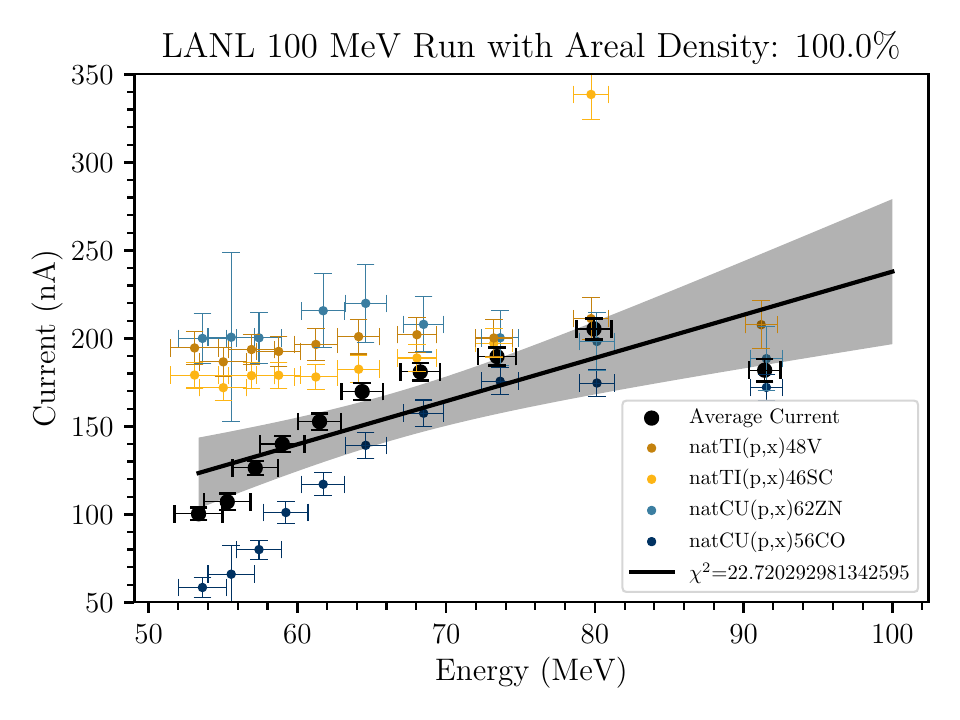}
    \label{plt:LANL_ad_1.0}
    }
    \vspace{0.01em} 
    \subfloat[]{
    \includegraphics[width=\linewidth]{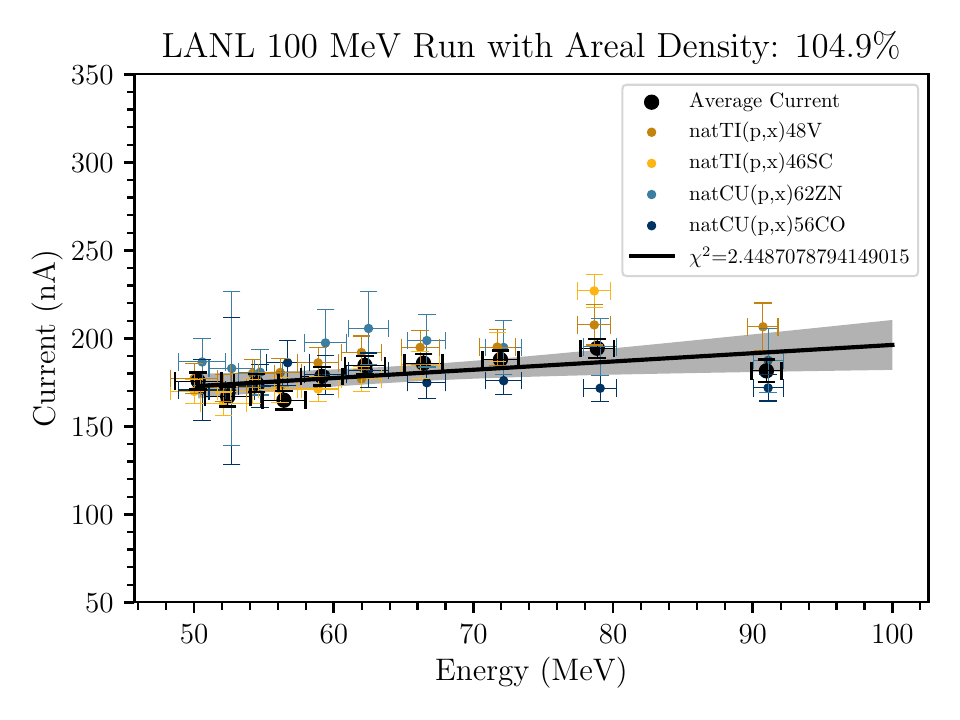}
    \label{plt:LANL_ad_1.049}
    }
    \caption{LANL calculated current based on monitor reactions, before and after variance minimization using a nominal areal density adjustment of 4.9\%.}
\end{figure}

\begin{figure}[ht]
        \centering
          \includegraphics[width=\linewidth]{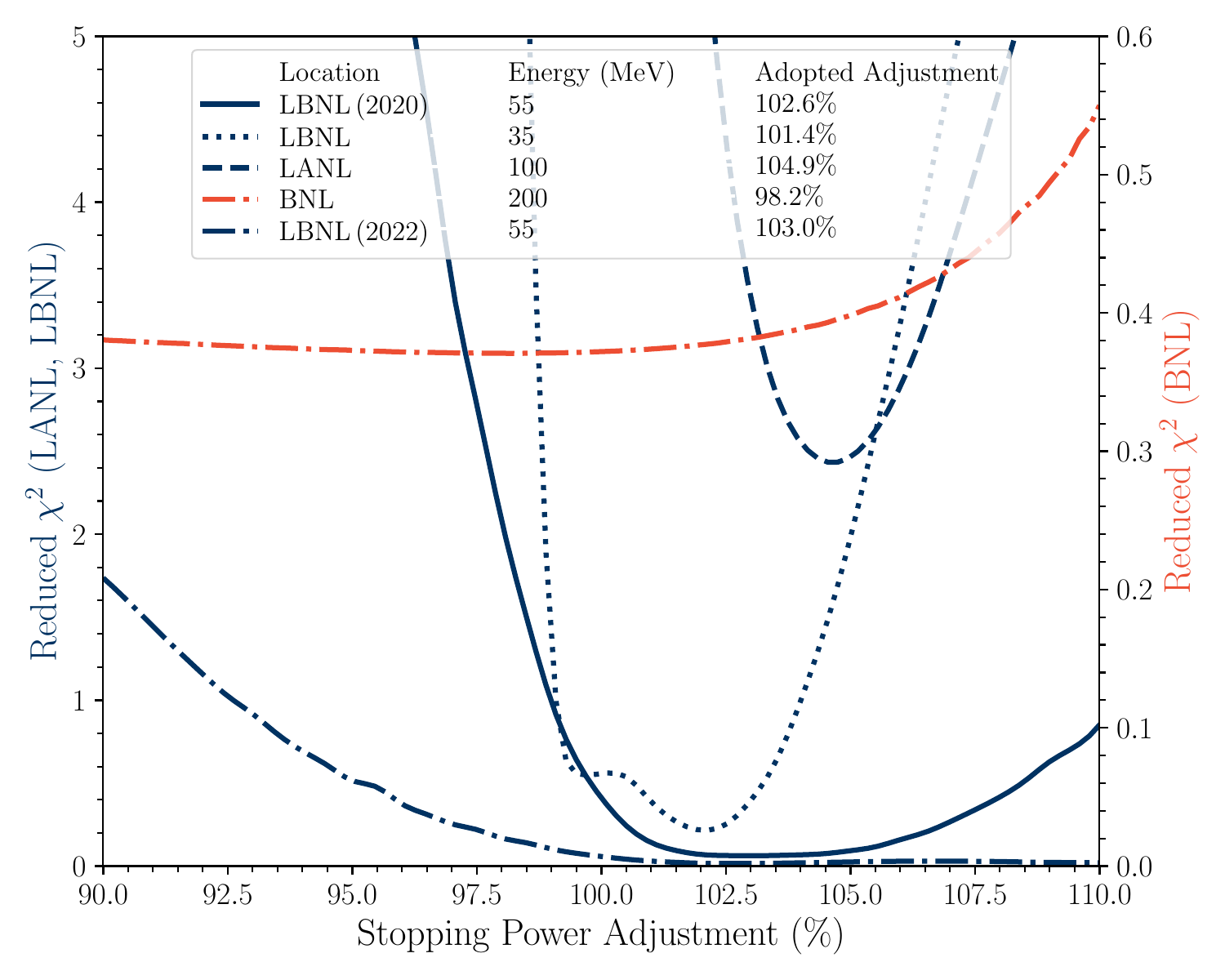}
            \caption{Local minima in $\mathrm{\chi^2}$ between the observed and modeled monitor activation data for the various experiments after exploring a $\pm10\%$ change in density.
            }
            \label{plt:Variance_Minimization_all}
        \end{figure}

        A comparison of the calculated currents from multiple monitor reactions diverges as the beam progresses through the stack as visualized in \autoref{plt:LANL_ad_1.0}.
        To address this, and following a procedure introduced in previous work \cite{graves_nuclear_2016, voyles_excitation_2018,fox_measurement_2021,morrell_measurement_2020}, a variance minimization technique was employed.
        This technique nominally adjusts the density of the materials in the stack using a global multiplier in the range of $\pm$ 10\%.
        The correlated uncertainties were quantified using a robust sandwich estimator \cite{huber_behavior_1967} using correlation metrics detailed in Voyles\,\etal \cite{voyles_proton-induced_2021}.
        The optimal value of this global multiplicative factor was calculated using a reduced $\mathrm{\chi^2}$ goodness-of-fit test, which resulted in a more consistent fit across the stack as seen in \autoref{plt:LANL_ad_1.049}.
        The minimum $\mathrm{\chi^2}$ values for each site are seen in \autoref{plt:Variance_Minimization_all}.
        

        The calculation of stopping power adjustment for the BNL experiment was based on a single monitor reaction, \ce{^{nat}Cu}\,(p,x)\,\ce{^{58}Co}. 
        The existing experimental data for \ce{^{nat}Cu}\,(p,x)\,\ce{^{56}Co} was leveraged as a validation check for the BNL results, but these data were not incorporated into the final calculations.
        This resulted in a relatively shallow minimum in reduced $\mathrm{\chi^2}$.

\begin{figure}[ht]
            \centering
            \includegraphics[width=\linewidth]{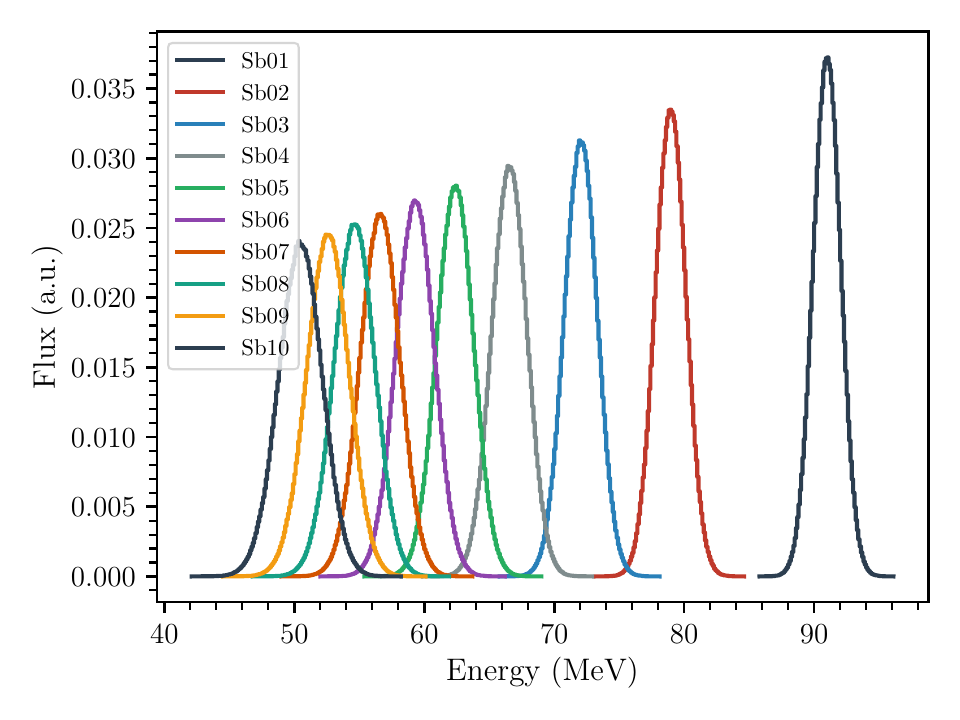}
            \caption{Energy and current distribution across the stack for \ce{^{nat}Sb} foils at LANL.}
            \label{fig:sb_flux_lanl}
        \end{figure}

        The application of this variance minimization technique provided the energy and flux for calculating the \ce{^{nat}Sb} residual product cross sections. An example of the \ce{^{nat}Sb} foil energies and flux is shown in \autoref{fig:sb_flux_lanl}.
        
    \subsection{\label{sec:Crosssections} Cross Section Measurements}
        With an optimized calculation of current and energy within each compartment, the cross sections of the residual products for proton-induced reactions were calculated using \autoref{eq:xs_current_r}.
        The cross section data presented in this paper are classified as independent \textit{(i)}, having been directly populated by the proton-induced reaction on the target material, or cumulative \textit{(c)}, where the total cross section includes direct production and contributions from the decay of co-produced isotopes. 
        For isotopes with measured isomeric states based on existing nuclear data, the naming convention specifies whether the measurement is specific to the ground state \textit{(g)}, the isomer \textit{(m)}, or both \textit{(m+g)}.
        
        A residual product labeled as both an independent channel and as an \textit{(m+g)} channel indicates that an isomeric state is present in the referenced nuclear structure data with a $t_{1/2}\,\geq\,1$\,s.
        The independent and cumulative cross section data for the \ce{^{nat}Sb}(p,x) residual products measured in these experiments are provided in \autoref{tab:sb_xs_data}.
        Additional cross section data for residual products via proton-induced reactions on \ce{^{nat}Nb}, \ce{^{nat}Ti}, and \ce{^{nat}Cu} are available in Appendix\,\ref{app:monitorXstable}.
The 13 independent cross sections and 9 cumulative residual product cross sections for \ce{^{nat}Sb}(p,x) reactions were calculated using characteristic decay gammas. 
In some cases, the decay gammas of daughter products in secular equilibrium with the residual product were used.
The measured \ce{^{nat}Sb}(p,x) cross sections are provided in \autoref{tab:sb_xs_data}.
Measured reactions on monitor foils are provided in Appendix \ref{app:monitorXstable}.

\begingroup
\squeezetable


\begin{tiny}

\begin{longtable*}{llllllll}

\caption{Independent and cumulative cross sections measured on Sb foils. Uncertainties are listed in NDS format. Cross sections are classified as independent \textit{(i)}, having been directly populated by the proton-induced reaction on the target material, or cumulative \textit{(c)}, where the total cross
section includes direct production and contributions from
the decay of co-produced isotopes}\\
\label{tab:sb_xs_data}\\

\endfirsthead
\hline \hline \\
\endhead

\hline \multicolumn{1}{r}{\raggedright Continued on next page}\\ \hline\hline
\endfoot

\hline \hline
\endlastfoot

\hline\hline
$E_p$\,(MeV)         & 188.511\,(2077) & 173.324\,(2192) & 159.924\,(2311) & 146.390\,(2451) & 131.800\,(2633) & 117.773\,(2848) & 102.724\,(3138) \Tstrut\Bstrut\\
Location             & BNL             & BNL             & BNL             & BNL             & BNL             & BNL             & BNL            \Bstrut \\\hline
\ce{^{89m}Nb_{(i)}} &                 &                 &                 & 0.698\,(88)     &                 &                 &                 \Tstrut\\
\ce{^{91m}Nb_{(c)}} & 0.068\,(42)     & 0.062\,(66)     & 0.048\,(25)     & 0.059\,(14)     & 0.062\,(47)     & 0.019\,(39)     &                 \\
\ce{^{106m}Ag_{(i)}} & 2.726\,(46)     & 1.969\,(34)     & 1.564\,(26)     & 1.044\,(27)     & 0.461\,(39)     & 0.263\,(50)     & 0.44\,(23)      \\
\ce{^{109(m+g)}In_{(c)}} & 19.08\,(86)     & 17.31\,(90)     & 14.08\,(46)     & 11.48\,(43)     &                 &                 & 2.9\,(30)       \\
\ce{^{111(m+g)}In_{(c)}} & 46.08\,(76)     & 42.20\,(57)     & 38.50\,(43)     & 32.85\,(38)     & 25.14\,(42)     & 20.87\,(33)     & 11.07\,(27)     \\
\ce{^{114m}In_{(i)}} & 11.73\,(20)     & 10.95\,(16)     & 10.72\,(12)     & 9.81\,(12)      & 9.15\,(12)      & 8.09\,(15)      & 7.22\,(18)      \\
\ce{^{109}Sn_{(c)}} & 17.72\,(76)     & 16.10\,(84)     & 13.06\,(55)     & 10.65\,(62)     &                 &                 & 2.73\,(67)      \\
\ce{^{110}Sn_{(c)}} & 9.5\,(18)       & 10.5\,(20)      & 7.51\,(27)      & 6.046\,(63)     & 4.5\,(16)       &                 &                 \\
\ce{^{111}Sn_{(c)}} & 45.68\,(75)     & 41.84\,(56)     & 38.16\,(42)     & 32.56\,(38)     & 24.92\,(42)     & 20.69\,(33)     & 10.97\,(27)     \\
\ce{^{113(m+g)}Sn_{(c)}} & 56.94\,(92)     & 57.17\,(77)     & 58.27\,(61)     & 54.91\,(58)     & 50.32\,(59)     & 38.17\,(58)     & 19.15\,(42)     \\
\ce{^{117m}Sn_{(c)}} & 19.99\,(53)     & 20.22\,(52)     & 21.59\,(29)     & 19.17\,(40)     & 19.72\,(41)     & 19.08\,(52)     & 17.20\,(58)     \\
\ce{^{116m}Sb_{(i)}} & 30.7\,(28)      & 35.0\,(39)      & 38.5\,(37)      & 42.5\,(32)      & 48.3\,(35)      & 53.5\,(44)      & 55.\,(12)       \\
\ce{^{118m}Sb_{(i)}} & 23.\,(12)       & 35.9\,(13)      & 34.2\,(72)      & 24.5\,(95)      & 28.9\,(81)      & 48.2\,(44)      & 36.\,(17)       \\
\ce{^{120m}Sb_{(i)}} & 25.65\,(52)     & 30.70\,(48)     & 32.51\,(39)     & 33.60\,(52)     & 37.50\,(59)     & 38.35\,(91)     & 42.7\,(13)      \\
\ce{^{122(m+g)}Sb_{(i)}} & 69.6\,(11)      & 74.13\,(99)     & 79.75\,(84)     & 84.10\,(89)     & 89.9\,(11)      & 94.9\,(14)      & 101.0\,(21)     \\
\ce{^{116}Te_{(i)}} & 14.1\,(52)      & 17.2\,(42)      & 18.7\,(45)      & 20.6\,(31)      & 28.7\,(29)      & 34.0\,(48)      & 59.0\,(31)      \\
\ce{^{117}Te_{(i)}} & 29.\,(15)       & 35.\,(18)       & 48.\,(30)       & 36.\,(15)       & 40.\,(18)       & 61.\,(26)       & 104.\,(38)      \\
\ce{^{118}Te_{(i)}} & 18.85\,(93)     & 17.67\,(70)     & 23.35\,(66)     & 23.41\,(46)     & 31.23\,(73)     & 35.9\,(12)      & 49.4\,(20)      \\
\ce{^{119g}Te_{(i)}} & 11.48\,(28)     & 11.79\,(21)     & 14.43\,(30)     & 14.31\,(42)     & 18.09\,(62)     & 19.45\,(34)     & 26.94\,(99)     \\
\ce{^{119m}Te_{(i)}} & 15.03\,(27)     & 16.99\,(25)     & 19.18\,(21)     & 21.44\,(26)     & 24.82\,(31)     & 28.59\,(46)     & 36.20\,(77)     \\
\ce{^{121g}Te_{(i)}} & 5.195\,(90)     & 6.112\,(96)     & 7.289\,(82)     & 7.642\,(87)     & 8.94\,(12)      & 9.62\,(15)      & 11.60\,(25)     \\
\ce{^{121m}Te_{(i)}} & 7.67\,(13)      & 8.95\,(16)      & 10.03\,(11)     & 10.84\,(12)     & 12.39\,(15)     & 13.87\,(22)     & 16.54\,(34)     \Bstrut\\\hline\hline
$E_p$\,(MeV)         & 91.038\,(1063)  & 78.958\,(1180)  & 72.045\,(1270)  & 66.546\,(1340)  & 62.366\,(1410)  & 59.283\,(1470)  & 56.576\,(1520)  \Tstrut\Bstrut\\
Location             & LANL            & LANL            & LANL            & LANL            & LANL            & LANL            & LANL            \Bstrut\\\hline
\ce{^{111(m+g)}In_{(c)}} & 5.15\,(14)      & 0.1596\,(51)    & 0.96\,(13)      & 1.05\,(14)      & 1.99\,(88)      & 4.47\,(65)      & 3.7\,(62)    \Tstrut\\   
\ce{^{114m}In_{(i)}} & 7.47\,(23)      & 7.02\,(47)      &                 & 3.501\,(90)     & 2.46\,(13)      & 1.592\,(98)     &                 \\
\ce{^{110}Sn_{(c)}} &                 & 0.07\,(59)      & 0.48\,(22)      & 0.16\,(20)      &                 &                 &                 \\
\ce{^{111}Sn_{(c)}} & 5.10\,(14)      & 0.1582\,(80)    & 0.95\,(11)      & 1.04\,(40)      & 2.0\,(27)       & 4.4\,(14)       & 4.\,(15)        \\
\ce{^{113m}Sn_{(i)}} &                 & 13.\,(31)       & 9.\,(26)        & 5.99\,(30)      & 2.515\,(70)     & 1.031\,(81)     &                 \\
\ce{^{113(m+g)}Sn_{(c)}} & 19.75\,(94)     & 16.\,(37)       & 19.\,(52)       & 12.05\,(61)     & 5.03\,(14)      & 2.04\,(16)      &                 \\
\ce{^{117m}Sn_{(c)}} & 15.50\,(72)     & 11.20\,(34)     & 10.25\,(26)     & 10.66\,(31)     & 10.22\,(13)     & 10.42\,(23)     & 9.67\,(12)      \\
\ce{^{115g}Sb_{(c)}} & 91.8\,(72)      & 51.9\,(43)      &                 &                 &                 &                 &                 \\
\ce{^{116m}Sb_{(i)}} & 34.4\,(41)      & 33.5\,(39)      & 29.2\,(35)      & 19.8\,(22)      & 10.67\,(58)     & 5.46\,(95)      &                 \\
\ce{^{118m}Sb_{(i)}} & 61.7\,(20)      & 66.3\,(17)      & 69.3\,(18)      & 63.2\,(17)      & 55.8\,(14)      & 52.0\,(15)      & 51.4\,(17)      \\
\ce{^{120m}Sb_{(i)}} & 41.0\,(10)      & 43.15\,(86)     & 46.65\,(79)     & 49.27\,(86)     & 55.40\,(58)     & 56.85\,(64)     & 58.16\,(72)     \\
\ce{^{122(m+g)}Sb_{(i)}} & 58.1\,(15)      & 59.0\,(11)      & 63.88\,(94)     & 64.76\,(91)     & 68.0\,(20)      & 71.\,(13)       & 68.3\,(64)      \\
\ce{^{116}Te_{(i)}} & 56.4\,(25)      & 72.3\,(31)      & 91.3\,(34)      & 77.5\,(28)      & 46.0\,(23)      & 21.7\,(12)      & 7.60\,(43)      \\
\ce{^{117}Te_{(i)}} & 96.\,(23)       & 115.\,(27)      & 127.\,(30)      & 131.\,(27)      & 158.\,(27)      & 194.\,(28)      & 180.\,(32)      \\
\ce{^{118}Te_{(i)}} & 55.7\,(27)      & 78.4\,(18)      & 107.9\,(21)     & 121.7\,(32)     & 132.6\,(16)     & 126.3\,(40)     & 125.8\,(16)     \\
\ce{^{119g}Te_{(i)}} & 26.8\,(25)      & 32.9\,(28)      & 38.2\,(41)      & 49.4\,(53)      & 74.2\,(73)      & 88.3\,(84)      & 99.\,(11)       \\
\ce{^{119m}Te_{(i)}} & 36.74\,(87)     & 45.41\,(76)     & 59.94\,(80)     & 80.1\,(10)      & 114.6\,(13)     & 140.1\,(16)     & 161.1\,(21)     \\
\ce{^{121g}Te_{(i)}} & 12.53\,(59)     & 13.51\,(37)     & 15.11\,(37)     & 18.76\,(65)     & 20.04\,(33)     & 20.86\,(52)     & 22.04\,(85)     \\
\ce{^{121m}Te_{(i)}} & 18.67\,(98)     & 20.01\,(96)     & 24.2\,(11)      & 26.31\,(91)     & 26.91\,(60)     & 32.0\,(12)      & 32.8\,(43)      \Bstrut\\\hline\hline
$E_p$\,(MeV)         & 54.606\,(1570)  & 53.981\,(564)   & 53.929\,(562)   & 52.563\,(1612)  & 51.952\,(579)   & 50.415\,(1661)  & 49.912\,(594)   \Tstrut\Bstrut\\
Location             & LANL            & LBNL        & LBNL            & LANL            & LBNL            & LANL            & LBNL            \Bstrut\\\hline
\ce{^{111(m+g)}In_{(c)}} & 3.3\,(71)       &                 &                 & 3.6\,(41)       &                 & 0.27\,(38)      &                 \Tstrut\\
\ce{^{111}Sn_{(c)}} & 4.\,(11)        &                 &                 & 4.\,(23)        &                 & 0.266\,(20)     &                 \\
\ce{^{117m}Sn_{(c)}} & 9.91\,(14)      & 9.20\,(49)      & 9.30\,(56)      & 10.23\,(15)     & 8.88\,(64)      & 10.48\,(16)     & 9.45\,(86)      \\
\ce{^{119m}Sn_{(c)}} &                 & 16.\,(12)       &                 &                 &                 &                 &                 \\
\ce{^{118m}Sb_{(i)}} & 51.4\,(18)      & 57.1\,(12)      & 39.6\,(46)      & 51.1\,(18)      & 52.0\,(22)      & 48.8\,(20)      & 50.1\,(22)      \\
\ce{^{120m}Sb_{(i)}} & 58.84\,(81)     & 61.88\,(60)     & 59.36\,(58)     & 58.81\,(83)     & 56.8\,(11)      & 57.53\,(89)     & 58.57\,(54)     \\
\ce{^{122(m+g)}Sb_{(i)}} & 73.6\,(66)      & 71.0\,(14)      & 67.23\,(95)     & 66.9\,(45)      & 64.6\,(14)      & 71.0\,(28)      & 67.13\,(77)     \\
\ce{^{116}Te_{(i)}} & 0.72\,(37)      &                 &                 &                 &                 &                 &                 \\
\ce{^{117}Te_{(i)}} & 165.\,(31)      & 200.0\,(56)     & 129.\,(23)      & 161.\,(30)      & 100.\,(20)      & 106.\,(24)      & 100.\,(16)      \\
\ce{^{118}Te_{(i)}} & 135.2\,(19)     & 98.\,(24)       & 143.\,(18)      & 156.3\,(22)     & 153.\,(12)      & 187.7\,(30)     & 206.\,(18)      \\
\ce{^{119g}Te_{(i)}} & 108.\,(11)      & 97.1\,(17)      & 90.8\,(41)      & 107.\,(10)      & 96.2\,(44)      & 96.0\,(83)      & 87.7\,(53)      \\
\ce{^{119m}Te_{(i)}} & 168.6\,(24)     & 131.5\,(25)     & 157.5\,(20)     & 166.9\,(25)     & 153.7\,(17)     & 153.5\,(24)     & 146.7\,(13)     \\
\ce{^{121g}Te_{(i)}} & 22.98\,(83)     & 22.0\,(71)      & 22.44\,(62)     & 23.81\,(81)     & 22.19\,(70)     & 24.23\,(87)     & 24.82\,(50)     \\
\ce{^{121m}Te_{(i)}} & 34.4\,(40)      & 34.0\,(35)      & 34.3\,(14)      & 36.1\,(40)      & 34.9\,(15)      & 39.4\,(28)      & 37.97\,(60)     \\
\ce{^{123m}Te_{(i)}} &                 & 3.78\,(44)      &                 &                 &                 &                 &                 \Bstrut\\\hline\hline
$E_p$\,(MeV)         & 49.426\,(601)   & 47.815\,(613)   & 46.176\,(633)   & 45.640\,(636)   & 43.380\,(661)   & 42.733\,(672)   & 40.580\,(696)   \Tstrut\Bstrut\\
Location             & LBNL        & LBNL            & LBNL        & LBNL            & LBNL            & LBNL        & LBNL            \Bstrut\\\hline
\ce{^{117m}Sn_{(c)}} & 10.24\,(32)     & 9.82\,(70)      & 9.74\,(71)      & 9.14\,(62)      & 7.86\,(57)      & 8.20\,(39)      & 6.18\,(48)      \Tstrut\\
\ce{^{119m}Sn_{(c)}} & 18.\,(14)       &                 & 11.\,(11)       &                 &                 & 3.6\,(65)       &                 \\
\ce{^{116m}Sb_{(i)}} & 1.6\,(22)       &                 & 0.18\,(19)      &                 &                 &                 &                 \\
\ce{^{118m}Sb_{(i)}} & 39.3\,(16)      & 25.8\,(66)      & 25.5\,(19)      & 37.6\,(18)      & 28.7\,(16)      & 22.1\,(11)      & 17.22\,(100)    \\
\ce{^{120m}Sb_{(i)}} & 60.07\,(41)     & 57.30\,(48)     & 53.4\,(12)      & 53.53\,(49)     & 48.00\,(39)     & 47.41\,(61)     & 40.88\,(35)     \\
\ce{^{122(m+g)}Sb_{(i)}} & 71.10\,(86)     & 67.76\,(99)     & 74.3\,(11)      & 67.05\,(96)     & 67.3\,(11)      & 71.0\,(12)      & 68.2\,(10)      \\
\ce{^{116}Te_{(i)}} & 0.11\,(49)      &                 &                 &                 &                 &                 &                 \\
\ce{^{117}Te_{(i)}} & 101.2\,(33)     & 69.7\,(24)      & 22.0\,(33)      & 10.5\,(21)      & 2.97\,(59)      & 2.17\,(31)      &                 \\
\ce{^{118}Te_{(i)}} & 214.4\,(82)     & 208.\,(51)      & 301.\,(29)      & 273.0\,(89)     & 267.\,(29)      & 286.\,(58)      & 261.\,(17)      \\
\ce{^{119g}Te_{(i)}} & 91.3\,(13)      & 80.8\,(46)      & 74.9\,(13)      & 71.7\,(34)      & 61.9\,(21)      & 62.32\,(94)     & 66.2\,(41)      \\
\ce{^{119m}Te_{(i)}} & 144.77\,(79)    & 129.9\,(11)     & 113.4\,(13)     & 109.82\,(97)    & 98.35\,(74)     & 96.2\,(10)      & 118.23\,(93)    \\
\ce{^{121g}Te_{(i)}} & 26.3\,(48)      & 26.84\,(56)     & 29.\,(15)       & 29.43\,(70)     & 32.97\,(70)     & 32.\,(12)       & 41.20\,(94)     \\
\ce{^{121m}Te_{(i)}} & 35.9\,(53)      & 41.40\,(71)     & 30.8\,(43)      & 46.06\,(81)     & 61.9\,(18)      & 59.4\,(78)      & 74.3\,(12)      \\
\ce{^{123m}Te_{(i)}} & 4.15\,(30)      &                 & 2.88\,(21)      &                 &                 & 4.86\,(56)      &                 \Bstrut\\\hline\hline
$E_p$\,(MeV)         & 39.040\,(721)   & 37.596\,(738)   & 35.074\,(784)   & 34.439\,(791)   & 33.339\,(376)   & 31.651\,(392)   & 29.891\,(409)   \Tstrut\Bstrut\\
Location             & LBNL        & LBNL            & LBNL        & LBNL            & LBNL            & LBNL            & LBNL            \Bstrut\\\hline
\ce{^{117m}Sn_{(c)}} & 5.23\,(20)      & 4.62\,(46)      & 4.13\,(17)      & 4.15\,(50)      & 4.11\,(50)      & 4.54\,(57)      & 5.20\,(62)      \Tstrut\\
\ce{^{118m}Sb_{(i)}} & 13.09\,(24)     & 7.41\,(64)      & 2.82\,(16)      & 1.80\,(23)      & 0.93\,(20)      & 0.515\,(81)     & 0.29\,(12)      \\
\ce{^{120m}Sb_{(i)}} & 37.68\,(27)     & 34.36\,(33)     & 31.54\,(24)     & 30.02\,(26)     & 29.11\,(37)     & 28.21\,(34)     & 27.93\,(33)     \\
\ce{^{122(m+g)}Sb_{(i)}} & 69.8\,(13)      & 68.1\,(10)      & 67.4\,(10)      & 67.2\,(12)      & 67.5\,(11)      & 66.77\,(90)     & 66.44\,(93)     \\
\ce{^{117}Te_{(i)}} & 0.223\,(97)     &                 &                 &                 &                 &                 &                 \\
\ce{^{118}Te_{(i)}} & 242.\,(15)      & 187.1\,(66)     & 85.9\,(96)      & 81.5\,(19)      & 40.6\,(75)      & 14.18\,(50)     &                 \\
\ce{^{119g}Te_{(i)}} & 82.3\,(15)      & 96.9\,(86)      & 154.4\,(22)     & 135.\,(17)      & 192.0\,(82)     & 88.\,(42)       & 230.\,(11)      \\
\ce{^{119m}Te_{(i)}} & 120.3\,(18)     & 183.1\,(16)     & 241.3\,(18)     & 266.2\,(22)     & 288.6\,(32)     & 275.3\,(90)     & 293.9\,(31)     \\
\ce{^{121g}Te_{(i)}} & 49.4\,(80)      & 58.8\,(13)      & 86.5\,(58)      & 94.0\,(22)      & 108.5\,(25)     & 138.0\,(28)     & 162.5\,(30)     \\
\ce{^{121m}Te_{(i)}} & 89.0\,(84)      & 107.5\,(17)     & 125.\,(19)      & 177.2\,(47)     & 210.9\,(54)     & 236.6\,(70)     & 255.1\,(51)     \\
\ce{^{123m}Te_{(i)}} & 5.20\,(54)      &                 & 5.35\,(43)      &                 &                 &                 &                 \Bstrut\\\hline\hline
$E_p$\,(MeV)         & 29.778\,(886)   & 28.051\,(430)   & 26.122\,(455)   & 24.025\,(484)   & 21.826\,(522)   & 19.431\,(571)   & 16.774\,(641)   \Tstrut\Bstrut\\
Location             & LBNL            & LBNL            & LBNL            & LBNL            & LBNL            & LBNL            & LBNL            \Bstrut\\\hline
\ce{^{117m}Sn_{(c)}} & 5.29\,(63)      & 5.54\,(56)      & 5.52\,(47)      & 4.62\,(23)      & 3.69\,(22)      & 2.83\,(23)      & 2.65\,(31)      \Tstrut\\
\ce{^{118m}Sb_{(i)}} & 0.333\,(76)     & 7.1\,(11)       &                 &                 &                 &                 &                 \\
\ce{^{120m}Sb_{(i)}} & 28.09\,(26)     & 26.58\,(27)     & 24.09\,(22)     & 18.42\,(15)     & 12.46\,(14)     & 6.355\,(75)     & 2.022\,(28)     \\
\ce{^{122(m+g)}Sb_{(i)}} & 66.3\,(11)      & 64.23\,(85)     & 61.97\,(66)     & 53.05\,(53)     & 42.63\,(46)     & 27.50\,(31)     & 13.31\,(17)     \\
\ce{^{119g}Te_{(i)}} & 181.\,(24)      & 218.\,(11)      & 186.5\,(98)     & 113.6\,(61)     & 24.5\,(10)      &                 &                 \\
\ce{^{119m}Te_{(i)}} & 293.4\,(27)     & 262.7\,(25)     & 209.6\,(18)     & 115.68\,(94)    & 22.36\,(26)     &                 &                 \\
\ce{^{121g}Te_{(i)}} & 164.0\,(37)     & 173.8\,(34)     & 168.8\,(29)     & 137.9\,(26)     & 82.2\,(12)      & 20.42\,(36)     & 22.93\,(33)     \\
\ce{^{121m}Te_{(i)}} & 258.9\,(80)     & 255.7\,(64)     & 233.9\,(36)     & 183.4\,(38)     & 100.3\,(15)     & 26.59\,(73)     & 29.82\,(59)     \Bstrut\\\hline\hline
$E_p$\,(MeV)         & \multicolumn{2}{l}{13.762\,(745)} &                 &                 &                 &                 &                 \Tstrut\Bstrut\\
Location             & LBNL            &                 &                 &                 &                 &                 &                 \Bstrut\\\hline
\ce{^{117m}Sn_{(c)}} & 5.47\,(71)      &                 &                 &                 &                 &                 &                 \Tstrut\\
\ce{^{120m}Sb_{(i)}} & 0.1568\,(26)    &                 &                 &                 &                 &                 &                 \\
\ce{^{122(m+g)}Sb_{(i)}} & 1.960\,(44)     &                 &                 &                 &                 &                 &                 \\
\ce{^{121g}Te_{(i)}} & 67.5\,(11)      &                 &                 &                 &                 &                 &                 \\
\ce{^{121m}Te_{(i)}} & 76.4\,(12)      &                 &                 &                 &                 &                 &                 \Bstrut \\ 

\end{longtable*}
\end{tiny}
\endgroup

    \section{\label{sec:TALYSModel}Reaction Modeling in TALYS\protect}
        These experimental results provide an opportunity to investigate the underlying physics of high-energy charged particle reactions and existing modeling capabilities. 
The choice of TALYS 1.95 as a modeling code for comparison was inspired by previous efforts in the group \cite{fox_investigating_2021,morrell_measurement_2024} as well as its relatively quick computational speed and deterministic nature. 
This enables rapid fine-tuning of parameters, of which approximately 40 were explored in this work.
TALYS 1.95 calculates ejectiles up to alphas in residual product calculations.
For the incident particle energies covered in these experiments, the total reaction cross section can be modeled as having four components: elastic, compound, pre-equilibrium, and direct emission. 
A residual product cross section with an atomic number $Z$ and neutron number $N$ and level $j$ can be defined as the sum of all individual cross sections where the ejectiles from the compound nucleus of the incident particle and target nucleus $(Z^*,N^*)$ produces the same final reaction product, as calculated in \autoref{eq:residproddetailed} \cite{koning_talys-195_2019}. 

\begin{equation}
\begin{aligned}
\sigma_{prod,j}(Z,N)=\sum_{i_n=0}^{\infty}\sum_{i_p=0}^{\infty}\sum_{i_d=0}^{\infty}\sum_{i_t=0}^{\infty}\sum_{i_h=0}^{\infty}\sum_{i_\alpha=0}^{\infty} \\ \sigma_j^{e x}\left(i_n,i_p,i_d,i_t,i_h,i_\alpha\right) \delta_N \delta_Z \\
\delta_N=
\begin{cases}
1 \text{ if } i_n+i_d+2i_t+i_h+2i_\alpha=N^*-N \\
0 \text{ otherwise}
\end{cases}\\
\delta_Z=
\begin{cases}
1 \text{ if } i_p+i_d+i_t+2i_h+2i_\alpha=Z^*-Z \\
0 \text{ otherwise}
\end{cases}
\end{aligned}
\label{eq:residproddetailed}
\end{equation}

\begin{figure}[htbp]
    \includegraphics[width=\linewidth]{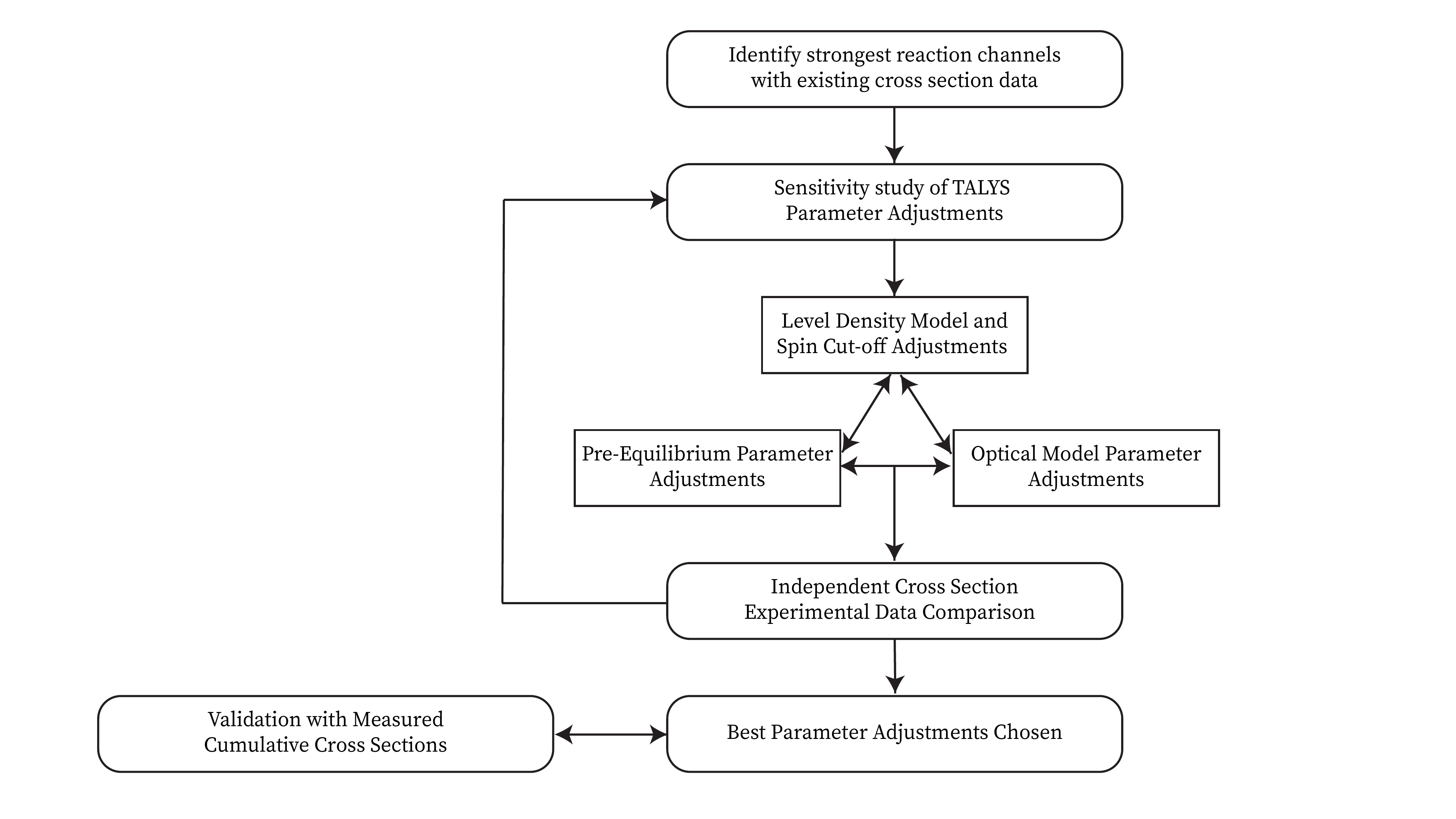}
    \caption{Flow chart of the iterative modeling process.}
    \label{fig:modeling_process}
    \centering
\end{figure}

Analysis focused on seven (p,$x$n) channels and specific long-lived isomers -- \ce{^{116(m+g)}Te}, \ce{^{117(m+g)}Te}, \ce{^{118(m+g)}Te}, \ce{^{119g}Te}, \ce{^{119m}Te}, \ce{^{121g}Te}, and \ce{^{121m}Te}, as well as other (p,x) channels -- \ce{^{116m}Sb}, \ce{^{118m}Sb}, \ce{^{120m}Sb}, \ce{^{122(m+g)}Sb}, and \ce{^{114m}In}.
Data above 100\,MeV for residual Sb isotopes were excluded from analysis due to potential contamination from secondary neutron production.
These 12 independent cross sections formed a comparative set to guide parameter adjustments.
An iterative, physically reasonable approach to level density, pre-equilibrium, and optical model potential parameter adjustments was performed with a qualitative and quantitative goodness-of-fit test.
The results were validated against nine cumulative channels and three smaller independent channels in a process outlined in \autoref{fig:modeling_process}.

Details for all parameter adjustments are provided in Appendices \ref{app:TALYSLDP}, \ref{app:TALYSPEP}, and \ref{app:TALYSOMP}.
Note that TALYS 1.95 defaults to logarithmic energy binning via the \texttt{equidistant} keyword, which facilitates a faster and more precise calculation of the evaporation peak in the neutron spectrum.
Logarithmic spacing is not effective for this analysis given the range of incident proton energies and significantly impacts the threshold energy for residual products. Therefore, this was set to \texttt{equidistant y} for linear spaced energy bins.
\subsection{\label{sec:LD_Adj}Level Density Adjustments}
\subsubsection{\label{sec:LD_Models} Level Density Model}
TALYS includes both phenomenological and microscopic level density models.
In comparison to these residual products, \texttt{ldmodel} 2 -- the Back-Shifted Fermi Gas Model (BFG), shifts the baseline energy level for calculating excitation energy and provides the best global fit as seen in \autoref{tab:ldmodels}.
\begin{table}
\centering
\caption{Available level density models in TALYS 1.95}
\label{tab:ldmodels}
\begin{tabular}{cc}
\hline\hline
\texttt{ldmodel} &
 Reduced $\mathrm{\chi^2}$\Tstrut\Bstrut\\ \hline
1  & 6.5135 \Tstrut\\
2  & 5.7367 \\ 
3 & 79.7651 \\
4 & 7.5453 \\
5 & 10.7160 \\
6 & 13.5216 \Bstrut\\ \hline \hline
\end{tabular}
\end{table}
\subsubsection{\label{sec:spincut} The Spin Cut-Off Parameter \texorpdfstring{$\sigma^2$}{sigma2}}
\begin{figure*}[htbp]
    \centering
    \subfloat[\ce{^{119}Te} Decay]{
        \includegraphics[width=0.32\linewidth]{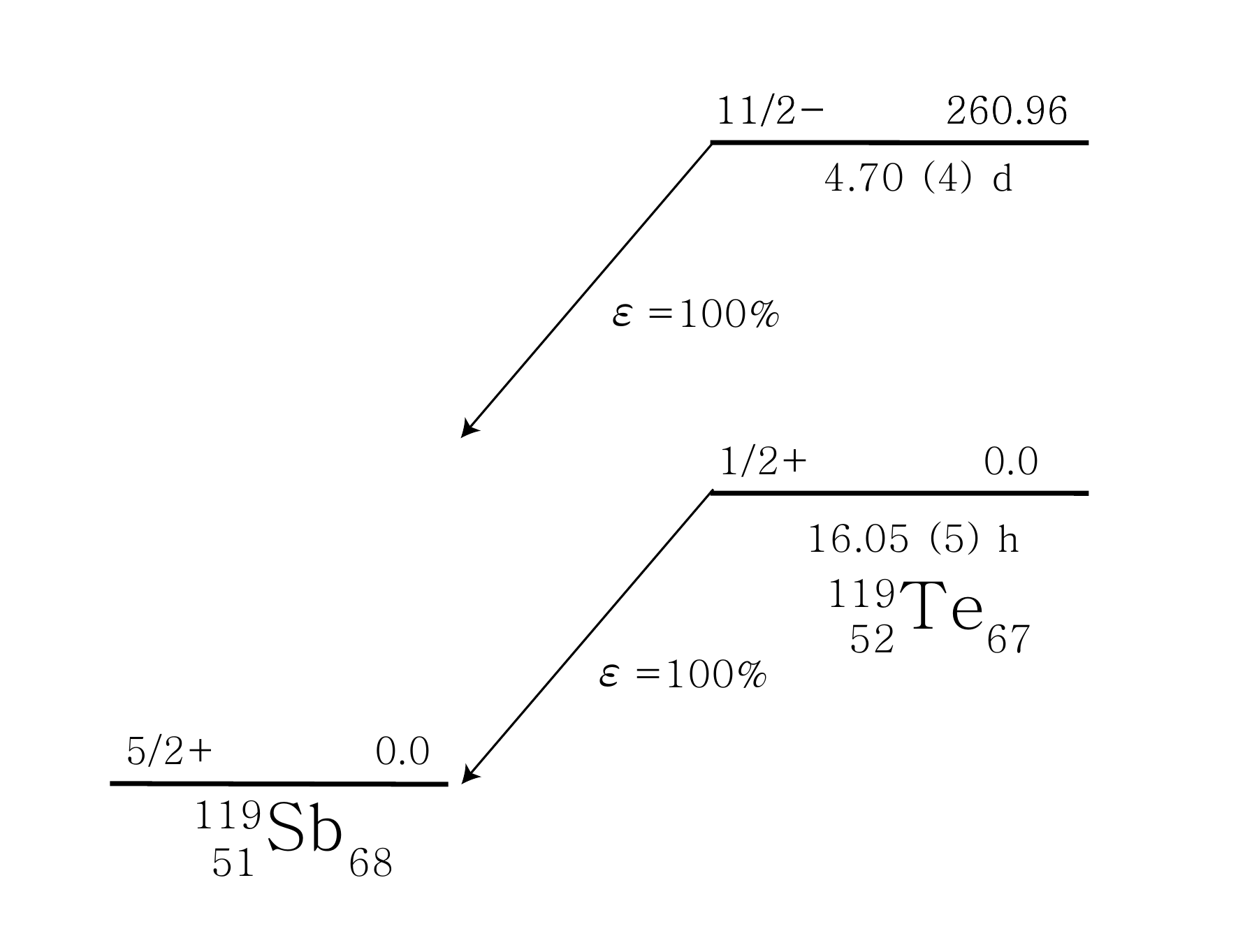}
        \label{fig:119Decay}
    }
    \subfloat[\ce{^{121}Te} Decay]{
        \includegraphics[width=0.32\linewidth]{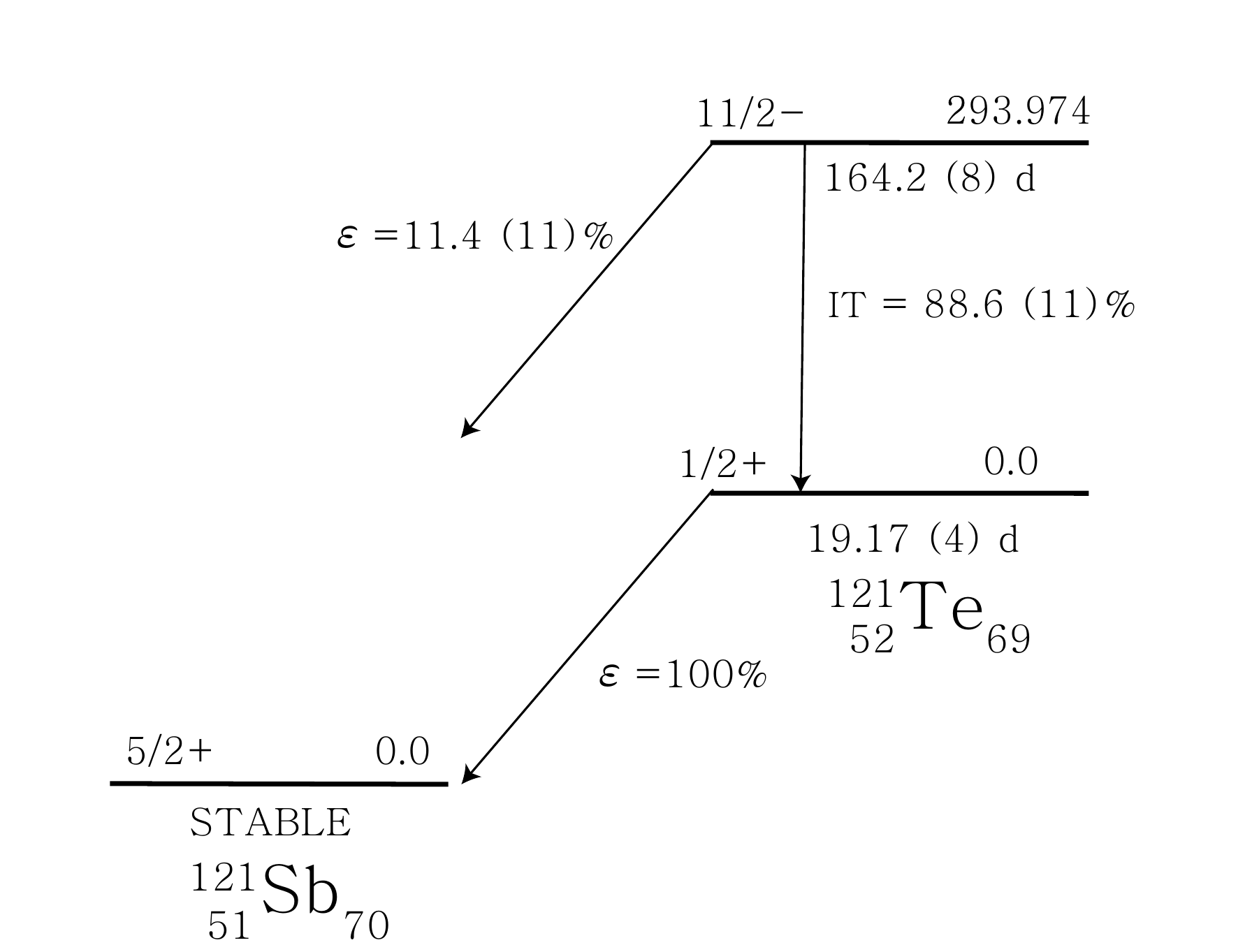}
        \label{fig:121Decay}
    }
    \subfloat[\ce{^{123}Te} Decay]{
        \includegraphics[width=0.32\linewidth]{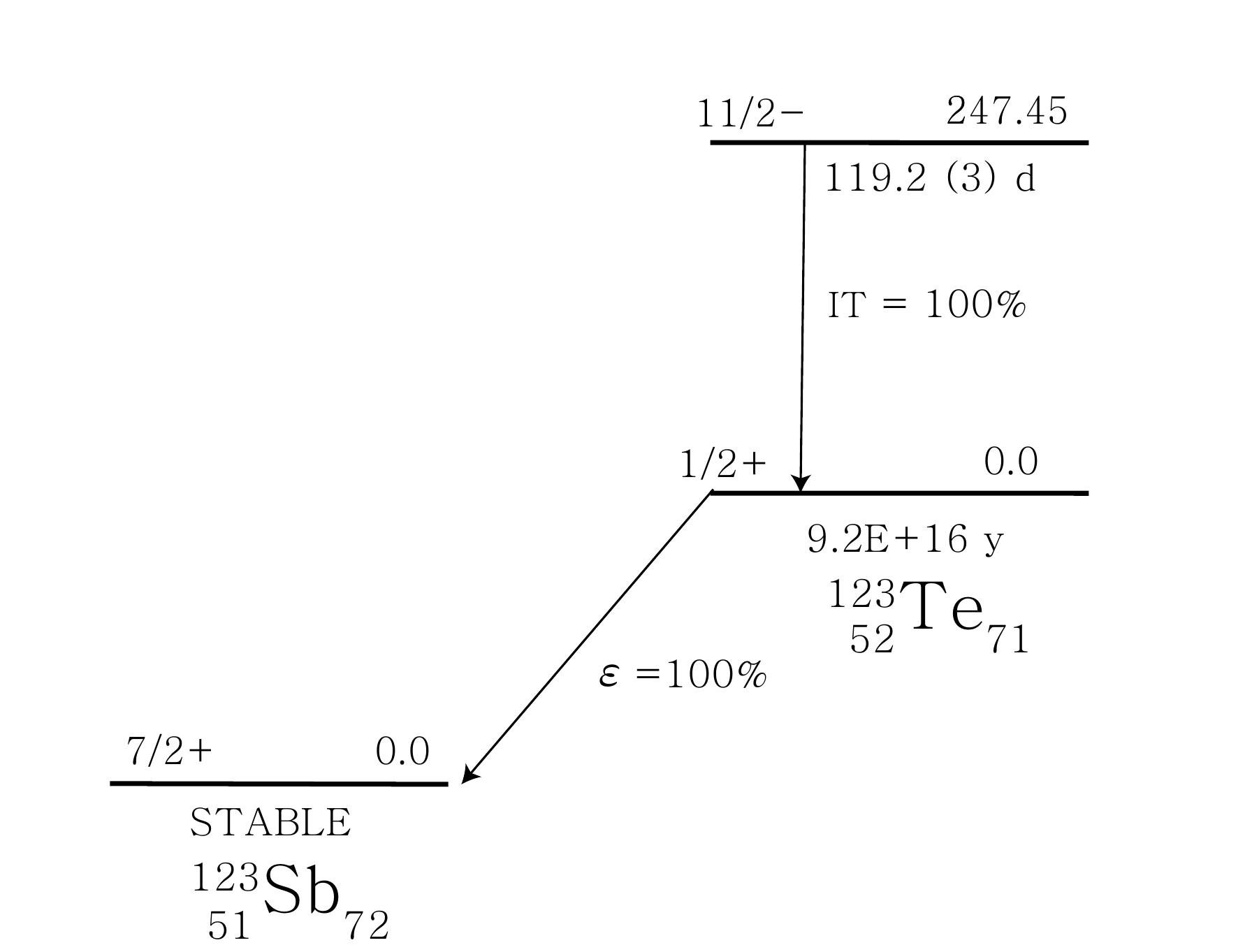}
        \label{fig:123Decay}
    }
    \caption{Decay schema for \ce{^{119}Te}, \ce{^{121}Te}, and \ce{^{123}Te}.}
    \label{img:119121isomerdecay}
\end{figure*}

The observation of two adjacent odd-A Te isotopes (\ce{^{119}Te} and \ce{^{121}Te}) and their long-lived isomeric states offered an opportunity to optimize reaction models for spin distribution values in unresolved multi-MeV nuclear states near shell closure.
The simplified decay schema for these isotopes is shown in \autoref{img:119121isomerdecay}.
The neighboring \ce{^{123}Te} isomer was used as validation due to its similar properties, seen in \autoref{fig:123Decay}.

\begin{figure}[htbp]
\centering
    \minipage{0.495\linewidth}     
        \includegraphics[width=\linewidth]{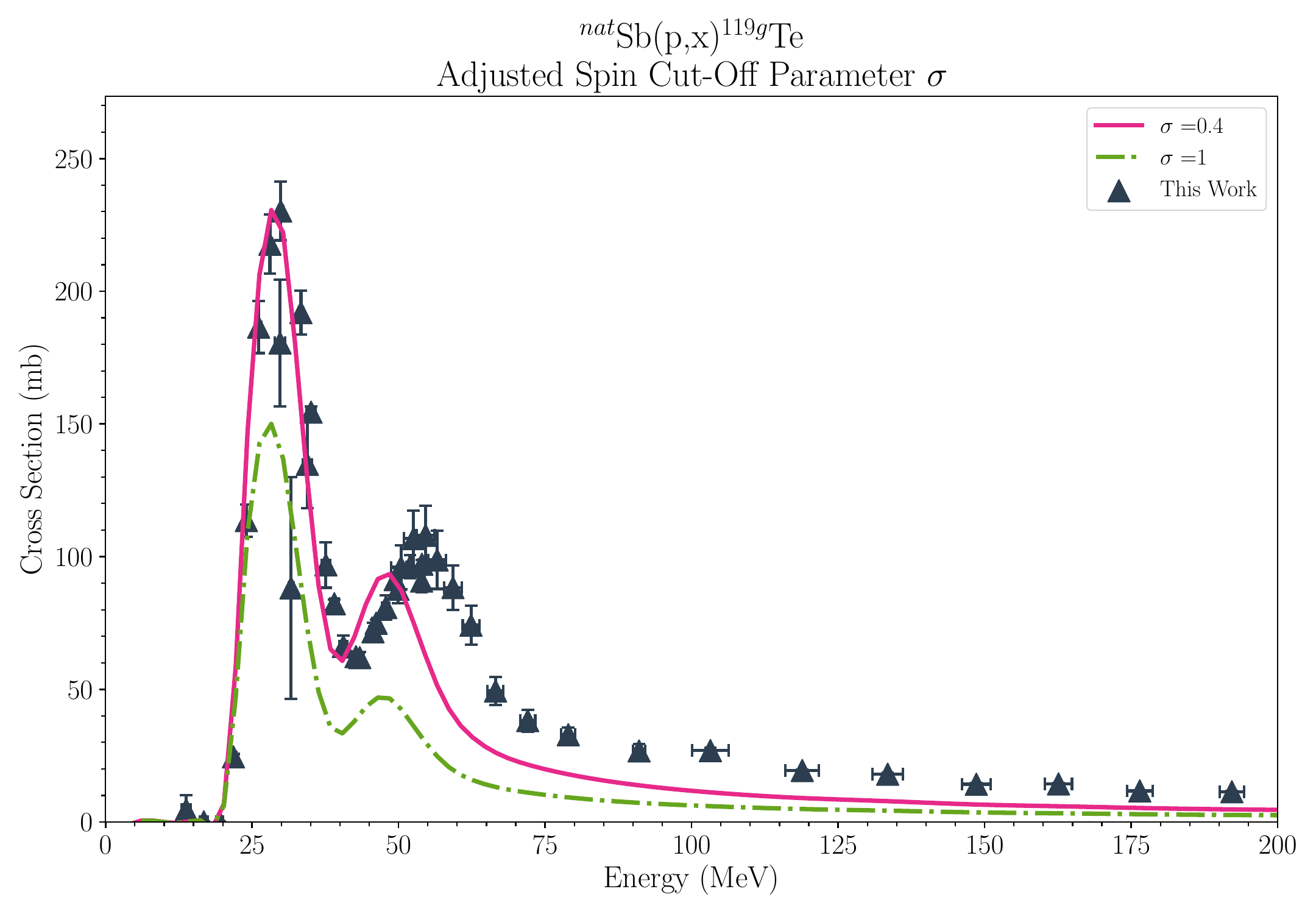}
    \endminipage \hfill 
    \minipage{0.495\linewidth}     
        \includegraphics[width=\linewidth]{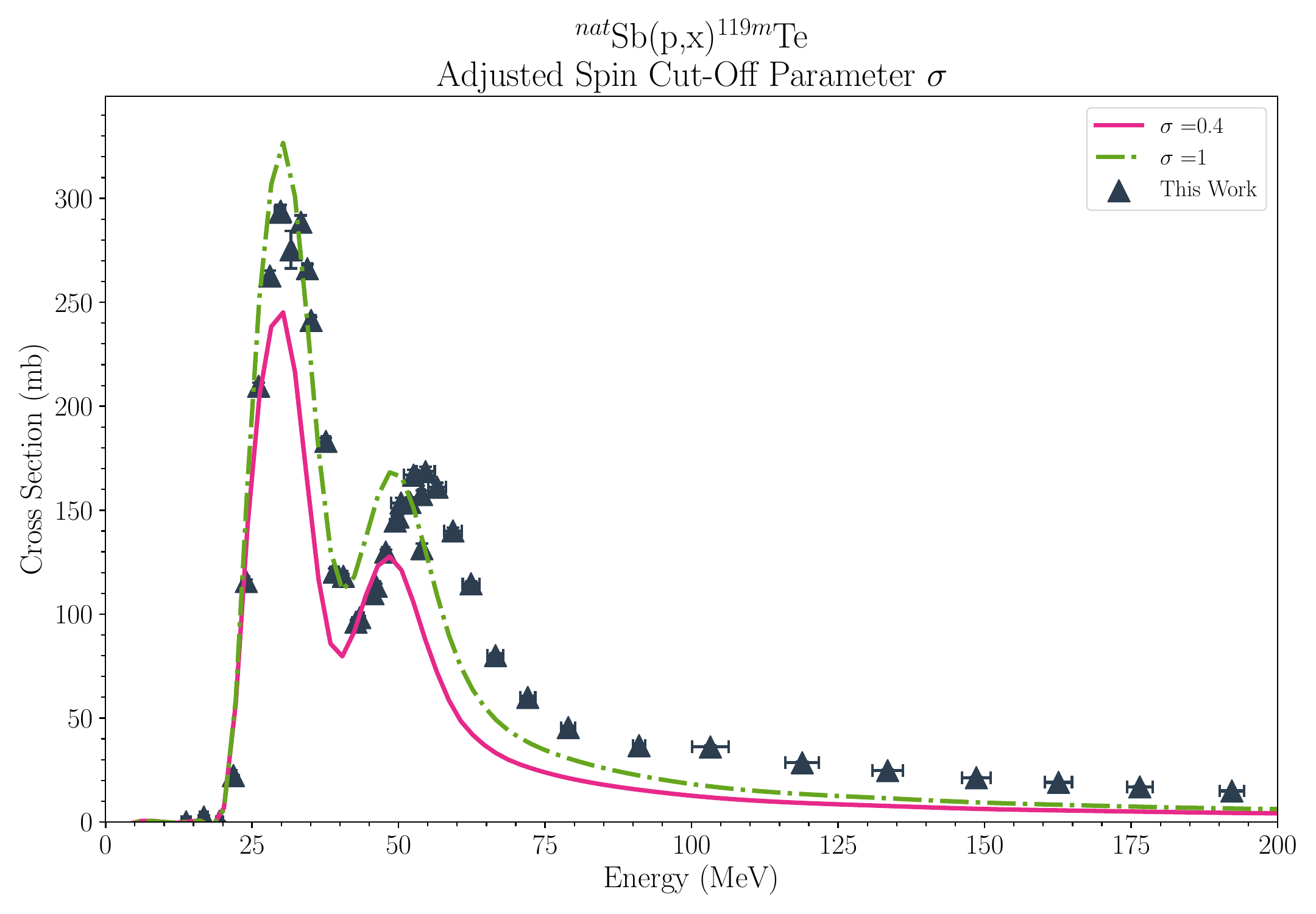}
    \endminipage \hfill 
    \minipage{0.495\linewidth}     
        \includegraphics[width=\linewidth]{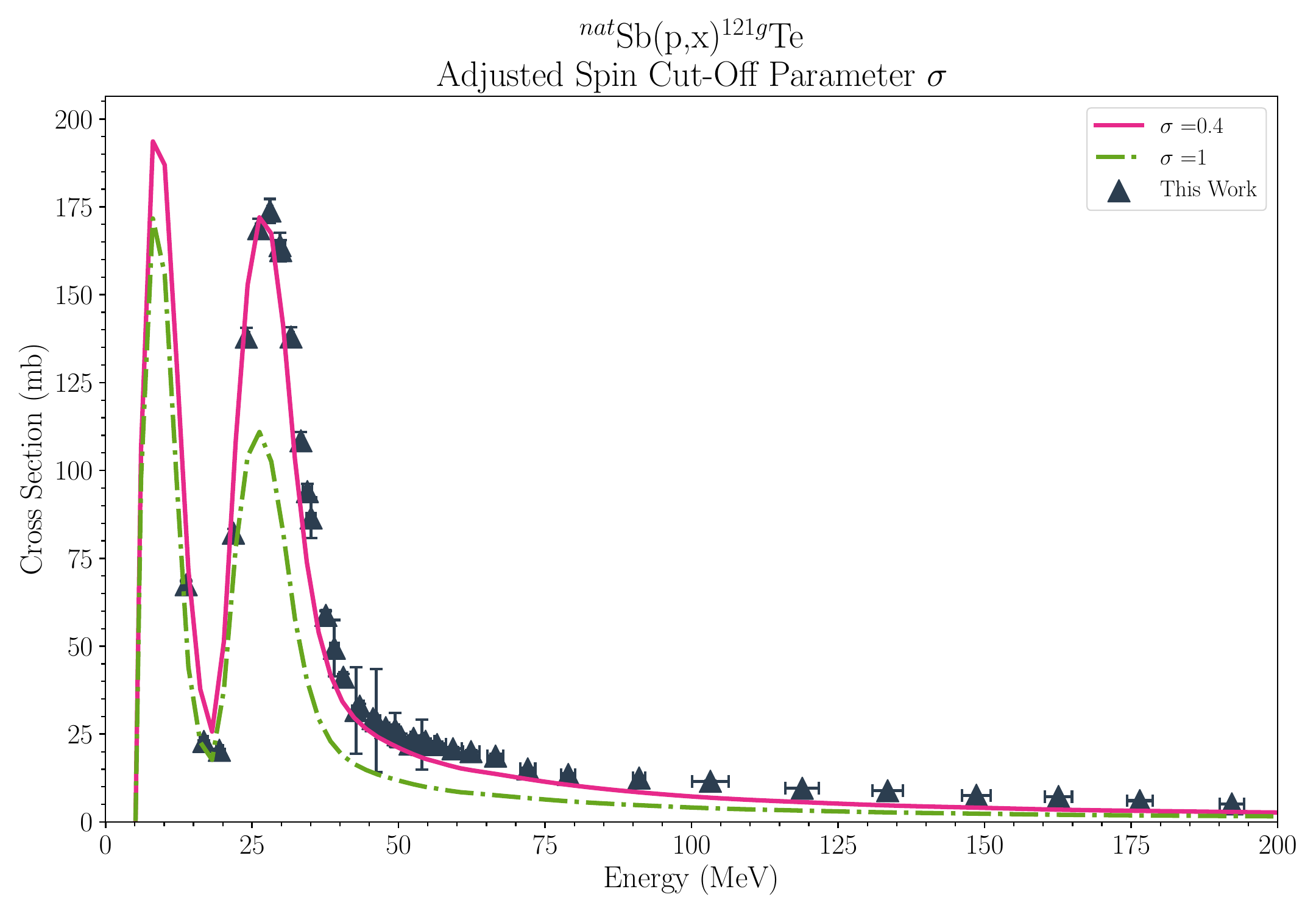}
    \endminipage \hfill 
    \minipage{0.495\linewidth}     
        \includegraphics[width=\linewidth]{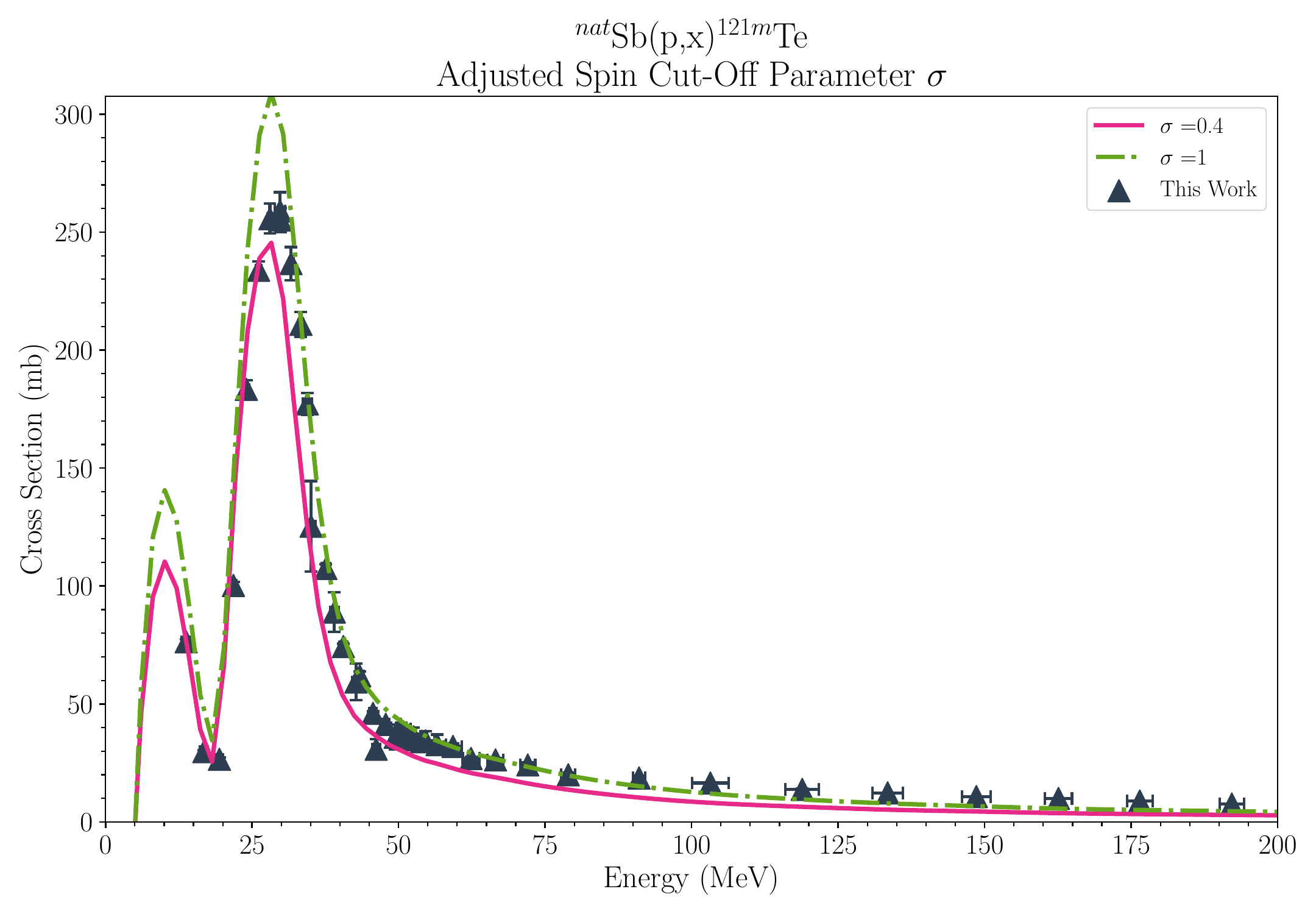}
    \endminipage \hfill 
    \caption{Default (1.0 - green-dashed curves) vs adjusted (0.4 - solid red line) TALYS value for \texttt{rspincut}}
    \label{fig:rspinxs}
\end{figure}

TALYS 1.95 systematically overestimated the population of the high-spin isomeric state and underestimated the population of the low-spin ground state compared to experimental results, as seen in \autoref{fig:rspinxs}.
The phenomenological level density models in TALYS 1.95 assume an energy-dependent Gaussian distribution $E$ for nuclear spin $J$. 
The spin cut-off parameter $\sigma^2$ is defined as the width of spin distribution, which in this case favors low-energy spin.

\begin{figure}[htbp]
\centering
    \includegraphics[width=\linewidth]{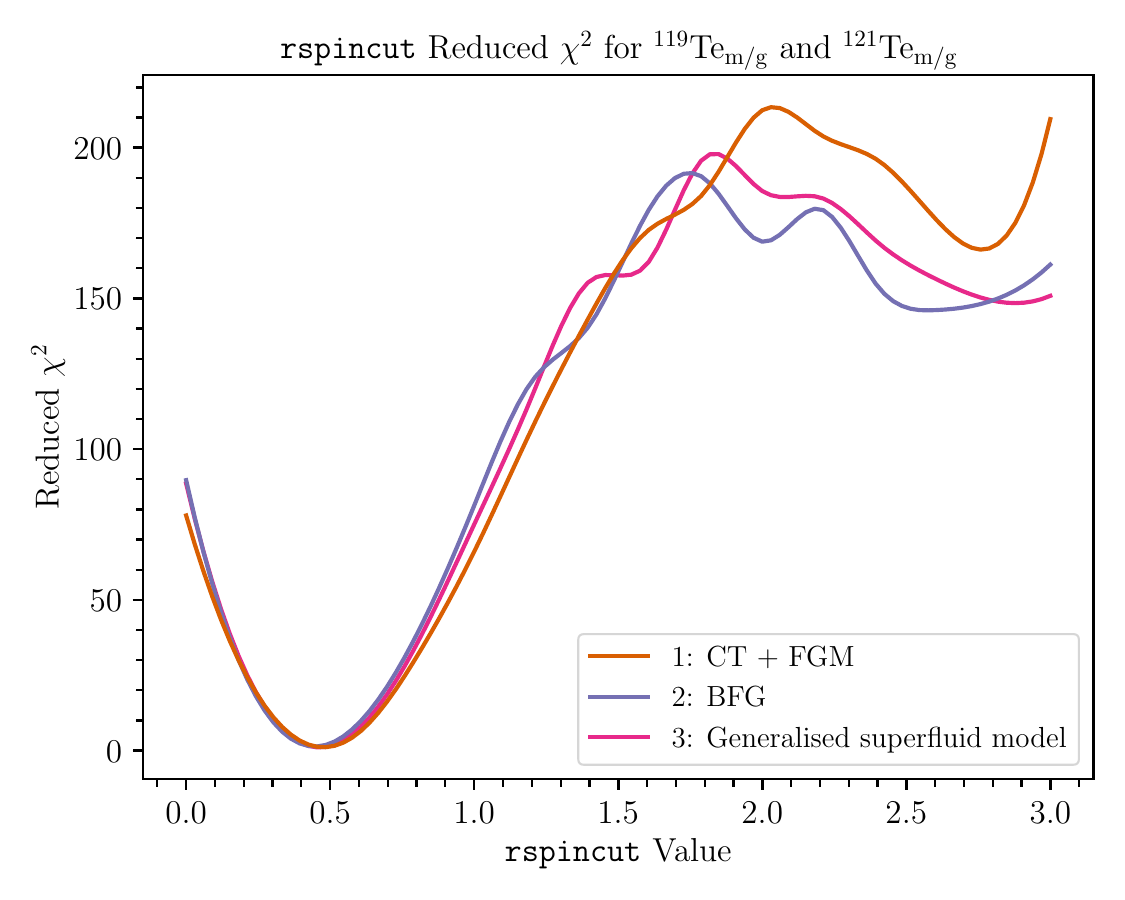} 
    \caption{Reduced $\chi^2$ minimization around the local minima.}
    \label{fig:rspinreducedchi}
\end{figure}

Adjustments to $\sigma^2$ in TALYS 1.95 use the \texttt{rspincut} keyword. This acts as a global multiplier to adjust the width of angular momentum, essentially increasing or decreasing the chance for a reaction to populate a certain level in the residual product.
A global adjustment to \texttt{rspincut=0.4} resulted in an improved fit for \ce{^{121}Te} and \ce{^{119}Te} and its isomers, with the local minimum for the reduced $\chi^2$ for the three phenomenological level density models shown in \autoref{fig:rspinreducedchi}. 
\autoref{tab:rspinreducedchi} shows a comparison of the reduced $\chi^2$ with \texttt{rspincut=0.4} against the default TALYS 1.95 value.
This is in agreement with results by Rodrigo\,\etal \cite{rodrigo_compilation_2023}.
\begin{table}[!!!htbp]
\centering
\caption{Phenomenological model reduced $\chi^2$ at \texttt{rspincut}\,=\,0.4 vs Default} 
        \label{tab:rspinreducedchi} 
\begin{ruledtabular}
\begin{tabular}{lll}
        \texttt{ldmodel} & \texttt{rspincut}\,=\,0.4 & \texttt{rspincut}\,=\,1.0 \\
        \hline
    1 & 2.528029 & 45.0339 \\
    2 & 1.407499 & 58.33051 \\
    3 & 2.079992 & 51.68439 \\
\end{tabular}
\end{ruledtabular}
\end{table}

The \texttt{spincutmodel} parameter can further modify the formula for isotopes with large shell effects. 
The m/g ratios for the neighboring Te isotopes were improved with \texttt{spincutmodel}=2, which neglects shell closure due to prolate deformation of the nucleus.

\begin{figure}[htbp]
\centering
        \includegraphics[width=\linewidth]{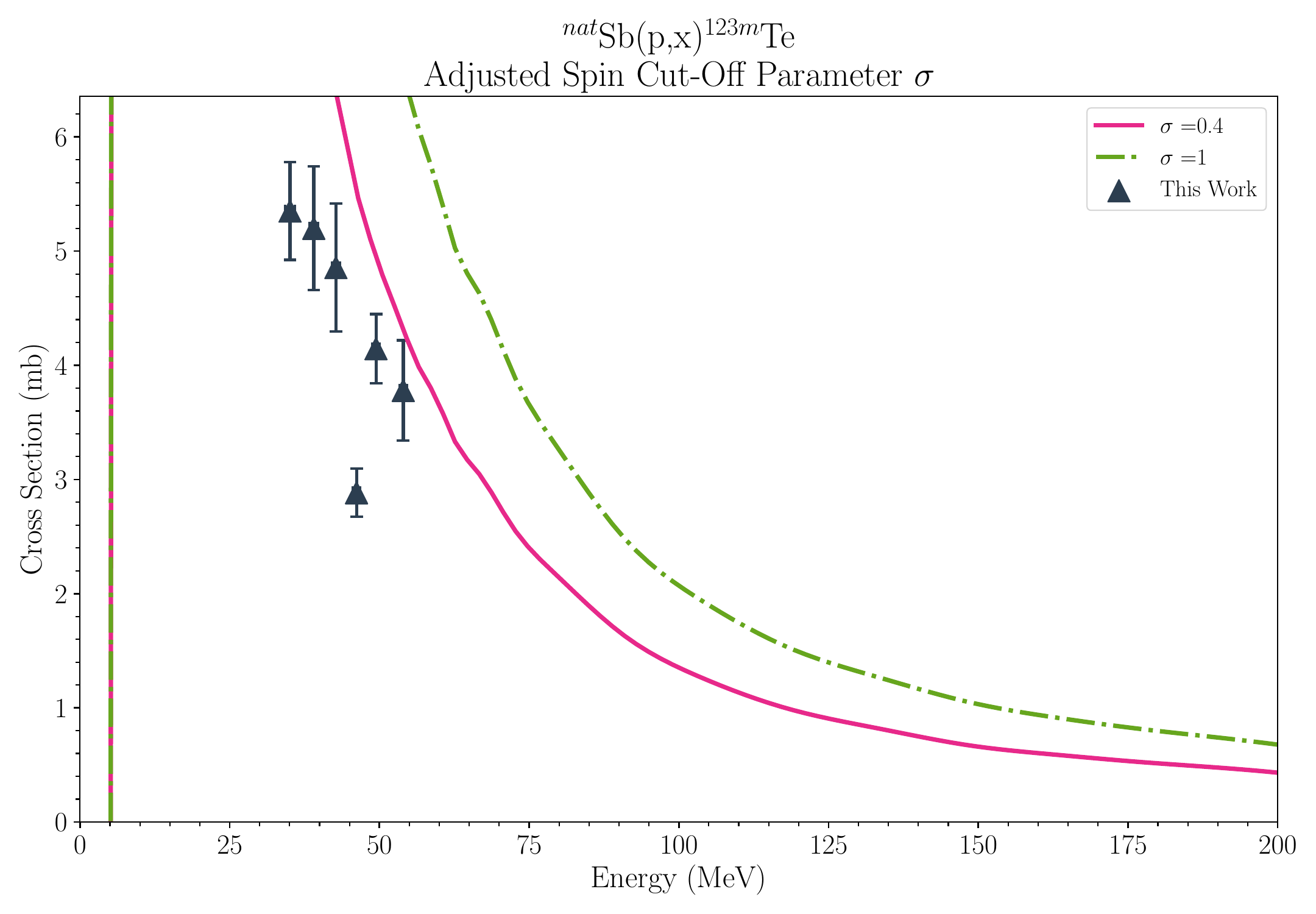} 
        \caption{Default vs adjusted TALYS value for \texttt{rspincut} for \ce{^{123m}Te}}
    \label{fig:rspinxs123}
\end{figure}

This reduction in rspincut was validated by analyzing the neighboring isomer of \ce{^{123}Te}.
While the ground state was not observed due to its long lifetime ($\mathrm{t_{1/2}\,>\,10^{16}\,y}$), the \texttt{rspincut} and \texttt{spincutmodel} adjustments improved the fit for the observed isomer as seen in \autoref{fig:rspinxs123}.

\subsection{\label{sec:Preequilibrium} Adjustments to Pre-Equilibrium Parameters}
In pre-equilibrium reactions, nuclear interactions are governed by the exciton model, where nucleon-hole pairs, known as excitons, can scatter off other nucleons and decrease the average energy of nucleons, resulting in compound nucleus.
The exciton scattering probabilities can be adjusted to best represent the measured data. 

The exciton model characterizes the nuclear state based on total energy $E_{tot}$ for protons ($\pi$) and neutrons ($\nu$) in the system with particles $p$ above and holes $h$ below the Fermi surface energy in the reaction.
The potential configurations of the shared excitation energy between these pairs are equally probabilistic and are expressed by an exciton number $n=p_\pi+h\pi+p\nu+n_\nu$\cite{koning_talys-195_2019}.

This applies to incident particles with energies above neutron separation energy. This is delineated into multiple compound decay as described by HF formalism and multiple pre-equilibrium decay, seen in \autoref{eq:HFlong}.
\begin{equation}
\begin{aligned}
\frac{d\sigma_k^{\mathrm{PE}}}{dE_k}=\sigma^{\mathrm{CF}}\sum_{p_\pi=p_\pi^0}^{p_\pi^{\max}}\sum_{p_\nu=p_\nu^0}^{p_\nu^{\max}}W_k(p_\pi,h_\pi,p_\nu,h_\nu, E_k) \\ 
\tau(p_\pi,h_\pi,p_\nu,h_\nu)P(p_\pi, h_\pi, p_\nu, h_\nu)
\end{aligned}
\label{eq:HFlong}
\end{equation}
where $\sigma^{\mathrm{CF}}$ is the compound nucleus cross section, $W_k(p_\pi,h_\pi,p_\nu,h_\nu,E_k)$ is the emission rate, derived by Cline and Blann \cite{cline_pre-equilibrium_1971}, $P(p_\pi,h_\pi,p_\nu,h_\nu)$ is the average pre-equilibrium population of residual nucleons, and $\tau(p_\pi,h_\pi,p_\nu,h_\nu)$ is the lifetime of the exciton.
The formal derivations for these components are detailed in the TALYS 1.95 manual \cite{koning_talys-195_2019}. 

Multiple pre-equilibrium emission transition rates are calculated with an effective square matrix $M^2$, with collision probability described in \autoref{eq:squarematrixfermi}. The corresponding reaction probability is calculated by swapping $\pi$ and $\nu$ in the equation. 
\begin{equation}
\begin{aligned}
\lambda_{\pi\pi}^{1p}(u)=\frac{2\pi}{\hbar}M_{\pi\pi}^2\omega(2,1,0,0,u)\\
\lambda_{\pi\pi}^{1h}(u)=\frac{2\pi}{\hbar}M_{\pi\pi}^2\omega(1,2,0,0,u) \\
\lambda_{\nu\pi}^{1p}(u)=\frac{2\pi}{\hbar}M_{\nu\pi}^2\omega(1,1,1,0,u) \\
\lambda_{\nu\pi}^{1h}(u)=\frac{2\pi}{\hbar}M_{\nu\pi}^2\omega(1,1,0,1,u)\\
\lambda_{\pi\nu}^{1p1h}(u)=\frac{2\pi}{\hbar}M_{\pi\nu}^2\omega(0,0,1,1,u)
\end{aligned}
\label{eq:squarematrixfermi}
\end{equation}
In TALYS 1.95, parameter adjustments can be made to the overall $M^2$, as well as to residual nucleon-nucleon interactions.
Following the work of Fox\,\etal \cite{fox_investigating_2021}, global adjustments to $M^2$ were explored using the \texttt{M2Constant}, \texttt{M2Limit}, and \texttt{M2Shift} parameters.
The best fit was achieved by increasing \texttt{m2constant} to 2, decreasing \texttt{m2limit} to 0.8, and increasing \texttt{m2shift} to 1.8. 
These changes are in contrast to the adjustments made by Fox \etal.
Although this improved the fit by adjusting the magnitude of the compound peak, the pre-equilibrium region for (p,$x$n) reactions was underestimated.
Adjustments to the residual nucleon-nucleon interaction parameters were explored. 
TALYS 1.95 facilitates these adjustments with the multiplying parameters \texttt{rpipi}, \texttt{rnunu}, \texttt{rpinu}, and \texttt{rnupi}, which are multipliers for individual exciton scattering probabilities.
An increase in \texttt{rpinu} and \texttt{rnupi} to 1.5 optimized fit.
An additional adjustment was made to TALYS 1.95 for pre-equilibrium reactions, in which the default model utilizes numerical transition rates in the exciton model; however, the best fit to experimental data was observed using an analytical transition rate. 
Additional parameter adjustments are covered in Appendix \ref{app:TALYSPEP}.

\subsection{\label{sec:OpticalModel} Adjustments to Optical Model Potential}

The default optical model potential in TALYS is based on the parameterization of Koning and Delaroche \cite{koning_local_2003}. 
The phenomenological optical model potential describes the interaction between the incident particle and the nucleons within the target nucleus, described in \autoref{eq:OMPtotal}.

\begin{equation}
\label{eq:OMPtotal}
\begin{aligned}
\mathcal{U}(r, E)= & -\mathcal{V}_V(r, E)-i \mathcal{W}_V(r, E)-i \mathcal{W}_D(r, E) \\
& +\mathcal{V}_{S O}(r, E) \cdot \mathrm{l} \cdot \sigma+i \mathcal{W}_{S O}(r, E) \cdot \mathrm{l} \cdot \sigma+\mathcal{V}_C(r)
\end{aligned}
\end{equation}

The optical model consists of real $\mathcal{V}$ and imaginary $\mathcal{W}$ components of the volume (V), surface (D), and spin-orbit (SO) potentials.
Both the proton and neutron potentials were investigated, with further details in Appendix \ref{app:TALYSOMP}. 
Although several parameters were explored, many, including adjustments to the real potential, were difficult to justify from a physical perspective. 
However, the imaginary potential was modified, specifically for the volume term of the imaginary neutron potential well.
 \texttt{w1adjust} was increased to 2.5, while \texttt{w2adjust} was decreased to 0.6.
These adjustments can be visualized as a deeper and narrower well that can affect the absorption and emission of neutrons in a nuclear reaction.
\subsection{\label{sec:Independentxsother} Comparison to Experimental Data}
TALYS parameters were explored and optimized using two approaches for goodness of fit ($\chi^2$). 
For a single residual product channel $c$, a goodness of fit was calculating using \autoref{eq:chisqTALYSchanneltotal}
\begin{equation}
\chi_c^2=\frac{1}{N_p} \sum_{i=1}^{N_p}\left(\frac{\sigma_T^i-\sigma_E^i}{\Delta \sigma_E^i}\right)^2
\label{eq:chisqTALYSchanneltotal}
\end{equation}

Where $N_p$ is the number of measured data points for a reaction channel, $\sigma_T$ is the calculated cross section from TALYS 1.95 at a given energy, $\sigma_E$ is the experimentally measured cross section, and $\Delta\sigma_E^i$ is the uncertainty of the experimental data.
The contribution to overall goodness-of-fit across all channels $\chi_{\mathrm{tot}}^2$ by a channel was calculated using \autoref{eq:chisqTALYSchannelind}

\begin{equation}
\chi_{\mathrm{tot}}^2=\frac{1}{N_c} \sum_{c=1}^{N_c} \chi_c^2 w_c
\label{eq:chisqTALYSchannelind}
\end{equation}

Here, $N_c$ is the number of residual product channels used in the analysis, $\chi^2$ is the goodness of fit for the channel, and $w_c$ is the weighting factor for that channel.

This does not address the surplus of measurement points below 55\,MeV and the lack of data for measurements above 200\,MeV.
Fitting of the pre-equilibrium region for larger-magnitude reactions as well as residual products with a reaction threshold near and above 100\,MeV is inherently weighted less.
However, alternative weighting within a channel would restrict the statistical impact of data points in the energy region below 55\,MeV, which is of interest for isotope production.

The weighting factor was considered from two approaches.
The first weighting factor is based on the summation of the channel cross section across the energy region and is referred to here as "Cumulative\,$\sigma \chi^2$"(\autoref{eq:cumweight}):
\begin{equation}
w_c=\frac{\sum_{i=1}^{N_p} \sigma_T^{c i}(E)}{\sum_{c=1}^{N_c} \sum_{i=1}^{N_p} \sigma_T^{c i}(E)}
\label{eq:cumweight}
\end{equation}

 The second approach is based on the maximum cross section value relative to other channels, here referred to as "Maximum\,$\sigma \chi^2$ " (\autoref{eq:maxweight})

\begin{equation}
w_c=\frac{\sigma_{T, \text { max }}^c}{\sum_{c=1}^{N_c} \sigma_{T, \text { max }}^c}
\label{eq:maxweight}
\end{equation}

These approaches have a trade-off for reaction modeling by prioritizing the number of data points: 26 of 43 possible data points were measured below 55\,MeV, while only 7 data points exist above 100\,MeV. 
While the 55\,MeV and below region is valuable from an isotope production standpoint, this weighting factor tends to neglect the fit for higher incident-energy protons.
This has a significant impact on the fit for pre-equilibrium parameters.

The majority of the independent residual product cross section channels, including all (p,$x$n) reactions, were used for optimization.
These showed an overall improvement in goodness of fit using the $\chi^2$ calculations referenced in Equations \ref{eq:chisqTALYSchanneltotal}, \ref{eq:chisqTALYSchannelind}, \ref{eq:cumweight}, and \ref{eq:maxweight}.
The tabulated results are presented in \autoref{tab:chisqtotalindependent}

\begin{table}[htbp]
\caption{Optimized $\chi^2$ values for independent cross sections used for modeling.}
\label{tab:chisqtotalindependent}
\begin{tabular}{cccc}

\hline \hline
 & TALYS 1.95 & Optimized & Improvement \Tstrut\\
 & Default & Parameters & Ratio \Bstrut\\
 \hline
Cumulative $\chi^2$ & 27.54 & 4.42 & 6.23 \Tstrut\\
Maximum $\chi^2$ & 31.43 & 4.15 & 7.58 \Bstrut\\
\hline \hline
\end{tabular}
\end{table}

\begin{figure*}[ht]
    \centering
    \subfloat[\ce{^{118}Te}]{\includegraphics[width=0.33\linewidth]{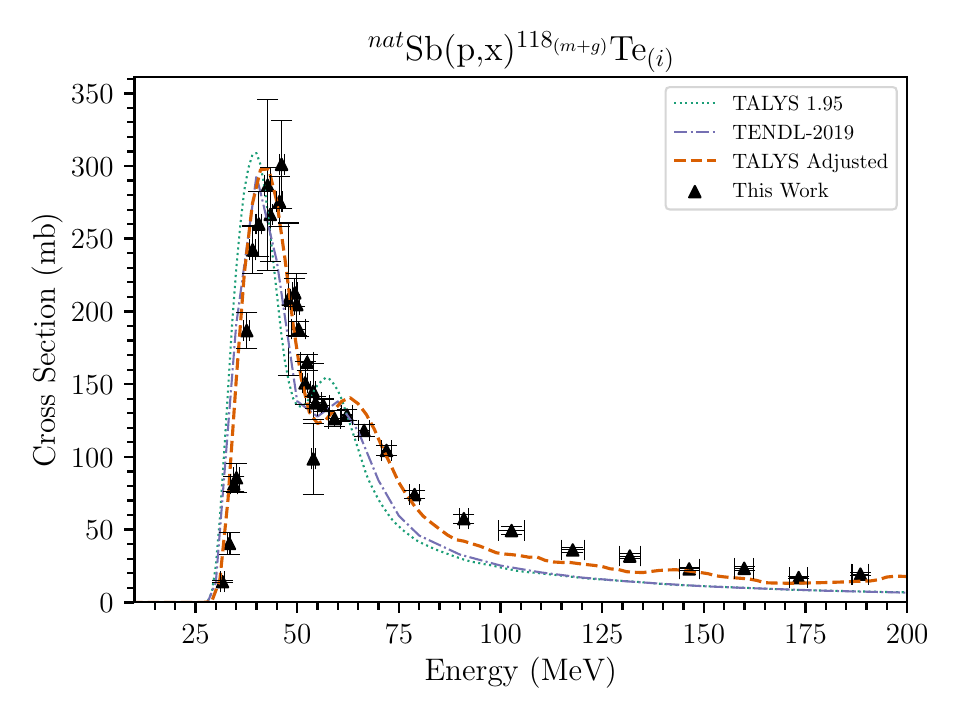} \label{img:118TETALYSOptimized}}
    \subfloat[\ce{^{119m}Te}]{\includegraphics[width=0.33\linewidth]{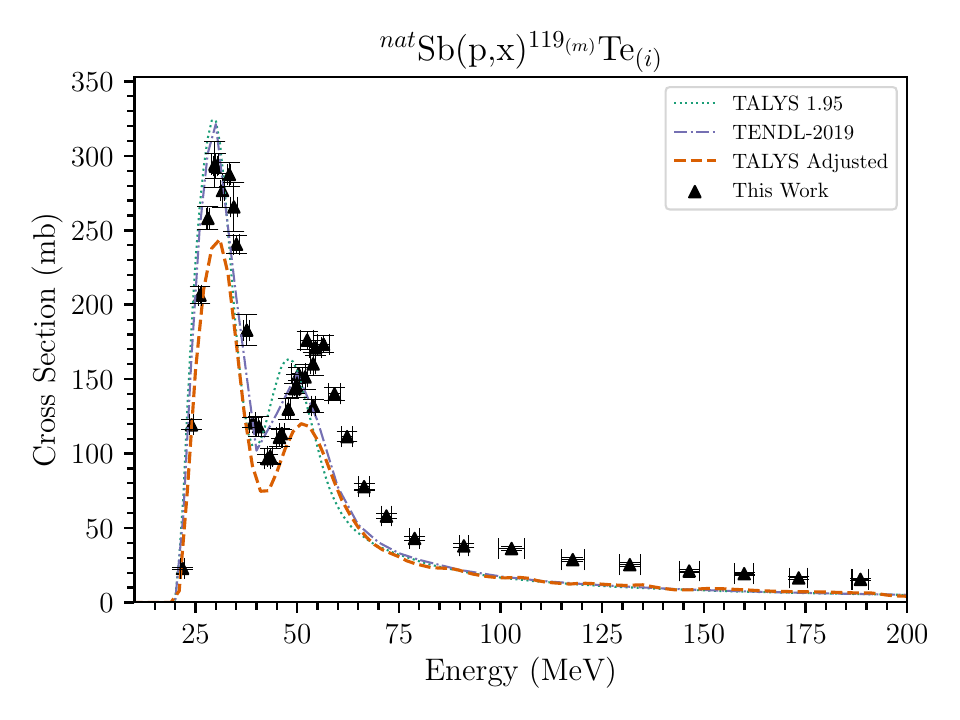} \label{img:119TEmTALYSOptimized}}
    \subfloat[\ce{^{123m}Te}]{\includegraphics[width=0.33\linewidth]{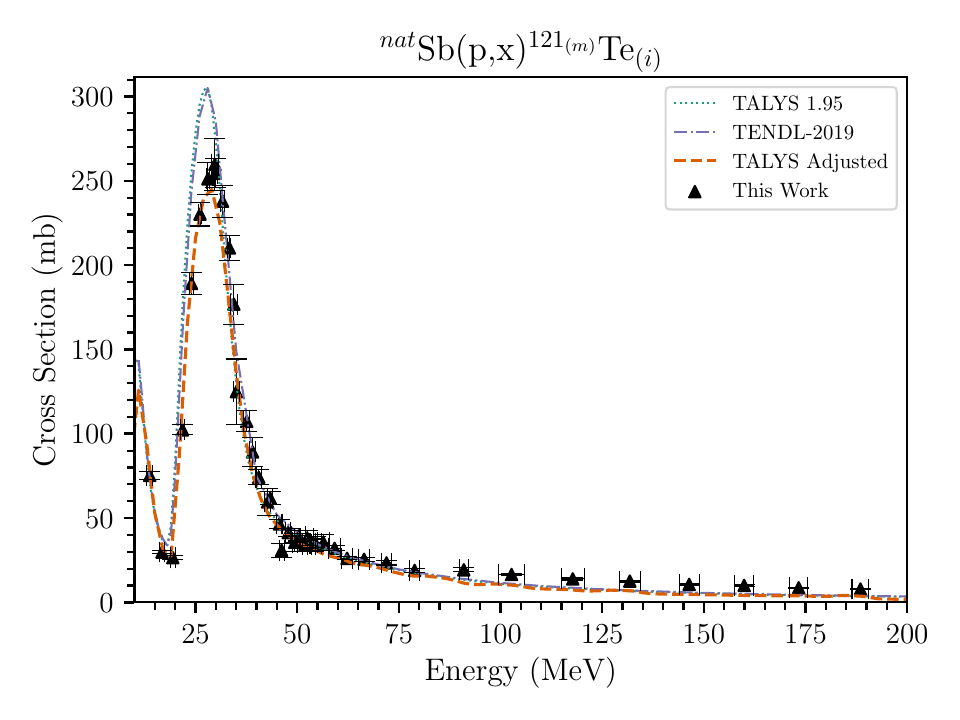} \label{img:121TEmTALYSOptimized}}

    \subfloat[\ce{^{119g}Te}]{\includegraphics[width=0.33\linewidth]{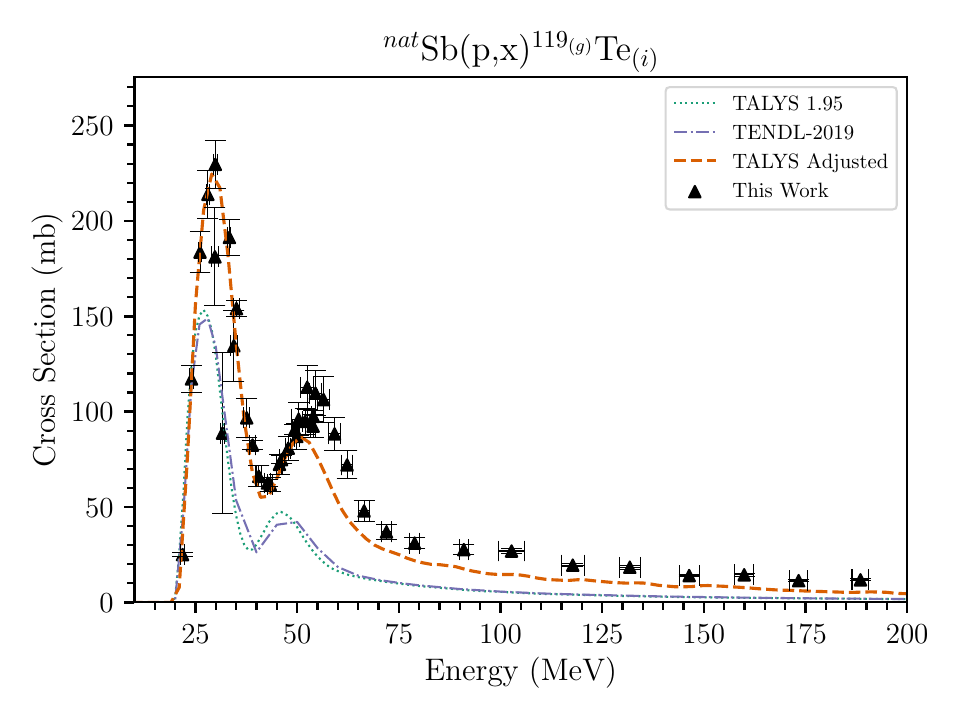} \label{img:119TETALYSOptimized}}
    \subfloat[\ce{^{117}Te}]{\includegraphics[width=0.33\linewidth]{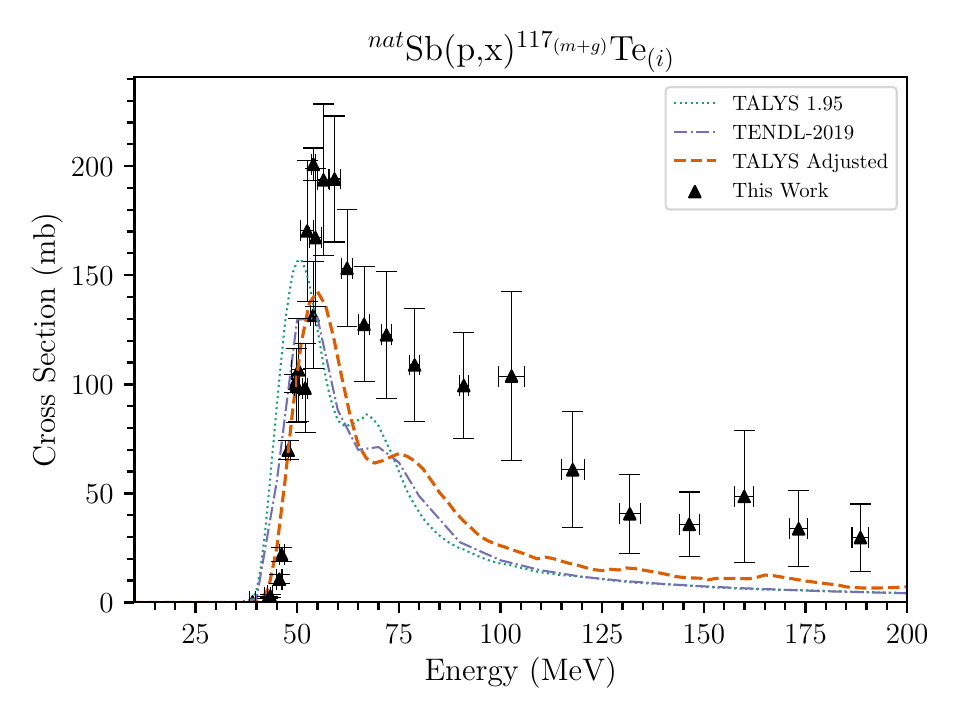} \label{img:117TETALYSOptimized}}
    \subfloat[\ce{^{121g}Te}]{\includegraphics[width=0.33\linewidth]{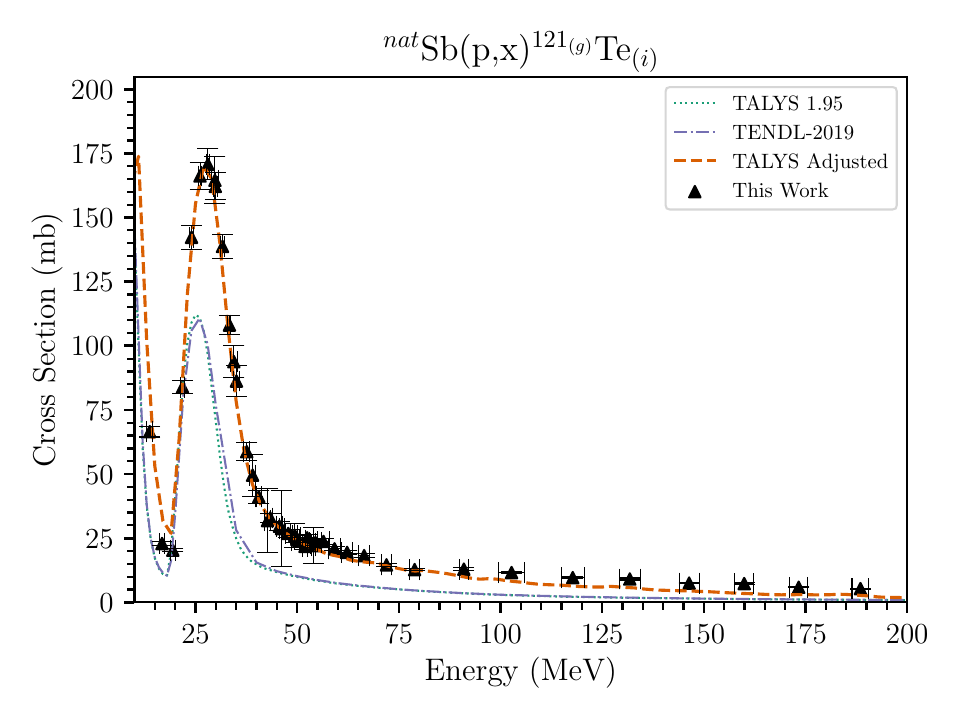} \label{img:121TETALYSOptimized}}

    \subfloat[\ce{^{116}Te}]{\includegraphics[width=0.33\linewidth]{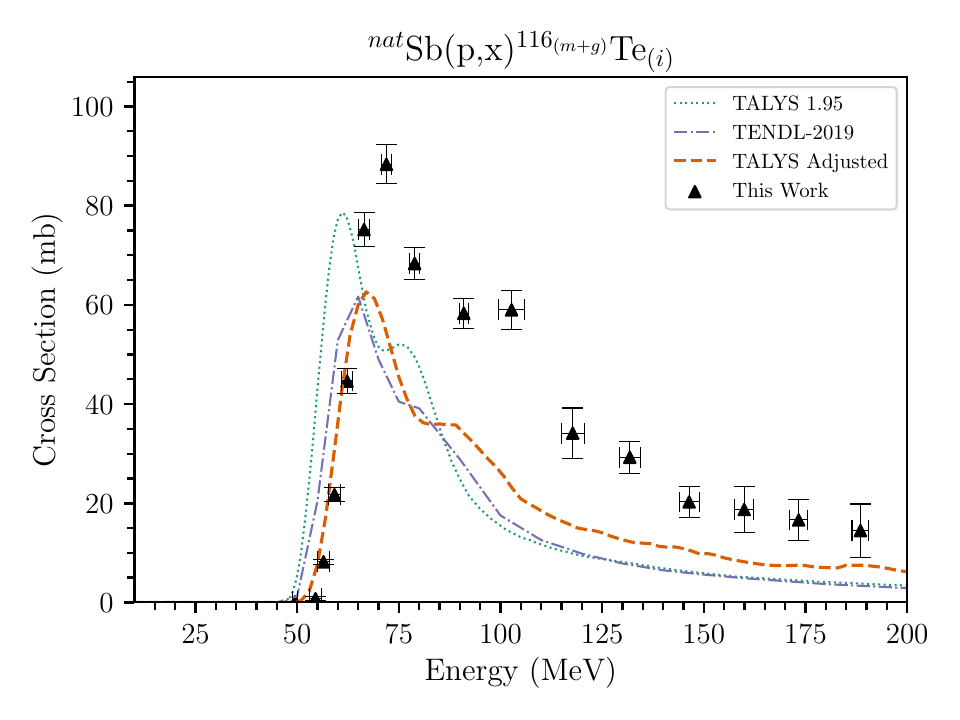} \label{img:116TETALYSOptimized}}
    \subfloat[\ce{^{122}Sb}]{\includegraphics[width=0.33\linewidth]{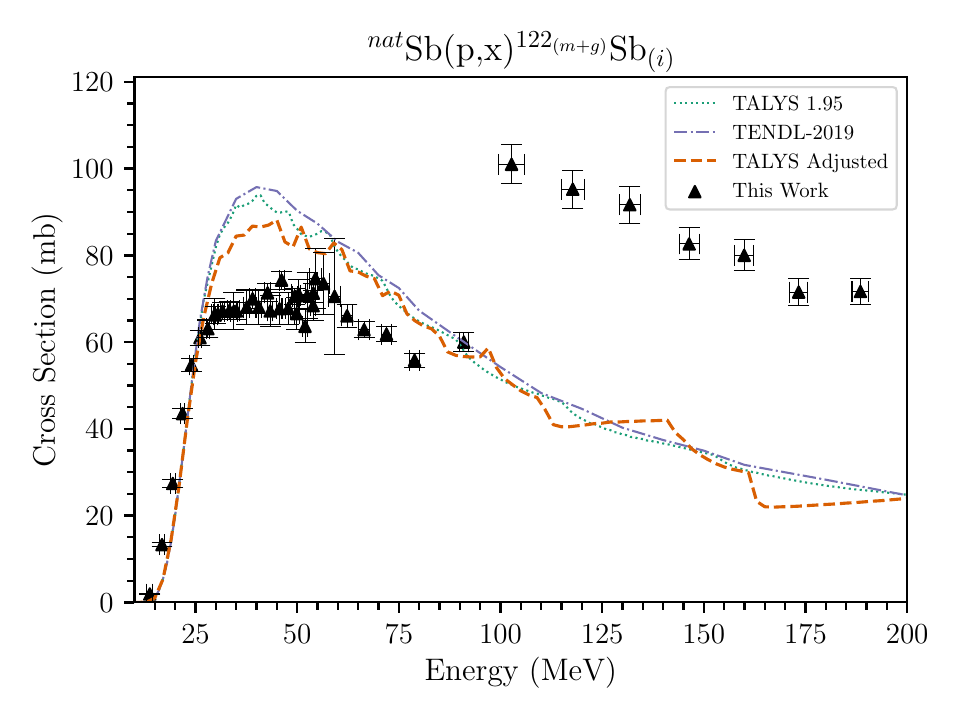} \label{img:122SBTALYSOptimized}}
    \subfloat[\ce{^{118m}Sb}]{\includegraphics[width=0.33\linewidth]{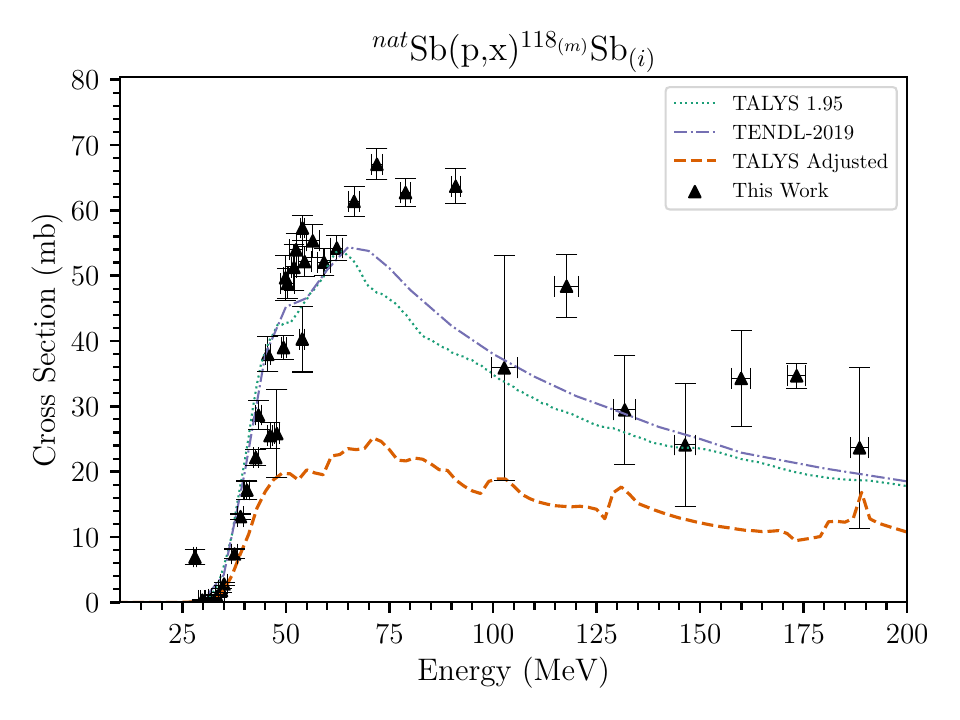} \label{img:118SBmTALYSOptimized}}

    \subfloat[\ce{^{120m}Sb}]{\includegraphics[width=0.33\linewidth]{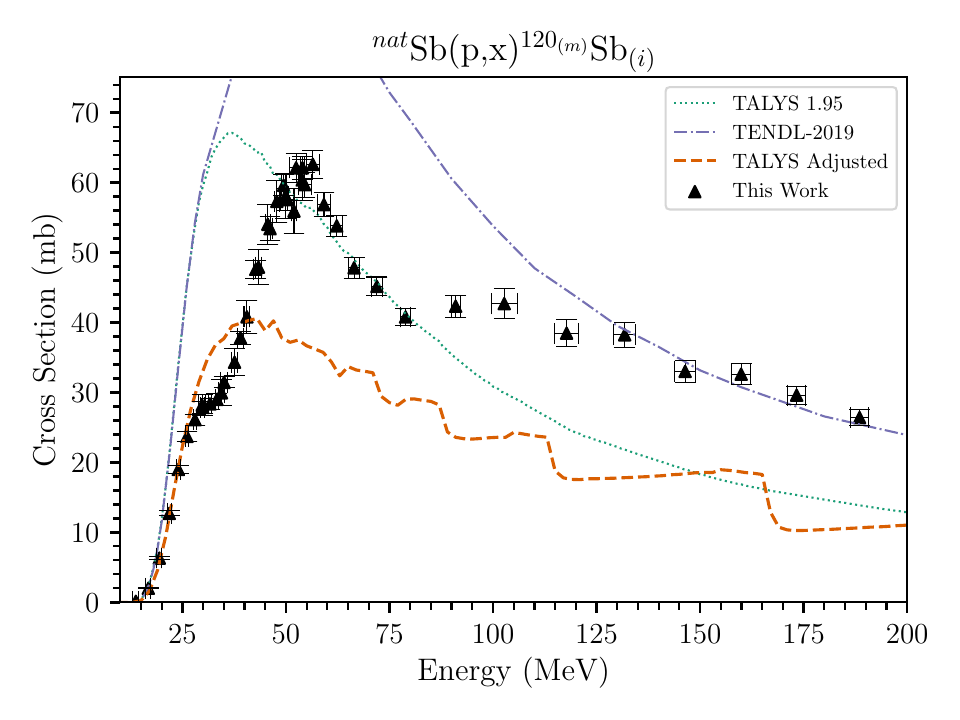} \label{img:120SBmTALYSOptimized}}
    \subfloat[\ce{^{116m}Sb}]{\includegraphics[width=0.33\linewidth]{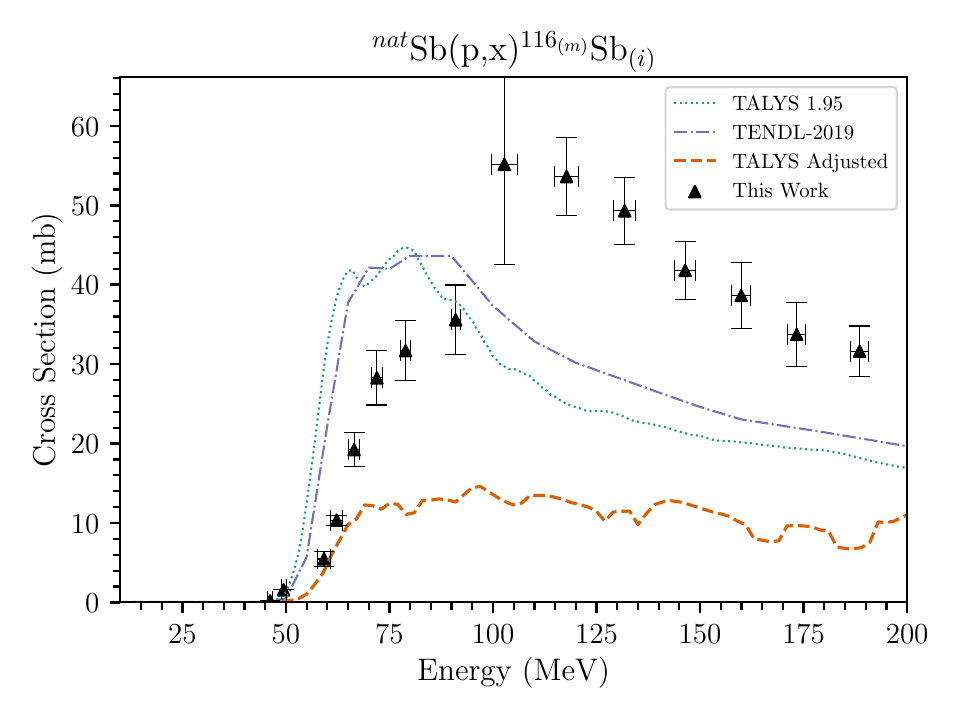} \label{img:116SBmTALYSOptimized}}
    \subfloat[\ce{^{114m}In}]{\includegraphics[width=0.33\linewidth]{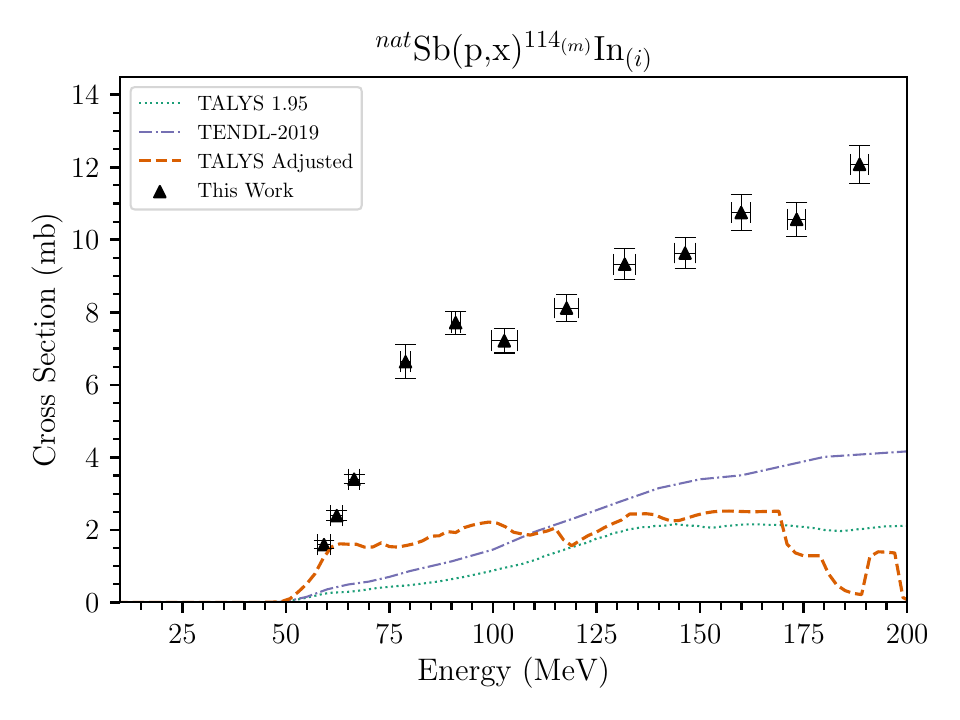} \label{img:114INmTALYSOptimized}}

    \caption{Experimental datapoints, along with baseline TALYS 1.95 values, TENDL 2019 results, and optimized TALYS 1.95 parameter adjustments.}
    \label{fig:TALYSchanneloptimized} 
\end{figure*}

 Twelve of the largest measured reaction channels were used to optimize TALYS 1.95 parameters. The plotted results are shown in \autoref{fig:TALYSchanneloptimized} with tabulated results by isotope in \autoref{tab:chisqindependentchannel}.

The \ce{^{nat}Sb}(p,$x$n)Te channels had the highest magnitude cross sections and, therefore, had the largest influence on the overall $\chi^2$ minimization procedure.
In contrast, smaller channels that resulted in the formation of Sb residual products showed mixed results. 
Some channels show a poorer overall fit between the measured and modeled residual product cross sections.
This systematic disagreement between optimized modeling for Sb(p,x)Sb channels, which could also be formed via \ce{^{nat}Sb}(n,$x$n)Sb reactions, further supports the idea that there is significant co-production of neutron-deficient Sb nuclei via secondary neutrons produced in the target stacks. 
Neutron-deficient products with the same Z as the target foil in high-energy stacked target activation experiments may be misinterpreted, and further studies of secondary neutron production should be performed to explore the role of the particles in isotope production. 
\begin{table}[ht]
\caption{Adjustments to pre-equilibrium to improve fit for Sb and In products}
\label{tab:SbInPreequil}
\begin{tabular}{ccc}
\hline \hline
\begin{tabular}[c]{@{}c@{}}TALYS \\ Parameter\end{tabular} & \begin{tabular}[c]{@{}c@{}}Best Fit\\ (Te)\end{tabular} & \begin{tabular}[c]{@{}c@{}}Best Fit\\ (Sb, In)\end{tabular} \Tstrut\Bstrut\\ \hline
\texttt{rpipi} & 1 & 0.25 \Tstrut\\
\texttt{rpinu} & 1.5 & 4 \\
\texttt{rnupi} & 1.5 & 0.75 \\
\texttt{rnunu} & 1.5 & 0.25 \Bstrut\\
\hline \hline
\end{tabular}
\end{table}

In order to further explore the potential contributions of (n,$x$n) reactions from secondary neutrons to the production of neutron-deficient Sb nuclei, a full review of the effect of TALYS parameter adjustments on these channels was performed.
Nevertheless, this does not fully explain why the reaction model adjustments do not improve the Sb and In residual channels used.
The best reaction model parameters from this second optimization were generally in agreement, except for adjustments to the residual nucleon-nucleon pre-equilibrium interactions \texttt{rpipi}, \texttt{rpinu}, \texttt{rnupi}, and \texttt{rnunu}, shown in \autoref{tab:SbInPreequil}.
The significant deviation in the magnitude of the parameters required to fit Sb and In excitation functions, including, most notably, a large increase in \texttt{rpinu} and decrease in \texttt{rnunu}. These adjustments would enhance proton emission while decreasing neutron emission but would also significantly reduce the compound peaks of residual Te products.

Lastly, among the parameters explored but not ultimately adjusted, improvements were observed when the optical model parameters for the real part of the potential well were adjusted. However, as discussed in \autoref{sec:OpticalModel}, it is difficult to justify changing these variables.
          
\begin{table*}[ht]
\centering
\begin{small}
\caption{Tabulated $\chi^2$ by Isotope}
\label{tab:chisqindependentchannel}
\tabcolsep=0.11cm
\begin{tabular}{ccc|ccc}
\hline \hline
Isotope & \begin{tabular}[c]{@{}c@{}}Sum of \\ Measurements\,(mb)\end{tabular} & \begin{tabular}[c]{@{}c@{}}Maximum Cross\\ Section\,(mb)\end{tabular} & \begin{tabular}[c]{@{}c@{}}$\chi^2$ \\ (Default)\end{tabular} & \begin{tabular}[c]{@{}c@{}}$\chi^2$\\ (Optimized)\end{tabular} & Improvement Ratio  \Tstrut\Bstrut
\\ \hline 
\ce{^{118}Te} & 4497.24 & 301.19 & 457.22 & 41.36 & 11.06 \Tstrut\\
\ce{^{119m}Te} & 5273.01 & 294.37 & 211.14 & 97.58 & 2.16 \\
\ce{^{121m}Te} & 3229.94 & 259.81 & 68.35 & 30.44 & 2.25 \\
\ce{^{119}Te} & 3424.78 & 229.65 & 148.76 & 22.78 & 6.53 \\
\ce{^{117}Te} & 2534.30 & 200.82 & 1148.87 & 20.24 & 56.78 \\
\ce{^{121}Te} & 2095.37 & 170.73 & 190.68 & 35.33 & 5.40 \\
\ce{^{116}Te} & 558.04 & 88.36 & 1293.79 & 50.70 & 25.52 \\
\ce{^{122}Sb} & 2228.27 & 74.65 & 71.19 & 54.11 & 1.33 \\
\ce{^{118m}Sb} & 1000.21 & 67.04 & 42.66 & 122.45 & 0.35 \\
\ce{^{120m}Sb} & 1463.03 & 62.62 & 381.31 & 74.97 & 5.09 \\
\ce{^{116m}Sb} & 132.36 & 35.58 & 345.40 & 15.75 & 21.93 \\
\ce{^{114m}In} & 89.42 & 12.08 & 328.37 & 246.69 & 1.33 \Bstrut\\
\hline \hline
\end{tabular}
\end{small}
\end{table*}
\subsection{\label{sec:Cumulativexs} Validation with Additional Cross Sections}

Isotopes produced both directly and indirectly via the decay of other reaction products are referred to as cumulative channels.
It is particularly challenging, or not possible, to deconvolute these cumulative channels into their independent counterparts. 
In lieu of this, the cumulative results add the cross sections of the parent product and the corresponding branching ratio along the drip line for the isotope. 
Because several products comprise these cross sections, the physics behind their modeling is significantly more complex. 
Using these products to guide reaction modeling is therefore not appropriate.
However, these channels provide an opportunity to validate the adjustments made to improve the fit to the independent channels. 

\begin{figure*}[!!!!htbp]
    \centering
    \subfloat[\ce{^{115}Sb}]{\includegraphics[width=0.33\linewidth]{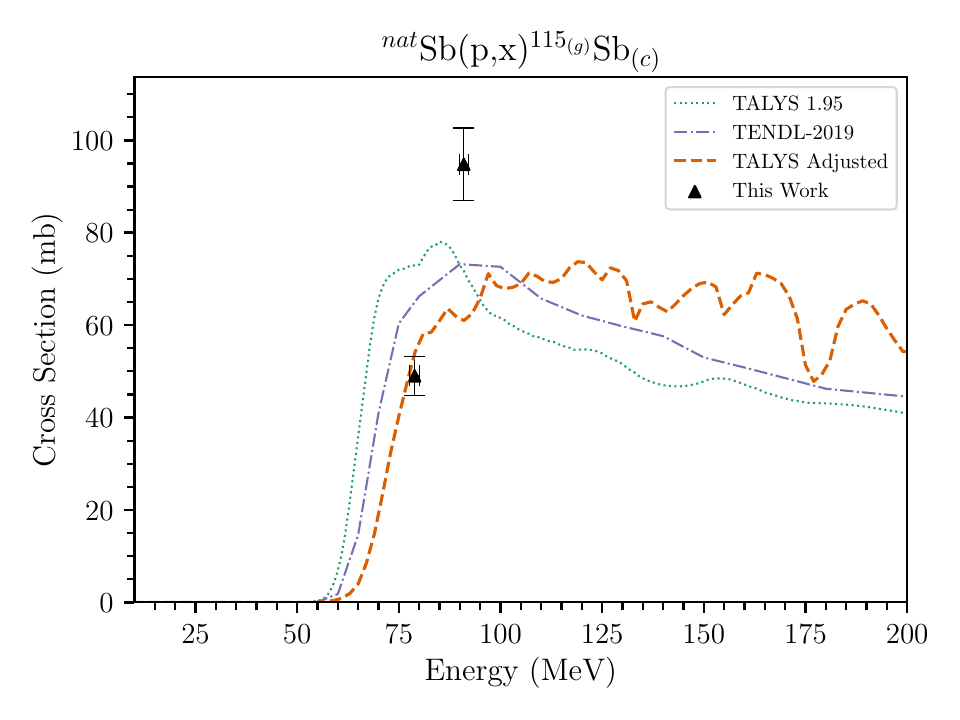} \label{img:115SBTALYSValidated}}
    \subfloat[\ce{^{113}Sn}]{\includegraphics[width=0.33\linewidth]{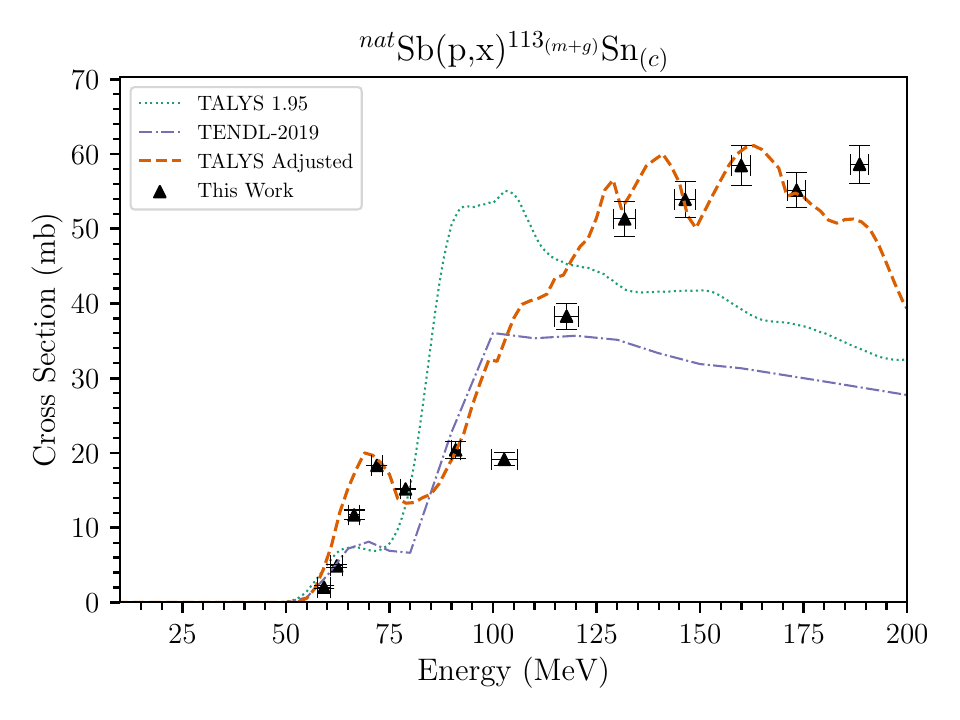} \label{img:113SNTALYSValidated}}
    \subfloat[\ce{^{111}In}]{\includegraphics[width=0.33\linewidth]{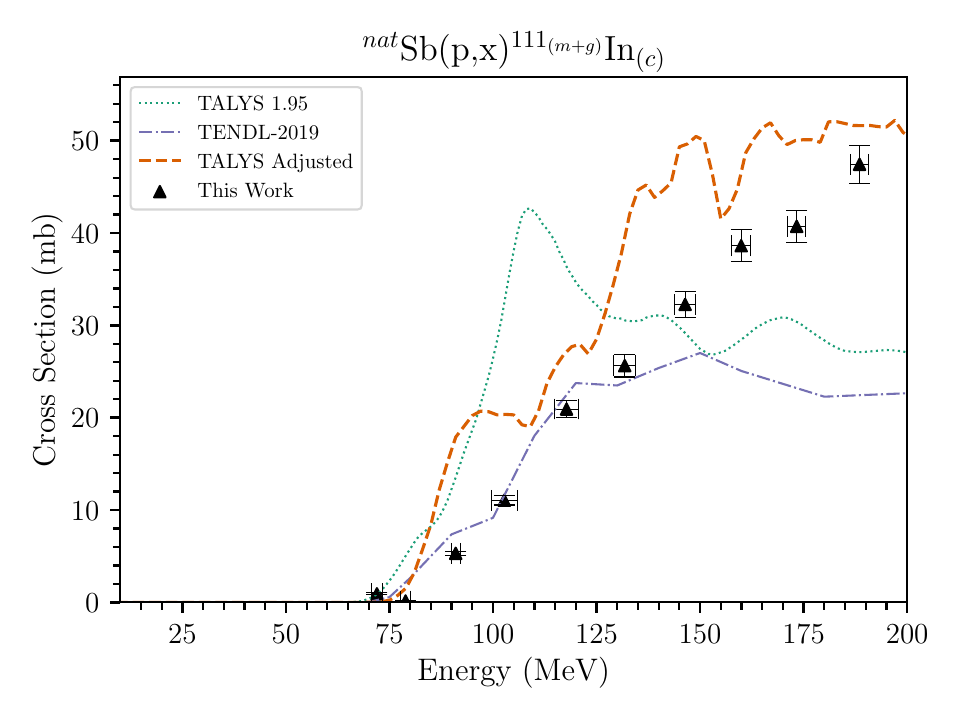} \label{img:111INTALYSValidated}}

    \subfloat[\ce{^{111}Sn}]{\includegraphics[width=0.33\linewidth]{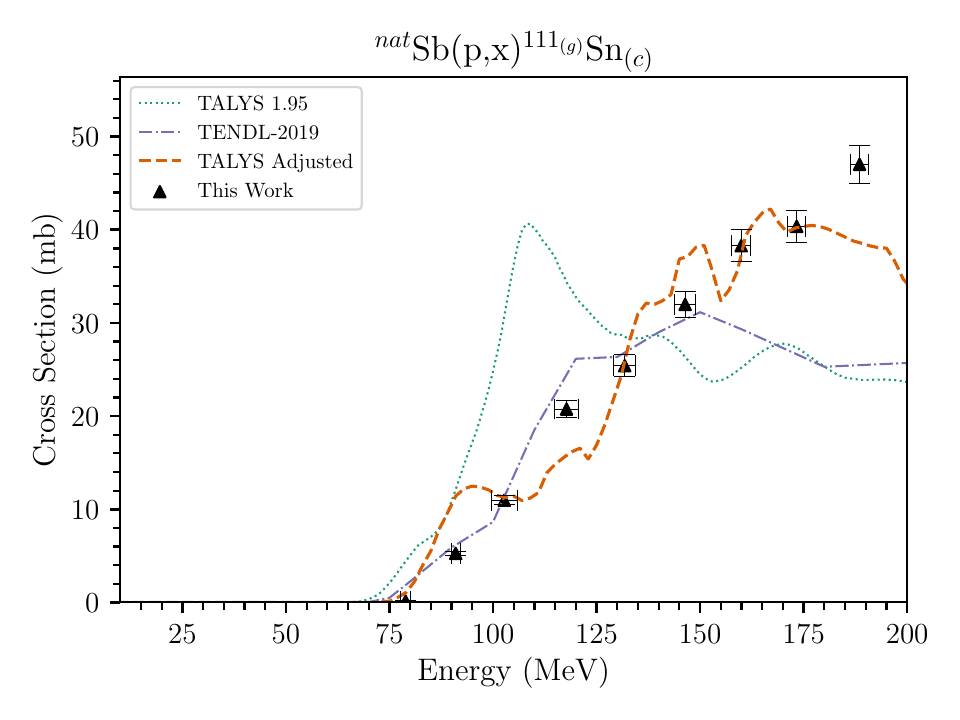} \label{img:111SNTALYSValidated}}
    \subfloat[\ce{^{117m}Sn}]{\includegraphics[width=0.33\linewidth]{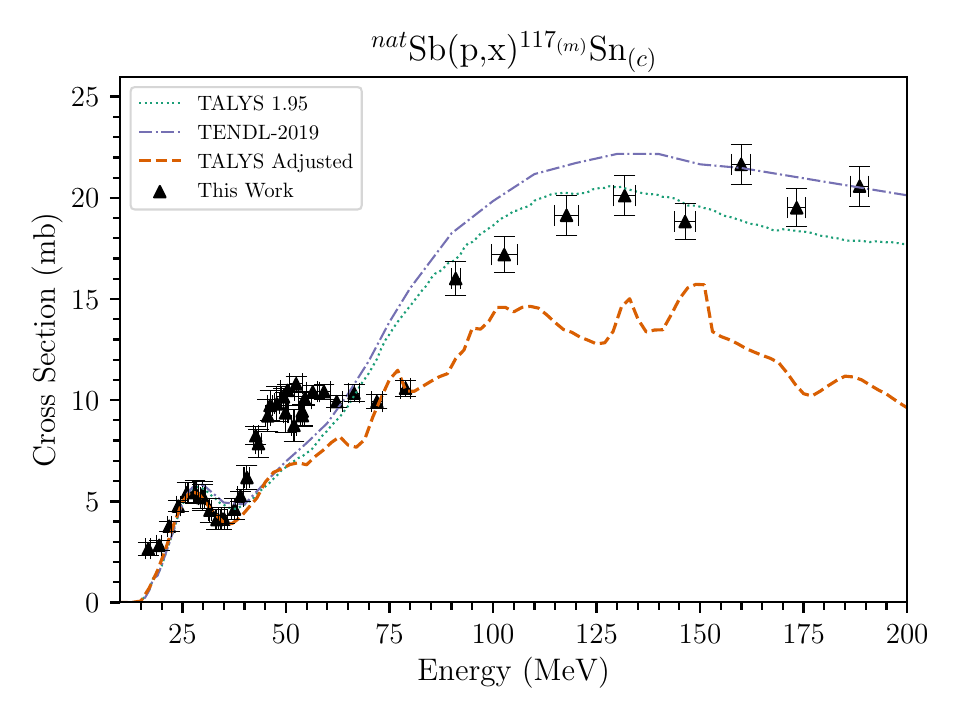} \label{img:117SNmTALYSValidated}}
    \subfloat[\ce{^{109}In}]{\includegraphics[width=0.33\linewidth]{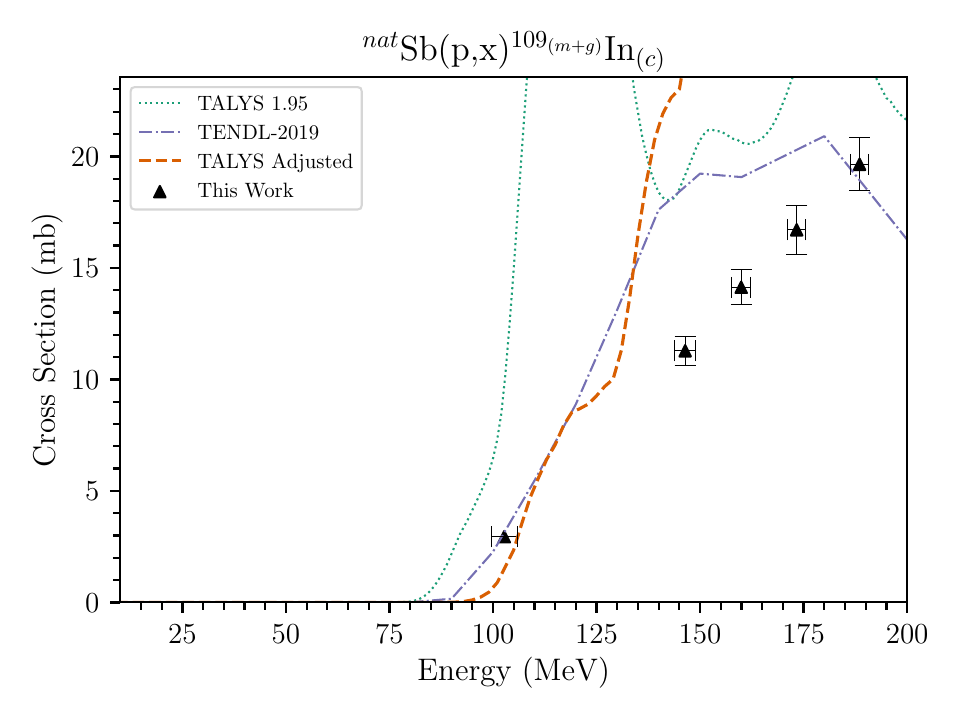} \label{img:109INTALYSValidated}}

    \subfloat[\ce{^{119m}Sn}]{\includegraphics[width=0.33\linewidth]{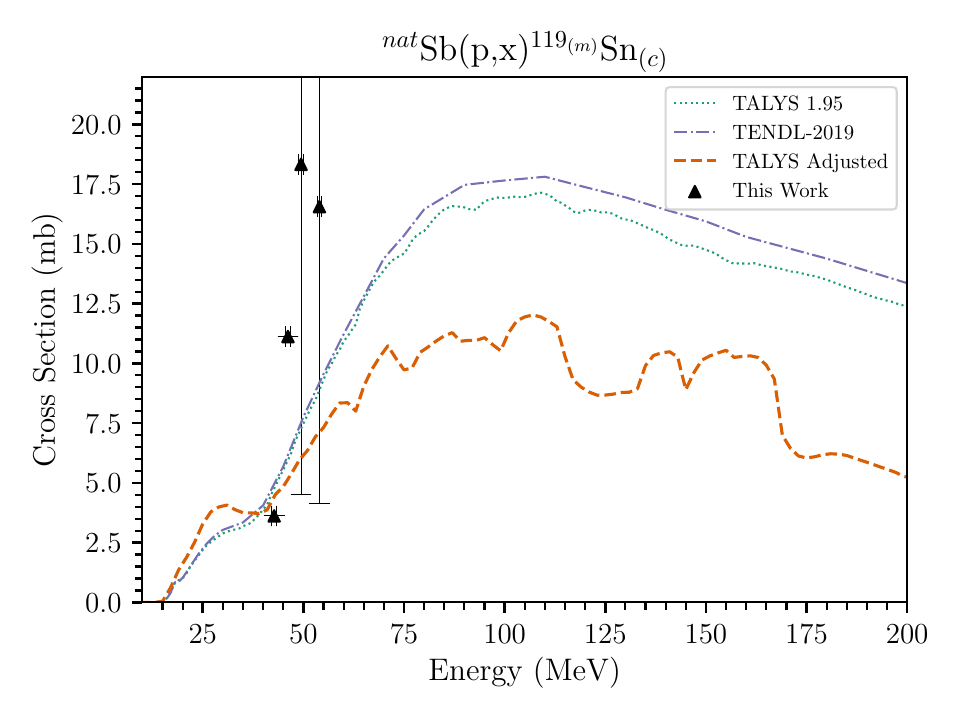} \label{img:119SNmTALYSValidated}}
    \subfloat[\ce{^{109}Sn}]{\includegraphics[width=0.33\linewidth]{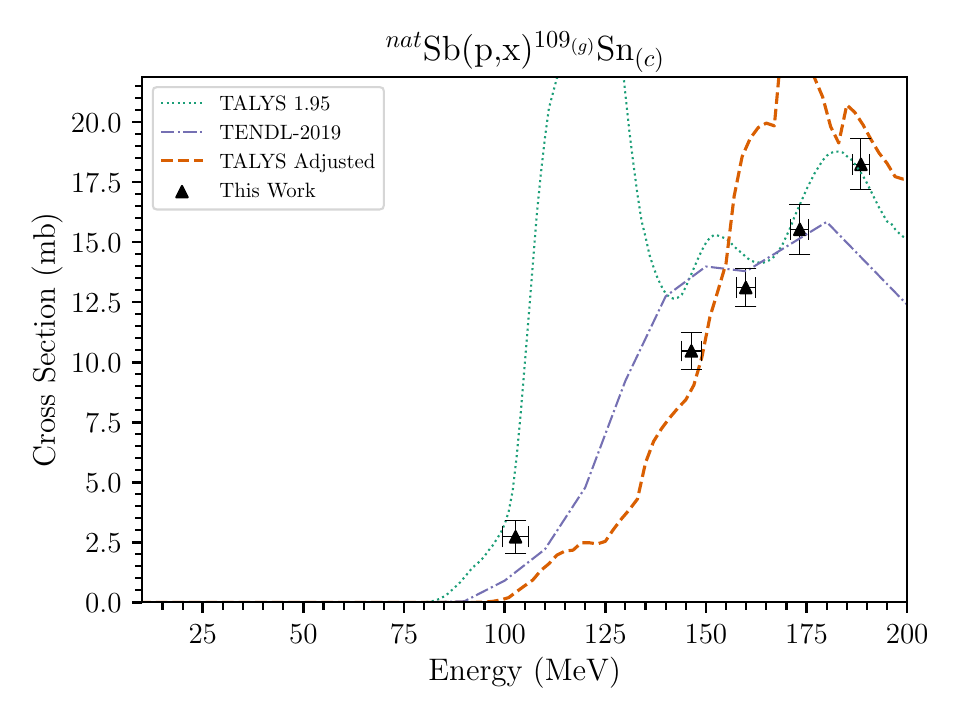} \label{img:109SNTALYSValidated}}
    \subfloat[\ce{^{113m}Sn}]{\includegraphics[width=0.33\linewidth]{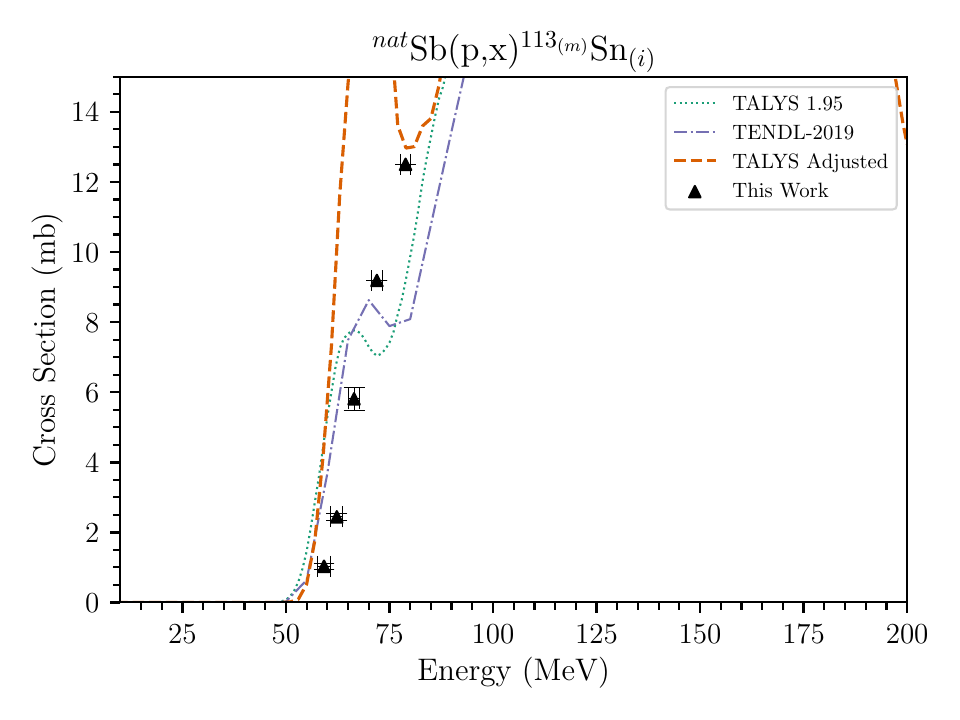} \label{img:113SNmTALYSValidated}}

    \subfloat[\ce{^{110}Sn}]{\includegraphics[width=0.33\linewidth]{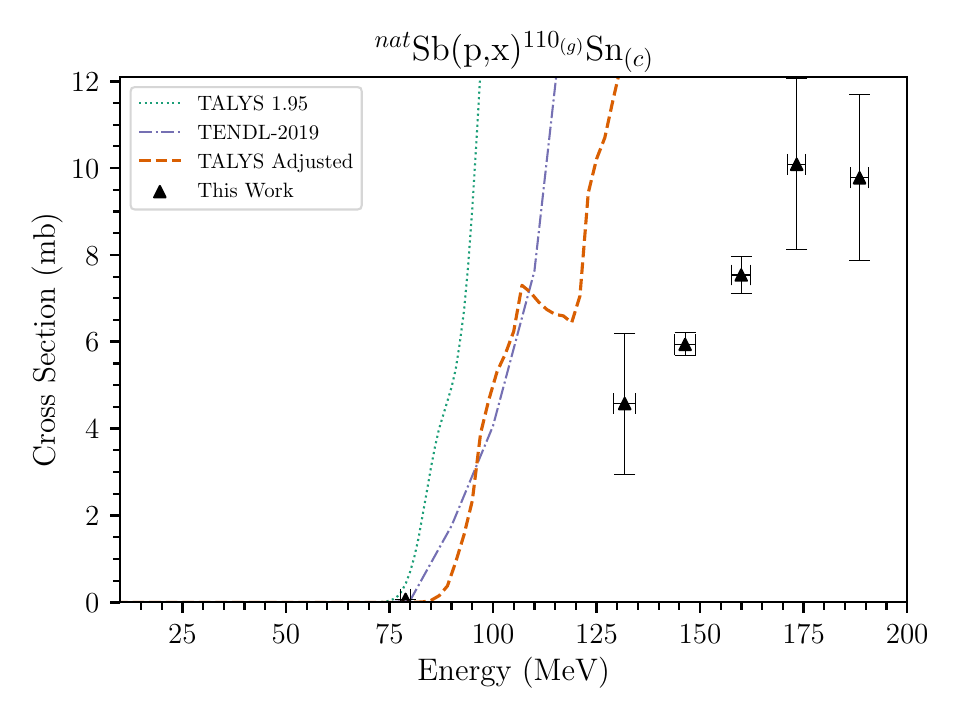} \label{img:110SNTALYSValidated}}
    \subfloat[\ce{^{123m}Te}]{\includegraphics[width=0.33\linewidth]{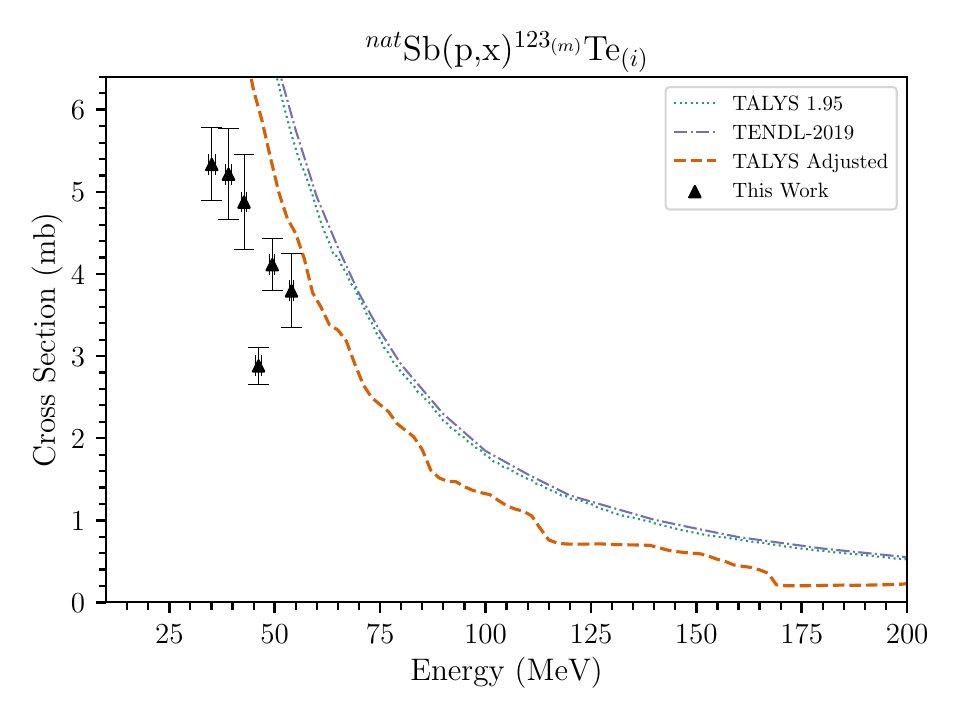} \label{img:123mTETALYSValidated}}
    \subfloat[\ce{^{106m}Ag}]{\includegraphics[width=0.33\linewidth]{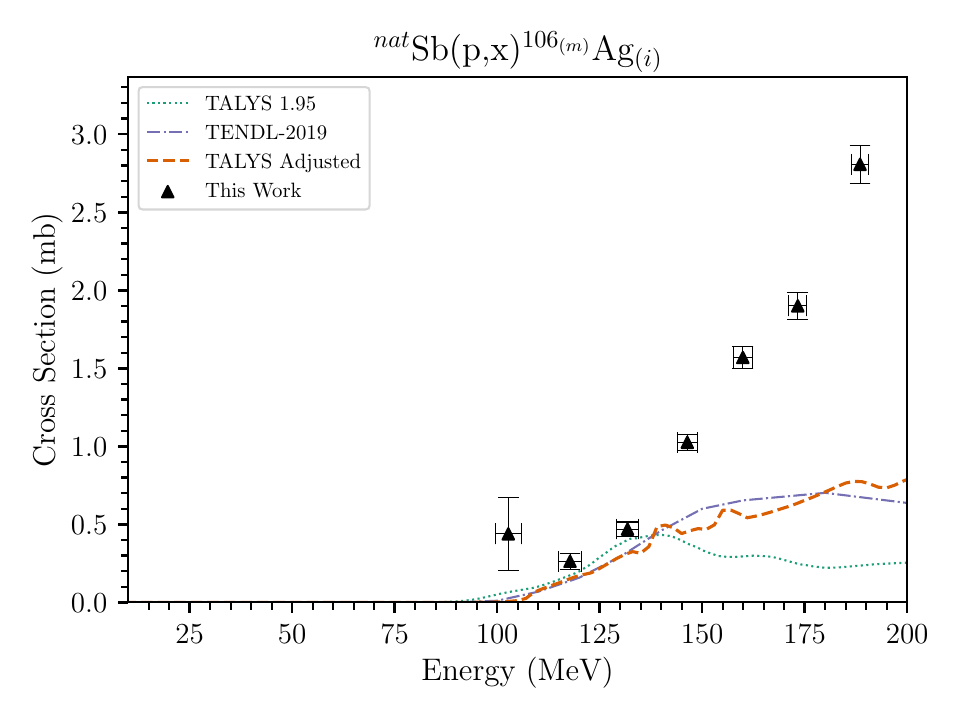} \label{img:106AGmTALYSValidated}}

    \caption{Experimental datapoints, along with baseline TALYS 1.95 values, TENDL 2019 results, and optimized TALYS 1.95 parameter adjustments used for validation.}
    \label{fig:TALYSchannelvalidated} 
\end{figure*}

Nine cumulative channels were used for validation: \ce{^{109g}In}, \ce{^{109g}Sn}, \ce{^{110g}Sn}, \ce{^{111g}In}, \ce{^{111g}Sn}, \ce{^{113g}Sn}, \ce{^{115g}Sb}, \ce{^{117m}Sn}, and \ce{^{119m}Sn}. 
\ce{^{91m}Nb} was excluded because the \texttt{equidistant y} command in TALYS 1.95 does not provide results for this residual product, despite the reaction threshold being within the incident proton energy of these experiments.
Validation also includes 3 independent channels: \ce{^{106m}Ag}, \ce{^{113m}Sn} and \ce{^{123m}Te}.
These were either weakly fed channels or had few experimental data and were therefore used for validation instead of sensitivity analysis.
Much like \ce{^{91m}Nb}, \ce{^{89m}Nb} was excluded since \texttt{equidistant y} in TALYS 1.95 does not provide results for this residual product.
These validation cross sections are illustrated in \autoref{fig:TALYSchannelvalidated} with default and optimized input parameters. 

The improvement in the goodness of fit to parameter adjustments for cumulative channels is shown in \autoref{tab:chisqtotalvalidation}.

\begin{table}[!!!htbp]
\centering
\caption{Default and optimized parameters for validation cross sections with improvement ratio}
\label{tab:chisqtotalvalidation}

\begin{tabular}{cccc}
\hline \hline
           & \begin{tabular}[c]{@{}c@{}}TALYS 1.95 \\ Default\end{tabular} & \begin{tabular}[c]{@{}c@{}}Optimized \\ Parameters\end{tabular} & \begin{tabular}[c]{@{}c@{}}Improvement \\ Ratio\end{tabular} \Tstrut\Bstrut\\
 \hline
Cumulative $\sigma$ & 1070.88 & 86.41 & 12.39 \Tstrut\\
Maximum $\sigma$& 1073.90 & 89.17 & 12.04 \Bstrut\\
\hline\hline
\end{tabular}
\end{table}

Again, several validation cross sections had a decrease in fit with the optimized parameters.
The optimized parameters involved adjustments to optical model parameters, which affect both the real and imaginary optical model potentials. 
\autoref{tab:validationadjustments} provides the optimized adjustments. 
\begin{table}[!!!htbp]
\centering
\caption{Optimal parameter adjustments for validation cross sections}
\label{tab:validationadjustments}
\begin{tabular}{lll} 
\hline \hline
\begin{tabular}[c]{@{}l@{}}TALYS\\ Parameter\end{tabular} & \begin{tabular}[c]{@{}l@{}}Best Fit\\  (Independent)\end{tabular} & \begin{tabular}[c]{@{}l@{}}Best Fit \\ (Validation)\end{tabular} \Tstrut\Bstrut\\
\hline
\texttt{rvadjust} & 1 & 2.5 \Tstrut\\
\texttt{rwadjust} & 1 & 0.5 \\
\texttt{avadjust} & 1 & 1.5 \\
\texttt{awadjust} & 1 & 1.5 \Bstrut\\
\hline \hline
\end{tabular}
\end{table}

\subsection{Conclusions and Takeaways from Reaction Modeling}
The reaction model fits were dominated by the largest independent reaction channels, specifically \ce{^{nat}Sb}(p,$x$n) channels for residual Te products. 
The optimized parameters are provided in \autoref{tab:finalTALYS}.

\begin{table}[!!!htbp]
\caption{Optimized TALYS Parameter Adjustments}
\label{tab:finalTALYS}
\begin{tabular}{cc}
\hline \hline
\begin{tabular}[c]{@{}c@{}}TALYS\\
Parameter\end{tabular} & Value \Tstrut\Bstrut\\
\hline
\texttt{ldmodel} & 2 \Tstrut\\
\texttt{strength} & 5 \\
\texttt{equidistant} & y \\
\texttt{rspincut} & 0.4 \\
\texttt{spincutmodel} & 2 \\
\texttt{w1adjust} & n 2.5 \\
\texttt{w2adjust} & n 0.6 \\
\texttt{preeqmode} & 1 \\
\texttt{m2constant} & 2 \\
\texttt{m2limit} & 0.8 \\
\texttt{m2shift} & 1.8 \\
\texttt{rpinu} & 1.5 \\
\texttt{rnupi} & 1.5 \\
\texttt{rpipi} & 1 \\
\texttt{rnunu} & 1.5 \Bstrut\\
\hline \hline
\end{tabular}
\end{table}

The overall fit for these channels was dramatically improved, even in isomer-to-ground-state ratios. However, this research raises several concerns. 
The fit in the pre-equilibrium region is still imperfect, as illustrated both by other independent cross sections as well as by the validation cross sections. 

Analyzing the remaining independent cross sections indicated a need for adjustment to the residual nucleon-nucleon interaction in the pre-equilibrium region, but these values were adjusted to such a significant degree that it was deemed not physically defensible. 
These adjustments concurrently flattened the compound peaks in the Te channels, producing worse fits for Te channels. 

The validation channels, 9 of which were cumulative and 3 independent, had an improved overall fit, but the largest reaction channels dominate this optimization. 
A review of the validation channels alone suggested changes to the form factor values in the optical potential model. 
This impacts both the real and imaginary components of the model, which would suggest a change in the physical volume of the nucleus for this specific reaction.


    \section{\label{sec:Conclusion}Conclusions\protect}
        
The work presented in this paper includes residual product cross sections measured using the stacked target activation technique for the medically relevant Auger-emitting radionuclide \ce{^{117m}Sn}, and \ce{^{119m}Te} - a generator for another radioisotope, \ce{^{119}Sb}. 
An additional 22 cross sections were measured; this includes measurements of \ce{^{113}Sn}, \ce{^{121m}Te}, \ce{^{121g}Te}, and \ce{^{123m}Te} contaminants, whose presence impacts the viability of these radionuclides for \emph{in vivo} use.
These measurements not only have medical importance but also provide essential guidance to the underlying physics in nuclear reaction modeling for proton-induced reactions.
This highlights the need to dramatically curtail spin distributions. 
This result is consistent with the recent compilation of Rodrigo\,\etal and has significant implications for residual nuclear reaction products with isomeric states hindered by significant differences in angular momentum.
This work also highlights the need to adjust the imaginary neutron optical potential and the parameters describing pre-equilibrium particle emission.  

Furthermore, this work highlights that results from a stacked target measurement cannot unambiguously assign the correct pre-equilibrium parameters independently. 
This included modifications to the exciton scattering matrix element parameters. 
This reinforces the earlier work of Fox\,\etal, which also indicates the need to adjust the exciton scattering matrix.
However, measurement of the outgoing neutron spectrum is needed to definitively assign the parameter choices that best represent the underlying physics.
This is merited by the observation of a large quantity of secondary-neutron-induced isotope production in higher incident energy proton beams.
Future experiments should include outgoing neutron spectral and angular information and total residual channel cross sections.
This uncertainty in pre-equilibrium modeling has significant implications, given that regional accelerator production facilities utilize high-energy proton beams.  

This work is the latest in a growing body of experimental work starting with Voyles\,\etal \cite{voyles_excitation_2018}, Morrell\,\etal \cite{morrell_measurement_2020}, and Fox\,\etal \cite{fox_measurement_2021, fox_investigating_2021} performed by the Tri-laboratory Effort in Nuclear Data, whose goal is an improved ability to model energetic proton-induced reactions for isotope production.  
The presented data highlights the value of improvements to reaction modeling. Improvements to reaction models can inform and direct future experiments. However, the acquisition of data for charged particle nuclear reactions is limited by time and resources. Nevertheless, these results suggest further experimental and computational research, encompassing a larger dataset, will improve reaction modeling holistically. 

The decay gamma spectroscopy used in this research is avalable for access. Upon publication, the experimental cross section datasets will be added to the EXFOR database.

    \begin{acknowledgments}
        This research was supported by the Department of Energy National Nuclear Security Administration through the Nuclear Science and Security Consortium under Award Number(s) DE-NA0003180, the DOE Isotope Program within the U.S. Department of Energy Office of Science, carried out under Lawrence Berkeley National Laboratory (Contract No. DE-AC02-05CH11231), Los Alamos National Laboratory (Contract No. 89233218CNA000001), and Brookhaven National Laboratory (Contract No. DEAC02-98CH10886).
The authors would like to acknowledge the support of the staff at the LBNL 88-Inch Cyclotron, including the operations, research, and facilities staff. The authors would also like to thank the team at the LANL IPF facility, the Accelerator Operations staff, and the staff of C-NR Countroom. We would also like to thank the group at the BNL LINAC and Isotope Research and Production group. The authors would also like to thank the radiological protection group staff present for each experiment for their support.
    \end{acknowledgments}

\appendix

    \section{\label{app:FoilCharacterization}Stacked Target Characterization}
\squeezetable

\begin{longtable}{@{}p{0.35\linewidth} p{0.3\linewidth} p{0.3\linewidth}@{}}
    \caption{Stack design for the LBNL 2020\,55 MeV run.}\\
    \label{tab:LBNLHEfoils2020}\\
    \hline \hline
    Target Layer & Areal Density & Areal density \Tstrut \\ 
     & (mg/cm$^2$) & uncertainty (\%)  \Bstrut \\
     \hline
    \endfirsthead
    \hline \hline
    Target Layer & Areal Density & Areal density \Tstrut  \\  
     & (mg/cm$^2$) & uncertainty (\%)  \Bstrut\\
     \hline
    \endhead
   \hline  \multicolumn{3}{r}{{Continued on next page}} \\  \hline\hline
\endfoot
\hline \hline
\endlastfoot
    SS Profile Monitor 2 & 100.20 & 0.07  \Tstrut\\
    Cu-SN1 & 22.28 & 0.22 \\
    Sb-SN1 & 15.04 & 0.64 \\
    Polyester Back-SN1 & 5.83 & 1.95 \\
    Nb-SN1 & 22.27 & 0.27 \\
    Ti-SN1 & 11.03 & 0.75 \\
    Al Degrader E1 & 68.31 & 0.07 \\
    Al Degrader E2 & 68.25 & 0.07 \\
    Cu-SN2 & 22.10 & 0.23 \\
    Sb-SN2 & 15.61 & 0.32 \\
    Polyester Back-SN2 & 5.83 & 1.95 \\
    Nb-SN2 & 22.39 & 0.38 \\
    Ti-SN2 & 11.26 & 0.44 \\
    Al Degrader E3 & 68.35 & 0.07 \\
    Al Degrader E4 & 68.29 & 0.07 \\
    Cu-SN3 & 22.18 & 0.22 \\
    Sb-SN3 & 15.19 & 0.53 \\
    Polyester Back-SN3 & 5.83 & 1.95 \\
    Nb-SN3 & 22.61 & 0.66 \\
    Ti-SN3 & 11.06 & 0.51 \\
    Al Degrader E5 & 68.24 & 0.07 \\
    Al Degrader E6 & 68.22 & 0.07 \\
    Cu-SN4 & 22.35 & 0.22 \\
    Sb-SN4 & 16.81 & 0.49 \\
    Polyester Back-SN4 & 5.83 & 1.95 \\
    Nb-SN4 & 22.41 & 0.25 \\
    Ti-SN4 & 11.12 & 0.86 \\
    Al Degrader E7 & 68.19 & 0.07 \\
    Al Degrader E8 & 68.16 & 0.07 \\
    Cu-SN5 & 22.32 & 0.45 \\
    Sb-SN5 & 15.95 & 0.51 \\
    Polyester Back-SN5 & 5.83 & 1.95 \\
    Nb-SN5 & 22.40 & 0.36 \\
    Ti-SN5 & 11.22 & 0.43 \\
    Al Degrader E9 & 68.24 & 0.07 \\
    Al Degrader E10 & 68.22 & 0.07 \\
    Cu-SN6 & 22.45 & 0.24 \\
    Sb-SN6 & 17.89 & 0.32 \\
    Polyester Back-SN6 & 5.83 & 1.95 \\
    Nb-SN6 & 22.25 & 0.22 \\
    Ti-SN6 & 10.99 & 0.47 \\
    Al Degrader D1 & 174.44 & 0.05 \\
    Cu-SN7 & 22.57 & 0.26 \\
    Sb-SN7 & 15.63 & 0.37 \\
    Polyester Back-SN7 & 5.83 & 1.95 \\
    Nb-SN7 & 22.41 & 0.23 \\
    Ti-SN7 & 10.74 & 0.54 \\
    Al Degrader D2 & 174.88 & 0.06 \\
    Cu-SN8 & 22.42 & 0.01 \\
    Sb-SN8 & 16.70 & 0.49 \\
    Polyester Back-SN8 & 5.83 & 1.95 \\
    Nb-SN8 & 22.50 & 0.26 \\
    Ti-SN8 & 10.79 & 1.27 \\
    Al Degrader D3 & 175.05 & 0.08 \\
    Cu-SN9 & 22.41 & 0.00 \\
    Sb-SN9 & 17.67 & 0.29 \\
    Polyester Back-SN9 & 5.83 & 1.95 \\
    Nb-SN9 & 22.27 & 0.00 \\
    Ti-SN9 & 11.02 & 5.02 \\
    Al Degrader C1 & 261.48 & 0.07 \\
    Cu-SN10 & 22.48 & 0.23 \\
    Sb-SN10 & 15.24 & 0.38 \\
    Polyester Back-SN10 & 5.83 & 1.95 \\
    Nb-SN10 & 22.44 & 0.22 \\
    Ti-SN10 & 10.79 & 0.50 \\
    SS Profile Monitor 2 & 100.87 & 0.07   \Bstrut\\ 
\end{longtable}

\begin{longtable}{@{}p{0.35\linewidth} p{0.3\linewidth} p{0.3\linewidth}@{}}
    \caption{Stack design for the LBNL 2020 35\,MeV run.}\\
    \label{tab:LBNLLEfoils2020}\\
    \hline \hline
    Target Layer & Areal Density & Areal density \Tstrut \\ 
     & (mg/cm$^2$) & uncertainty (\%)  \Bstrut \\
     \hline
    \endfirsthead
    \hline \hline
    Target Layer & Areal Density & Areal density \Tstrut  \\  
     & (mg/cm$^2$) & uncertainty (\%)  \Bstrut\\
     \hline
    \endhead
   \hline  \multicolumn{3}{r}{{Continued on next page}} \\  \hline\hline
\endfoot
\hline \hline
    SS Profile Monitor 4 & 101.25 & 0.07 \Tstrut\\
    Cu-SN11 & 22.40 & 0.22 \\
    Sb-SN11 & 17.50 & 0.33 \\
    Polyester Back-SN11 & 5.83 & 1.95 \\
    Ti-SN11 & 10.93 & 0.48 \\
    Al Degrader E8 & 68.16 & 0.07 \\
    Cu-SN12 & 22.42 & 0.27 \\
    Sb-SN12 & 17.57 & 0.55 \\
    Polyester Back-SN12 & 5.83 & 1.95 \\
    Ti-SN12 & 11.06 & 0.88 \\
    Al Degrader E9 & 68.18 & 0.07 \\
    Cu-SN13 & 22.15 & 0.26 \\
    Sb-SN13 & 17.68 & 0.55 \\
    Polyester Back-SN13 & 5.83 & 1.95 \\
    Ti-SN13 & 10.93 & 0.89 \\
    Al Degrader E1 & 68.31 & 0.07 \\
    Cu-SN14 & 22.11 & 0.26 \\
    Sb-SN14 & 16.11 & 0.51 \\
    Polyester Back-SN14 & 5.83 & 1.95 \\
    Ti-SN14 & 11.07 & 0.74 \\
    Al Degrader E2 & 68.25 & 0.07 \\
    Cu-SN15 & 22.22 & 0.24 \\
    Sb-SN15 & 16.11 & 0.31 \\
    Polyester Back-SN15 & 5.83 & 1.95 \\
    Ti-SN15 & 10.88 & 0.54 \\
    Al Degrader E3 & 68.35 & 0.07 \\
    Cu-SN16 & 22.44 & 0.00 \\
    Sb-SN16 & 16.53 & 0.30 \\
    Polyester Back-SN16 & 5.83 & 1.95 \\
    Ti-SN16 & 10.86 & 0.75 \\
    Al Degrader E4 & 68.29 & 0.07 \\
    Cu-SN17 & 22.27 & 0.23 \\
    Sb-SN17 & 16.08 & 0.36 \\
    Polyester Back-SN17 & 5.83 & 1.95 \\
    Ti-SN17 & 11.07 & 0.00 \\
    Al Degrader E5 & 68.24 & 0.07 \\
    Cu-SN18 & 22.40 & 0.35 \\
    Sb-SN18 & 17.83 & 0.28 \\
    Polyester Back-SN18 & 5.83 & 1.95 \\
    Ti-SN18 & 10.87 & 0.45 \\
    Al Degrader E6 & 68.22 & 0.07 \\
    Cu-SN19 & 22.51 & 0.24 \\
    Sb-SN19 & 16.55 & 0.49 \\
    Polyester Back-SN19 & 5.83 & 1.95 \\
    Ti-SN19 & 11.10 & 0.87 \\
    Al Degrader E7 & 68.19 & 0.07 \\
    Cu-SN20 & 22.37 & 0.00 \\
    Sb-SN20 & 15.74 & 0.32 \\
    Polyester Back-SN20 & 5.83 & 1.95 \\
    Ti-SN20 & 11.13 & 0.44 \\
    SS Profile Monitor 5 & 100.57 & 0.07 \Bstrut\\ 
\end{longtable}

\begin{longtable}{@{}p{0.35\linewidth} p{0.3\linewidth} p{0.3\linewidth}@{}}
    \caption{Stack design for the LBNL 2022 55\,MeV run.}\\
    \label{tab:LBNLHEfoils2022}\\
    \hline \hline
    Target Layer & Areal Density & Areal density \Tstrut \\ 
     & (mg/cm$^2$) & uncertainty (\%)  \Bstrut \\
     \hline
    \endfirsthead
    \hline \hline
    Target Layer & Areal Density & Areal density \Tstrut  \\  
     & (mg/cm$^2$) & uncertainty (\%)  \Bstrut\\
     \hline
    \endhead
   \hline  \multicolumn{3}{r}{{Continued on next page}} \\  \hline\hline
\endfoot
\hline \hline
SS Profile Monitor 1 & 100.20 & 0.07 \Tstrut\\
Cu-SN01 & 22.06 & 0.05 \\
Sb-SN01 & 16.52 & 0.10 \\
Polyester Back-SN01 & 5.83 & 1.95 \\
Ti-SN01 & 11.09 & 0.00 \\
Al Degrader B1 & 415.02 & 0.07 \\
Cu-SN02 & 21.92 & 0.06 \\
Sb-SN02 & 16.19 & 0.00 \\
Polyester Back-SN02 & 5.83 & 1.95 \\
Ti-SN02 & 11.19 & 0.00 \\
Al Degrader C1 & 261.48 & 0.07 \\
Cu-SN03 & 22.04 & 0.00 \\
Sb-SN03 & 17.43 & 0.06 \\
Polyester Back-SN03 & 5.83 & 1.95 \\
Ti-SN03 & 11.19 & 0.00 \\
Al Degrader C2 & 261.65 & 0.07 \\
Cu-SN04 & 21.72 & 0.06 \\
Sb-SN04 & 16.56 & 0.05 \\
Polyester Back-SN04 & 5.83 & 1.95 \\
Ti-SN04 & 11.12 & 0.05 \\
Al Degrader C3 & 261.64 & 0.07 \\
Cu-SN05 & 21.90 & 0.05 \\
Sb-SN05 & 17.19 & 0.05 \\
Polyester Back-SN05 & 5.83 & 1.95 \\
Ti-SN05 & 11.06 & 0.05 \\
Al Degrader C4 & 261.71 & 0.07 \\
Cu-SN06 & 22.00 & 0.05 \\
Sb-SN06 & 15.98 & 0.05 \\
Polyester Back-SN06 & 5.83 & 1.95 \\
Ti-SN06 & 11.08 & 0.05 \\
SS Profile Monitor 3 & 100.48 & 0.07 \Bstrut\\
\end{longtable}


\begin{longtable}{@{}p{0.35\linewidth} p{0.3\linewidth} p{0.3\linewidth}@{}}
    \caption{Stack design for the LANL 100\,MeV run.}  \\
    \label{tab:LANLfoils}\\
    \hline \hline
    Target Layer & Areal Density & Areal density \Tstrut \\ 
     & (mg/cm$^2$) & uncertainty (\%)  \Bstrut \\
     \hline
    \endfirsthead
    \hline \hline
    Target Layer & Areal Density & Areal density \Tstrut  \\  
     & (mg/cm$^2$) & uncertainty (\%)  \Bstrut\\
     \hline
    \endhead
   \hline  \multicolumn{3}{r}{{Continued on next page}} \\  \hline\hline
\endfoot
\hline \hline
    SS Profile Monitor 1 & 101.78 & 1.18 \Tstrut\\
    Cu-SN01 & 22.07 & 0.26 \\
    Sb-SN01 & 14.35 & 0.35 \\
    Polyester Back-SN01 & 5.83 & 1.95 \\
    Nb-SN01 & 22.52 & 0.00 \\
    Ti-SN01 & 10.88 & 0.48 \\
    Al Degrader D1 & 1709.20 & 0.00 \\
    Cu-SN02 & 22.12 & 0.01 \\
    Sb-SN02 & 17.33 & 0.29 \\
    Polyester Back-SN02 & 5.83 & 1.95 \\
    Nb-SN02 & 22.13 & 0.23 \\
    Ti-SN02 & 10.79 & 0.46 \\
    Al Degrader D2 & 863.27 & 0.04 \\
    Cu-SN03 & 22.07 & 0.22 \\
    Sb-SN03 & 17.68 & 0.00 \\
    Polyester Back-SN03 & 5.83 & 1.95 \\
    Nb-SN03 & 22.30 & 0.00 \\
    Ti-SN03 & 10.89 & 0.45 \\
    Al Degrader D3 & 624.52 & 0.11 \\
    Cu-SN04 & 22.01 & 0.24 \\
    Sb-SN04 & 17.14 & 0.30 \\
    Polyester Back-SN04 & 5.83 & 1.95 \\
    Nb-SN04 & 22.28 & 0.26 \\
    Ti-SN04 & 10.99 & 0.46 \\
    Al Degrader D4 & 428.52 & 0.08 \\
    Cu-SN05 & 22.03 & 0.00 \\
    Sb-SN05 & 18.15 & 0.27 \\
    Polyester Back-SN05 & 5.83 & 1.95 \\
    Nb-SN05 & 22.24 & 0.23 \\
    Ti-SN05 & 11.04 & 0.44 \\
    Al Degrader D5 & 280.15 & 0.12 \\
    Cu-SN06 & 22.08 & 0.26 \\
    Sb-SN06 & 18.04 & 0.28 \\
    Polyester Back-SN06 & 5.83 & 1.95 \\
    Nb-SN06 & 22.55 & 0.26 \\
    Ti-SN06 & 10.99 & 0.00 \\
    Al Degrader D6 & 226.22 & 0.43 \\
    Cu-SN07 & 21.98 & 0.36 \\
    Sb-SN07 & 15.08 & 0.38 \\
    Polyester Back-SN07 & 5.83 & 1.95 \\
    Nb-SN07 & 22.42 & 0.26 \\
    Ti-SN07 & 11.02 & 0.45 \\
    Al Degrader D7 & 139.16 & 0.19 \\
    Cu-SN08 & 22.06 & 0.00 \\
    Sb-SN08 & 15.86 & 0.52 \\
    Polyester Back-SN08 & 5.83 & 1.95 \\
    Nb-SN08 & 22.29 & 0.00 \\
    Ti-SN08 & 11.07 & 0.00 \\
    Al Degrader D8 & 140.17 & 0.86 \\
    Cu-SN09 & 21.96 & 0.00 \\
    Sb-SN09 & 18.06 & 0.32 \\
    Polyester Back-SN09 & 5.83 & 1.95 \\
    Nb-SN09 & 22.35 & 0.36 \\
    Ti-SN09 & 11.45 & 0.43 \\
    Al Degrader D9 & 140.31 & 0.18 \\
    Cu-SN10 & 21.22 & 0.24 \\
    Sb-SN10 & 15.99 & 0.32 \\
    Polyester Back-SN10 & 5.83 & 1.95 \\
    Nb-SN10 & 22.14 & 0.45 \\
    Ti-SN10 & 10.99 & 0.45 \\
    SS Profile Monitor 2 & 101.52 & 1.18 \Bstrut\\
\end{longtable}

\begin{longtable}{@{}p{0.35\linewidth} p{0.3\linewidth} p{0.3\linewidth}@{}}
    \caption{Stack design for the BNL 200\,MeV run.}\\
    \label{tab:BNLfoils}\\
    \hline \hline
    Target Layer & Areal Density & Areal density \Tstrut \\ 
     & (mg/cm$^2$) & uncertainty (\%)  \Bstrut \\
     \hline
    \endfirsthead
    \hline \hline
    Target Layer & Areal Density & Areal density \Tstrut  \\  
     & (mg/cm$^2$) & uncertainty (\%)  \Bstrut\\
     \hline
    \endhead
   \hline  \multicolumn{3}{r}{{Continued on next page}} \\  \hline\hline
\endfoot
\hline \hline
    SS Profile Monitor 1 & 96.80 & 0.08 \Tstrut\\
    Cu-SN01 & 22.15 & 0.26 \\
    Sb-SN01 & 16.12 & 0.31 \\
    Polyester Back-SN01 & 5.83 & 1.95 \\
    Nb-SN01 & 22.42 & 0.38 \\
    Al Degrader D1 & 4668.98 & 0.40 \\
    Cu-SN02 & 22.25 & 0.26 \\
    Sb-SN02 & 17.14 & 0.34 \\
    Polyester Back-SN02 & 5.83 & 1.95 \\
    Nb-SN02 & 22.72 & 0.00 \\
    Al Degrader D2 & 3892.53 & 0.46 \\
    Cu-SN03 & 22.27 & 0.23 \\
    Sb-SN03 & 15.98 & 0.00 \\
    Polyester Back-SN03 & 5.83 & 1.95 \\
    Nb-SN03 & 22.49 & 0.37 \\
    Al Degrader D3 & 3716.27 & 1.09 \\
    Cu-SN04 & 22.17 & 0.26 \\
    Sb-SN04 & 16.75 & 0.34 \\
    Polyester Back-SN04 & 5.83 & 1.95 \\
    Nb-SN04 & 22.52 & 0.22 \\
    Al Degrader D4 & 3754.24 & 0.43 \\
    Cu-SN05 & 22.17 & 0.26 \\
    Sb-SN05 & 18.12 & 0.32 \\
    Polyester Back-SN05 & 5.83 & 1.95 \\
    Nb-SN05 & 22.54 & 0.35 \\
    Al Degrader D5 & 3340.26 & 0.75 \\
    Cu-SN06 & 22.18 & 0.26 \\
    Sb-SN06 & 15.34 & 0.33 \\
    Polyester Back-SN06 & 5.83 & 1.95 \\
    Nb-SN06 & 22.42 & 0.22 \\
    Al Degrader D6 & 3285.91 & 0.43 \\
    Cu-SN07 & 22.16 & 0.37 \\
    Sb-SN07 & 17.06 & 0.29 \\
    Polyester Back-SN07 & 5.83 & 1.95 \\
    Nb-SN07 & 22.52 & 0.22 \\
    SS Profile Monitor 2 & 96.80 & 0.08 \Bstrut\\
\end{longtable}

        
    \section{\label{app:Sb_product_data}Decay data for Sb Products}

\begingroup
\squeezetable
\begin{longtable}{llll}

\caption{Decay data for cross sections measured on Sb foils}\\
\label{tab:Sb_gamma_data}\\

\hline \hline
Isotope & Half Life & Energy (keV) & $I_\gamma$ (\%) \Tstrut\Bstrut\\ \hline
\endfirsthead
\hline \hline
Isotope & Half Life & Energy (keV) & $I_\gamma$ (\%) \Tstrut\Bstrut\\ \hline
\endhead
\hline \multicolumn{4}{r}{{Continued on next page}} \\ \hline\hline
\endfoot
\hline \hline
\endlastfoot
\ce{^{89m}Nb} \cite{singh_nuclear_2013} & 1.100\,(33)\,h & 588 & 95.57\,(13)\,\% \Tstrut\\
\ce{^{91m}Nb} \cite{baglin_nuclear_2013} & 60.86\,(22)\,d & 603.5 & 1.30\,(13)\,\%  \\
 &  & 1082.6 & 1.09\,(11)\,\%  \\
 &  & 1204.67 & 2.0\,(3)\,\%  \\
 &  & 1790.6 & 36.3\,(12)\,\%  \\
 &  & 1984.6 & 62.6\,(15)\,\%  \\
\ce{^{106m}Ag} \cite{de_frenne_nuclear_2008} & 8.28\,(2)\,d & 221.701 & 6.6\,(3)\,\%\\
 &  & 228.633 & 2.1\,(1)\,\%\\
 &  & 328.463 & 1.14\,(5)\,\%  \\
 &  & 406.182 & 13.4\,(4)\,\%  \\
 &  & 429.646 & 13.2\,(4)\,\%  \\
 &  & 450.976 & 28.2\,(7)\,\%  \\
 &  & 601.17 & 1.61\,(9)\,\%  \\
 &  & 680.42 & 1.54\,(8)\,\%  \\
 &  & 717.34 & 28.9\,(8)\,\%  \\
 &  & 748.36 & 20.6\,(6)\,\%  \\
 &  & 793.17 & 5.9\,(3)\,\% \\
 &  & 804.28 & 12.4\,(5)\,\%  \\
 &  & 808.36 & 4.0\,(4)\,\%  \\
 &  & 824.69 & 15.3\,(4)\,\%  \\
 &  & 847.03 & 2.8\,(6)\,\%  \\
 &  & 847.27 & 1.6\,(5)\,\%  \\
 &  & 1019.72 & 1.04\,(16)\,\%  \\
 &  & 1045.83 & 29.6\,(10)\,\%  \\
 &  & 1128.02 & 11.8\,(5)\,\%  \\
 &  & 1199.39 & 11.2\,(5)\,\%  \\
 &  & 1222.88 & 7.0\,(4)\,\%  \\
 &  & 1394.35 & 1.49\,(18)\,\%  \\
 &  & 1527.65 & 16.3\,(13)\,\%  \\
 &  & 1572.35 & 6.6\,(5)\,\%  \\
 &  & 1722.76 & 1.40\,(18)\,\%  \\
 &  & 1839.05 & 2.0\,(3)\,\%  \\
\ce{^{110m}Ag} \cite{gurdal_nuclear_2012} & 249.76\,(4)\,d & 620.3553 & 2.73\,(8)\,\%  \\
 &  & 657.76 & 95.61\,(10)\,\% \\
 &  & 677.6217 & 10.70\,(5)\,\%  \\
 &  & 687.0091 & 6.53\,(3)\,\%  \\
 &  & 706.676 & 16.69\,(7)\,\%  \\
 &  & 744.2755 & 4.77\,(3)\,\%  \\
 &  & 818.0244 & 7.43\,(4)\,\%  \\
 &  & 884.6781 & 75.0\,(11)\,\%  \\
 &  & 937.485 & 35.0\,(3)\,\%  \\
 &  & 1384.2931 & 25.1\,(5)\,\%  \\
 &  & 1475.7792 & 4.08\,(5)\,\%  \\
 &  & 1505.028 & 13.33\,(15)\,\%  \\
 &  & 1562.294 & 1.22\,(3)\,\%  \\
\ce{^{109}In} \cite{kumar_nuclear_2016} & 4.167\,(18)\,h & 288.1 & 1.51\,(11)\,\%  \\
 &  & 426.3 & 4.05\,(12)\,\%  \\
 &  & 623.8 & 5.64\,(23)\,\%  \\
 &  & 1148.5 & 4.67\,(15)\,\%  \\
 &  & 1419 & 1.25\,(5)\,\%  \\
 &  & 1622.3 & 2.08\,(8)\,\%  \\
\ce{^{110}In} \cite{gurdal_nuclear_2012} & 4.9\,(1)\,h & 120.154 & 1.41\,(4)\,\%  \\
 &  & 584.21 & 6.49\,(23)\,\%  \\
 &  & 641.68 & 26.0\,(8)\,\%  \\
 &  & 707.4 & 29.5\,(11)\,\%  \\
 &  & 708.12 & 1.64\,(17)\,\%  \\
 &  & 759.87 & 3.15\,(12)\,\%  \\
 &  & 844.667 & 3.24\,(12)\,\%  \\
 &  & 997.16 & 10.5\,(3)\,\%  \\
 &  & 1117.36 & 4.23\,(13)\,\%  \\
\ce{^{110m}In} \cite{gurdal_nuclear_2012} & 69.102\,(498)\,m & 2129.4 & 2.15\,(3)\,\%  \\
 &  & 2211.33 & 1.74\,(3)\,\%  \\
\ce{^{111}In} \cite{blachot_nuclear_2009} & 2.8047\,(4)\,d & 171.28 & 90.7\,(9)\,\%  \\
 &  & 245.35 & 94.1\,(10)\,\%  \\
\ce{^{114m}In} \cite{blachot_nuclear_2012-1} & 49.51\,(1)\,d & 190.27 & 15.56\,(15)\,\%  \\
 &  & 558.43 & 4.4\,(6)\,\%  \\
 &  & 725.24 & 4.4\,(6)\,\%  \\
\ce{^{113}Sn} \cite{blachot_nuclear_2010-1} & 115.09\,(3)\,d & 255.134 & 2.11\,(8)\,\%  \\
 &  & 391.698 & 64.97\,(17)\,\%  \\
\ce{^{117m}Sn} \cite{blachot_nuclear_2002} & 13.60\,(4)\,d & 156.02 & 2.113\,(12)\,\%  \\
 &  & 158.56 & 86.4\,(4)\,\% \\
\ce{^{119m}Sn} \cite{symochko_nuclear_2009} & 293\,(1)\,d & 23.875 & 16.5\,(2)\,\%  \\
\ce{^{115}Sb} \cite{blachot_nuclear_2012} & 32.1\,(3)\,m & 489.3 & 1.3\,(3)\,\%  \\
 &  & 497.31 & 97.9\,(4)\,\% \\
\ce{^{116m}Sb} \cite{blachot_nuclear_2010} & 1.01\,(1)\,h & 99.802 & 100\,(4)\,\%  \\
 &  & 135.511 & 28.5\,(12)\,\%  \\
 &  & 436.666 & 3.58\,(16)\,\%  \\
 &  & 542.867 & 48.1\,(200)\,\%  \\
 &  & 844.001 & 11.2\,(5)\,\%  \\
 &  & 972.573 & 100\,(4)\,\%  \\
 &  & 1072.373 & 25.5\,(11)\,\%  \\
 &  & 1293.557 & 100\,(4)\,\%  \\
\ce{^{118m}Sb} \cite{kitao_nuclear_1995} & 5.00\,(2)\,h & 253.678 & 99\,(6)\,\%  \\
 &  & 984 & 1.5\,(5)\,\%  \\
 &  & 1050.69 & 97\,(5)\,\% \\
 &  & 1091.51 & 3.6\,(3)\,\%  \\
 &  & 1229.65 & 100\,(5)\,\%\\
\ce{^{120m}Sb} \cite{kitao_nuclear_2002} & 5.76\,(2)\,d & 197.3 & 87.0\,(11)\,\%  \\
 &  & 1023.3 & 99.4\,(3)\,\%  \\
 &  & 1171.7 & 100\,(0)\,\%  \\
\ce{^{122}Sb} \cite{tamura_nuclear_2007} & 2.7238\,(2)\,d & 564.24 & 70.67\,(18)\,\% \\
 &  & 692.65 & 3.85\,(13)\,\%  \\
\ce{^{116}Te} \cite{blachot_nuclear_2010} & 2.49\,(4)\,h & 93.7 & 33.1\,(21)\,\%  \\
 &  & 103 & 1.98\,(14)\,\%  \\
 &  & 628.7 & 3.21\,(15)\,\%\\
 &  & 931.84 & 24.8\,(19)\,\%   \textsuperscript{\ref{Sb_decay_data_table_116Sb_footnote}}\\
 &  & 2225 & 14.6\,(13)\,\%  \textsuperscript{\ref{Sb_decay_data_table_116Sb_footnote}}\\
\ce{^{117}Te} \cite{blachot_nuclear_2002} & 61.99\,(198)\,m & 719.7 & 64.7\,(14)\,\% \\
 &  & 886.7 & 1.49\,(20)\,\%  \\
 &  & 923.9 & 6.2\,(7)\,\%  \\
 &  & 996.7 & 3.9\,(4)\,\%  \\
 &  & 1090.7 & 6.9\,(7)\,\%  \\
 &  & 1716.4 & 15.9\,(17)\,\%  \\
 &  & 2300 & 11.2\,(12)\,\%   \\
\ce{^{118}Te} \cite{kitao_nuclear_1995} & 6.00\,(2)\,d & 1229.33 & 2.5\,(3)\,\%  \textsuperscript{\ref{Sb_decay_data_table_118Sb_footnote}}\\
\ce{^{119}Te} \cite{symochko_nuclear_2009} & 16.05\,(5)\,h & 644.01 & 84.1\,(5)\,\% \\
 &  & 699.85 & 10.1\,(5)\,\% \\
 &  & 1413.19 & 1.09\,(8)\,\%  \\
 &  & 1749.65 & 4.0\,(3)\,\%  \\
\ce{^{119m}Te} \cite{symochko_nuclear_2009} & 4.70\,(4)\,d & 153.59 & 66\,(3)\,\% \\
 &  & 164.34 & 1.30\,(5)\,\%  \\
 &  & 270.53 & 28.0\,(4)\,\%  \\
 &  & 912.6 & 6.24\,(8)\,\%  \\
 &  & 942.21 & 5.08\,(6)\,\%  \\
 &  & 976.37 & 2.71\,(7)\,\%  \\
 &  & 979.29 & 3.01\,(7)\,\%  \\
 &  & 1013.2 & 1.7\,(3)\,\%  \\
 &  & 1048.44 & 3.19\,(5)\,\%  \\
 &  & 1081.35 & 1.59\,(3)\,\%  \\
 &  & 1095.75 & 2.23\,(3)\,\%  \\
 &  & 1136.75 & 7.65\,(7)\,\%  \\
 &  & 1212.73 & 66.1\,(3)\,\% \\
 &  & 1366.39 & 1.064\,(20)\,\%  \\
 &  & 2089.57 & 4.68\,(6)\,\%  \\
\ce{^{121}Te} \cite{ohya_nuclear_2010} & 19.17\,(4)\,d & 470.472 & 1.41\,(4)\,\%  \\
 &  & 573.139 & 80.4\,(14)\,\%  \\
\ce{^{121m}Te} \cite{ohya_nuclear_2010} & 164.2\,(8)\,d & 212.189 & 81.5\,(11)\,\% \\
 &  & 1102.149 & 2.5\,(3)\,\%  \\
\ce{^{123m}Te} \cite{chen_nuclear_2021} & 119.2\,(3)\,d & 159 & 84.3\,(3)\,\% \Bstrut

\end{longtable}
\endgroup

\textcolor{white}{\footnote{\label{Sb_decay_data_table_116Sb_footnote}These are decay gammas from the progeny isotope, \ce{^{116}Sb}}}
\textcolor{white}{\footnote{\label{Sb_decay_data_table_118Sb_footnote}These are decay gammas from the progeny isotope, \ce{^{118}Sb}}}

    

    \section{\label{app:TALYSLDP}TALYS Level Density Parameters Explored}
        \begin{turnpage}
\begingroup
\squeezetable

\begin{longtable*}{@{}llllll@{}}

\caption{TALYS Level Density Parameters Explored}\\
\label{tab:TALYSLDP}\\
\hline \hline
TALYS Parameter & Description & Range & Default & Explored \textsuperscript{\ref{TALYSLD_table_header_footnote}} & Chosen \Tstrut\Bstrut\\
\hline
\endfirsthead
\hline \hline
TALYS Parameter & Description & Range & Default & Explored \textsuperscript{\ref{TALYSLD_table_header_footnote}} & Chosen \Tstrut\Bstrut\\
\hline
\endhead
\endfoot
\endlastfoot
\texttt{ldmodel} & Model for level densities & \begin{tabular}[c]{@{}l@{}}1: Constant temperature\,+\,Fermi \\gas model\\ 2: Back-shifted Fermi gas model\\ 3: Generalized superfluid model\\ 4: Microscopic level densities \\(Skyrme force) from \\ Goriely’s tables\\ 5: Microscopic level densities \\(Skyrme force) from Hilaire’s \\combinatorial tables\\ 6: Microscopic level densities \\(temperature dependent HF, \\Gogny force) from Hilaire’s \\combinatorial tables\end{tabular} & 1 & 1-6 & 2 \Tstrut\\
\texttt{spincutmodel} & Model for spin cut-off parameter & 1,2 & 1 & 1-2 & 2 \\
\texttt{rspincut} &\begin{tabular}[c]{@{}l@{}} Adjustable constant for \\spin cut-off parameter \end{tabular}& $\mathrm{0 \,<\,x\,\le\, 10}$ & 1 & 0.01-10 & 0.4 \Bstrut\\
\hline\hline
\end{longtable*}

\endgroup
\textcolor{white}{\footnote{\label{TALYSLD_table_header_footnote}These values were explored in different bin sizes to determine effects on results}}
\textcolor{white}{\footnote{\label{TALYSLD_parameter_header_footnote}Six additional level density parameters were explored, including \texttt{ldmodelracap}, \texttt{colenhance}, \texttt{colldamp}, \texttt{parity}, \texttt{pglobal}, and \texttt{cglobal}.}}
\end{turnpage}
    
    \section{\label{app:TALYSPEP}TALYS Pre-Equilibrium Parameters Explored}
\begingroup
\squeezetable 

\begin{longtable*}{llllll}
\caption{TALYS Pre-Equilibrium Parameters Explored}\\
\label{tab:TALYSPE}\\
\hline \hline
TALYS Parameter & Description & Range & Default & Explored\textsuperscript{\ref{TALYS_table_header_footnote}} & Chosen \Tstrut\Bstrut\\\hline
\endfirsthead
\hline \hline

TALYS Parameter\textsuperscript{\ref{TALYS_parameter_header_footnote}} & Description & Range & Default & Explored \textsuperscript{\ref{TALYS_table_header_footnote}}& Chosen \Tstrut\Bstrut\\\hline
\endhead
\texttt{preeqmode}& Model for pre-eq reactions & \begin{tabular}[c]{@{}l@{}}1: Exciton model: Analytical \\transition rates\\ 2: Exciton model: Numerical \\transition rates\\ 3: Exciton model: Numerical \\transition rates with optical \\ model for collision probability.\\ 4: Multi-step direct/compound\\ model\end{tabular}
& 2 & 1-4 & 1 \Tstrut\\

\texttt{m2constant} & \begin{tabular}[c]{@{}l@{}}Overall constant for the matrix \\ element in exciton model\end{tabular} & $\mathrm{0.1 \,\le\,x\,\le\, 100}$ & 1 & 0.1-10.0 & 2 \\
\texttt{m2limit} & \begin{tabular}[c]{@{}l@{}}Constant to scale asymptotic \\ value of matrix element\end{tabular} & $\mathrm{0.1 \,\le\,x\,\le\, 100}$ & 1 & 0.1-10.0 & 0.8 \\
\texttt{m2shift} & \begin{tabular}[c]{@{}l@{}}Constant to scale energy shift \\ of the matrix element\end{tabular} & $\mathrm{0.1 \,\le\,x\,\le\, 100}$ & 1 & 0.1-10.0 & 1.8 \\
\texttt{rpinu} & \begin{tabular}[c]{@{}l@{}}Proton-neutron ratio in \\ two-component exciton model\end{tabular} & $\mathrm{0.1 \,\le\,x\,\le\, 100}$ & 1 & 0.1-50.0  & 1.5 \\
\texttt{rnupi} & \begin{tabular}[c]{@{}l@{}}Neutron-proton ratio in \\ two-component exciton model\end{tabular} & $\mathrm{0.1 \,\le\,x\,\le\, 100}$ & 1 & 0.1-50.0 & 1.5 \Bstrut\\
\hline\hline
\end{longtable*}

\endgroup
\textcolor{white}{\footnote{\label{TALYS_table_header_footnote}These values were explored in different bin sizes to determine effects on results}}
\textcolor{white}{\footnote{\label{TALYS_parameter_header_footnote}Eight additional pre-equilibrium parameters were explored -- \texttt{cknock}, \texttt{cstrip}, and \texttt{cbreak} for both proton and alpha particles, and \texttt{mpreeqmode}, \texttt{preeqspin}, \texttt{rgamma}, \texttt{rpipi}, and \texttt{rnunu}. Ultimately, there was little improvement in the reaction model fit to justify changing these parameters from their default value.}}

    
    \section{\label{app:TALYSOMP} TALYS Optical Model Potential Parameters Explored}
\begingroup
\squeezetable 
\begin{longtable*}{@{}llllll@{}}

\caption{TALYS Optical Model Parameters Explored}\\
\label{tab:TALYSOMP}\\
\hline \hline
TALYS Parameter & Description & Range & Default & Explored \textsuperscript{\ref{TALYS_table_header_footnote}} & Chosen \Tstrut\Bstrut\\
\hline
\endfirsthead
\hline \hline
TALYS Parameter & Description & Range & Default & Explored \textsuperscript{\ref{TALYS_table_header_footnote}} & Chosen \Tstrut\Bstrut\\
\hline
\endhead

\texttt{w1adjust} (n) & w1 term for neutrons & $\mathrm{0.1 \,\le\,x\,\le\, 10}$ & 1 & 0.1-5.0 & 2.5 \Tstrut\\
\texttt{w2adjust} (n) & w2 term for neutrons & $\mathrm{0.1 \,\le\,x\,\le\, 10}$ & 1 & 0.1-5.0 & 0.6 \Bstrut\\
\hline\hline
\end{longtable*}

\endgroup
\textcolor{white}{\footnote{\label{TALYSOOMP_table_header_footnote}These values were explored in different bin sizes to determine effects on results}}
\textcolor{white}{\footnote{\label{TALYSOOMP_parameter_header_footnote}Thirteen additional optical model parameters were explored for protons and neutrons:  \texttt{avadjust}, \texttt{avdadjust}, \texttt{avdadjust}, \texttt{avsoadjust}, \texttt{awadjust}, \texttt{awdadjust}, \texttt{awsoadjust}, \texttt{rvadjust}, \texttt{rvdadjust}, \texttt{rvsoadjust}, \texttt{rwadjust}, \texttt{rwdadjust}, and \texttt{rwsoadjust}.
Four additional parameters were explored: \texttt{w1adjust} and \texttt{w2adjust} for adjusting the proton imaginary optical potential model, and \texttt{w3adjust} and \texttt{w4adjust} for neutrons to adjust the imaginary optical potential model above 180\,MeV.}}

    \section{\label{app:monitor_product_data} Decay data for Monitor foils}
        

\begingroup
\squeezetable
\begin{longtable}{llll}

\caption{Decay data for cross sections measured on Cu foils}\\
\label{tab:Cu_gamma_data}\\

\hline \hline
Isotope & Half Life & Energy (keV) & $I_\gamma$ (\%) \Tstrut\Bstrut\\ \hline
\endfirsthead
\hline \hline
Isotope & Half Life & Energy (keV) & $I_\gamma$ (\%) \Tstrut\Bstrut\\ \hline
\endhead

\hline \multicolumn{4}{r}{{Continued on next page}} \\ \hline\hline
\endfoot

\hline \hline
\endlastfoot
\ce{^{51}Cr} \cite{wang_nuclear_2017} & 27.7010\,(11)\,d & 320.0824 & 9.91\,(1)\,\%  \Tstrut\\
\ce{^{52}Mn} \cite{dong_nuclear_2015} & 5.591\,(3)\,d & 744.233 & 90.0\,(12)\,\%  \\
 &  & 848.18 & 3.32\,(4)\,\%  \\
 &  & 935.544 & 94.5\,(13)\,\%  \\
 &  & 1246.278 & 4.21\,(7)\,\%  \\
 &  & 1434.092 & 100.0\,(14)\,\%  \\
\ce{^{54}Mn} \cite{dong_nuclear_2014} & 312.05\,(4)\,d & 834.848 & 99.976\,(1)\,\%  \\
\ce{^{56}Mn} \cite{junde_nuclear_2011} & 2.5789\,(1)\,h & 1810.726 & 26.9\,(4)\,\%  \\
 &  & 2113.092 & 14.2\,(3)\,\%  \\
 &  & 2523.06 & 1.018\,(20)\,\%  \\
\ce{^{59}Fe} \cite{basunia_nuclear_2018} & 44.495\,(9)\,d & 142.651 & 1.02\,(4)\,\%  \\
 &  & 192.343 & 3.08\,(12)\,\%  \\
 &  & 1099.245 & 56.5\,(18)\,\%  \\
 &  & 1291.59 & 43.2\,(14)\,\%  \\
\ce{^{55}Co} \cite{junde_nuclear_2008} & 17.53\,(3)\,h & 91.9 & 1.16\,(9)\,\%  \\
 &  & 411.5 & 1.07\,(8)\,\%  \\
 &  & 477.2 & 20.2\,(17)\,\%  \\
 &  & 803.7 & 1.87\,(15)\,\%  \\
 &  & 931.1 & 75.0\,(35)\,\% \\
 &  & 1316.6 & 7.1\,(3)\,\%  \\
 &  & 1370 & 2.9\,(3)\,\%  \\
 &  & 1408.5 & 16.9\,(8)\,\% \\
\ce{^{56}Co} \cite{junde_nuclear_2011} & 77.233\,(27)\,d & 846.77 & 99.939\,(28)\,\%\\
 &  & 977.372 & 1.421\,(6)\,\%  \\
 &  & 1037.843 & 14.05\,(4)\,\%  \\
 &  & 1175.101 & 2.252\,(6)\,\%  \\
 &  & 1238.288 & 66.46\,(12)\,\%  \\
 &  & 1360.212 & 4.283\,(12)\,\%  \\
 &  & 1771.357 & 15.41\,(6)\,\%  \\
 &  & 2015.215 & 3.016\,(12)\,\%  \\
 &  & 2034.791 & 7.77\,(3)\,\%  \\
 &  & 2598.5 & 16.97\,(4)\,\%  \\
\ce{^{57}Co} \cite{bhat_nuclear_1998} & 271.74\,(6)\,d & 14.4129 & 9.16\,(15)\,\%  \\
 &  & 122.06065 & 85.60\,(17)\,\%  \\
 &  & 136.47356 & 10.68\,(8)\,\%  \\
\ce{^{58}Co} \cite{nesaraja_nuclear_2010} & 70.86\,(6)\,d & 810.7593 & 99.45\,(1)\,\% \\
 &  & 863 & 0.686\,(10)\,\%  \\
 &  & 1674.725 & 0.517\,(10)\,\%  \\
\ce{^{60}Co} \cite{browne_nuclear_2013} & 5.27113\,(38)\,y & 1173.228 & 99.85\,(3)\,\%  \\
 &  & 1332.492 & 99.9826\,(6)\,\%  \\
\ce{^{61}Co} \cite{zuber_nuclear_2015} & 1.649\,(5)\,h & 917.5 & 3.6\,(3)\,\%  \\
\ce{^{56}Ni} \cite{junde_nuclear_2011} & 6.08\,(1)\,d & 158.38 & 98.8\,(10)\,\%  \\
 &  & 269.5 & 36.5\,(8)\,\%  \\
 &  & 480.44 & 36.5\,(8)\,\%  \\
 &  & 749.95 & 49.5\,(12)\,\%  \\
 &  & 1561.8 & 14.0\,(6)\,\%  \\
\ce{^{57}Ni} \cite{bhat_nuclear_1998} & 35.60\,(6)\,h & 127.164 & 16.7\,(5)\,\%  \\
 &  & 1377.63 & 81.7\,(24)\,\%  \\
 &  & 1757.55 & 5.75\,(20)\,\%  \\
 &  & 1919.52 & 12.3\,(4)\,\%  \\
\ce{^{61}Cu} \cite{zuber_nuclear_2015} & 3.333\,(5)\,h & 282.956 & 12.7\,(20)\,\% \\
 &  & 373.05 & 2.1\,(4)\,\%  \\
 &  & 588.605 & 1.17\,(21)\,\%  \\
 &  & 656.008 & 10.8\,(20)\,\%  \\
 &  & 908.631 & 1.1\,(2)\,\%  \\
 &  & 1185.234 & 3.7\,(7)\,\%  \\
\ce{^{64}Cu} \cite{singh_nuclear_2021} & 12.701\,(2)\,h & 1345.77 & 0.472\,(4)\,\%  \\
\ce{^{60}Cu} \cite{browne_nuclear_2013} & 23.7\,(4)\,m & 467.3 & 3.52\,(18)\,\% \\
 &  & 497.9 & 1.67\,(9)\,\% \\
 &  & 826.4 & 21.7\,(11)\,\%  \\
 &  & 952.4 & 2.73\,(18)\,\%  \\
 &  & 1035.2 & 3.70\,(18)\,\% \\
 &  & 1110.5 & 1.06\,(18)\,\%  \\
 &  & 1293.7 & 1.85\,(18)\,\%  \\
\ce{^{62}Zn} \cite{nichols_nuclear_2012} & 9.26\,(2)\,h & 40.85 & 25.5\,(24)\,\%  \\
 &  & 243.36 & 2.52\,(23)\,\%  \\
 &  & 246.95 & 1.90\,(18)\,\%  \\
 &  & 260.43 & 1.35\,(13)\,\%  \\
 &  & 394.03 & 2.24\,(17)\,\%  \\
 &  & 548.35 & 15.3\,(14)\,\%  \\
 &  & 596.56 & 26\,(2)\,\% \\
\ce{^{63}Zn} \cite{erjun_nuclear_2001} & 38.47\,(5)\,m & 669.62 & 8.2\,(3)\,\% \\
 &  & 962.06 & 6.5\,(4)\,\%  \\
\ce{^{65}Zn} \cite{browne_nuclear_2010} & 243.93\,(9)\,d & 1115.539 & 50.04\,(10)\,\% \Bstrut
\end{longtable}
\endgroup

        \begingroup
\squeezetable
\begin{longtable}{llll}

\caption{Decay data for cross sections measured on Nb foils}\\
\label{tab:Nb_gamma_data}\\

\hline \hline
Isotope & Half Life & Energy (keV) & $I_\gamma$ (\%)\Tstrut\Bstrut\\ \hline
\endfirsthead
\hline \hline
Isotope & Half Life & Energy (keV) & $I_\gamma$ (\%) \Tstrut\Bstrut\\ \hline
\endhead

\hline \multicolumn{4}{r}{{Continued on next page}} \\ \hline\hline
\endfoot

\hline \hline
\endlastfoot
\ce{^{69}Ge} \cite{nesaraja_nuclear_2014} & 39.0504\,(1008)\,h & 1106.77 & 36\,(4)\,\% \Tstrut\\
 &  & 1336.6 & 4.5\,(6)\,\%  \\
\ce{^{71}As} \cite{singh_nuclear_2021}
& 65.2992\,(696)\,h & 174.954 & 82.4\,(20)\,\%  \\
 &  & 1095.49 & 4.10\,(12)\,\%  \\
\ce{^{72}As} \cite{abriola_nuclear_2010} & 25.9992\,(1008)\,h & 1050.75 & 1.00\,(3)\,\%  \\
 &  & 1464 & 1.13\,(3)\,\%  \\
\ce{^{76}As} \cite{singh_nuclear_1995} & 26.2392\,(912)\,h & 1212.92 & 1.44\,(11)\,\%  \\
 &  & 1216.08 & 3.42\,(24)\,\%  \\
 &  & 1228.52 & 1.22\,(10)\,\%  \\
\ce{^{77}Br} \cite{singh_nuclear_2012} & 57.0360\,(48)\,h & 87.59 & 1.40\,(4)\,\%  \\
 &  & 238.98 & 23.1\,(5)\,\% \\
 &  & 439.47 & 1.56\,(5)\,\%  \\
\ce{^{82}Br} \cite{tuli_nuclear_2019} & 35.2824\,(72)\,h & 698.374 & 28.3\,(4)\,\%  \\
 &  & 776.517 & 83.4\,(12)\,\%  \\
 &  & 1007.59 & 1.276\,(21)\,\%  \\
 &  & 1044.002 & 28.3\,(4)\,\%  \\
 &  & 1317.473 & 26.8\,(4)\,\%  \\
 &  & 1474.88 & 16.60\,(23)\,\%  \\
\ce{^{76}Kr} \cite{singh_nuclear_1995} & 14.8\,(1)\,h & 252 & 6.2\,(8)\,\%  \\
\ce{^{79}Kr} \cite{singh_nuclear_2016} & 1.4600\,(42)\,d & 261.29 & 12.7\,(4)\,\% \\
 &  & 306.47 & 2.60\,(13)\,\% \\
 &  & 397.54 & 9.3\,(4)\,\%  \\
\ce{^{83}Rb} \cite{mccutchan_nuclear_2015} & 86.2\,(1)\,d & 520.3991 & 45\,(3)\,\%  \\
 &  & 529.5945 & 29.3\,(21)\,\%  \\
 &  & 552.5512 & 16.0\,(11)\,\%  \\
\ce{^{84}Rb} \cite{abriola_nuclear_2009} & 32.82\,(7)\,d & 881.6041 & 68.9\,(21)\,\% \\
\ce{^{86}Rb} \cite{negret_nuclear_2015} & 18.631\,(18)\,d & 1077 & 8.64\,(4)\,\% \\
\ce{^{83}Sr} \cite{mccutchan_nuclear_2015} & 32.4096\,(312)\,h & 381.17 & 1.79\,(22)\,\%  \\
 &  & 381.53 & 14.0\,(11)\,\%  \\
 &  & 418.37 & 4.2\,(3)\,\%  \\
 &  & 423.63 & 1.44\,(11)\,\%  \\
 &  & 762.65 & 26.7\,(22)\,\%  \\
 &  & 778.44 & 1.76\,(14)\,\%  \\
 &  & 1147.33 & 1.14\,(8)\,\%  \\
 &  & 1159.97 & 1.36\,(11)\,\%  \\
 &  & 1562.51 & 1.60\,(12)\,\%  \\
\ce{^{85m}Sr} \cite{singh_nuclear_2014} & 67.632\,(42)\,m & 151.194 & 12.8\,(4)\,\%  \\
\ce{^{85m}Y} \cite{singh_nuclear_2014} & 4.86\,(13)\,h & 231.7 & 22.8\,(22)\,\%  \\
 &  & 504.4 & 1.51\,(13)\,\%  \\
 &  & 535.6 & 3.5\,(3)\,\%  \\
 &  & 568.4 & 1.67\,(14)\,\%  \\
 &  & 787.9 & 1.57\,(13)\,\%  \\
 &  & 1123.2 & 1.78\,(15)\,\%  \\
 &  & 1220.5 & 1.98\,(17)\,\%  \\
 &  & 1404.6 & 3.1\,(3)\,\%  \\
 &  & 2123.8 & 5.0\,(5)\,\%  \\
\ce{^{86g}Y} \cite{negret_nuclear_2015} & 14.74\,(2)\,h & 187.87 & 1.26\,(4)\,\%  \\
 &  & 190.8 & 1.01\,(3)\,\%  \\
 &  & 443.13 & 16.9\,(5)\,\%  \\
 &  & 580.57 & 4.78\,(14)\,\%  \\
 &  & 608.29 & 2.01\,(15)\,\%  \\
 &  & 627.72 & 32.6\,(10)\,\%  \\
 &  & 644.82 & 2.2\,(3)\,\%  \\
 &  & 645.87 & 9.2\,(11)\,\%  \\
 &  & 703.33 & 15.4\,(4)\,\%  \\
 &  & 709.9 & 2.62\,(8)\,\%  \\
 &  & 740.81 & 1.36\,(5)\,\%  \\
 &  & 767.63 & 2.4\,(3)\,\%  \\
 &  & 777.37 & 22.4\,(6)\,\%  \\
 &  & 835.67 & 4.4\,(6)\,\%  \\
 &  & 955.35 & 1.04\,(4)\,\%  \\
 &  & 1024.04 & 3.79\,(17)\,\%  \\
 &  & 1076.63 & 82.5\,(4)\,\% \\
 &  & 1153.05 & 30.5\,(9)\,\%  \\
 &  & 1163.03 & 1.18\,(4)\,\%  \\
 &  & 1253.11 & 1.53\,(5)\,\%  \\
 &  & 1349.15 & 2.95\,(9)\,\%  \\
 &  & 1801.7 & 1.65\,(5)\,\%  \\
 &  & 1854.38 & 17.2\,(5)\,\%  \\
 &  & 1920.72 & 20.8\,(7)\,\%  \\
 &  & 2567.97 & 2.25\,(11)\,\% \\
 &  & 2610.11 & 1.24\,(7)\,\%  \\
\ce{^{87g}Y} \cite{johnson_nuclear_2015} & 3.325\,(125)\,d & 388.5276 & 82.2\,(7)\,\% \\
 &  & 484.805 & 89.8\,(9)\,\%  \\
\ce{^{88g}Y} \cite{mccutchan_nuclear_2014} & 106.626\,(21)\,d & 898.042 & 93.7\,(3)\,\%  \\
 &  & 1836.063 & 99.2\,(3)\,\%  \\
\ce{^{90m}Y} \cite{basu_nuclear_2020} & 3.19\,(6)\,h & 479.51 & 90.74\,(5)\,\%  \\
\ce{^{84g}Y} \cite{abriola_nuclear_2009} & 39.5\,(8)\,m & 703.6 & 5.8\,(6)\,\%  \\
 &  & 793.1 & 98.3\,(4)\,\%\\
 &  & 974.3 & 78\,(4)\,\% \\
 &  & 2309.5 & 1.18\,(20)\,\%  \\
\ce{^{86m}Y} \cite{negret_nuclear_2015} & 48\,(1)\,m & 208.1 & 93.8\,(4)\,\%  \\
\ce{^{91g}Y} \cite{baglin_nuclear_2013} & 58.51\,(6)\,d & 1204.8 & 0.26\,(4)\,\%  \\
\ce{^{87m}Y} \cite{johnson_nuclear_2015} & 13.37\,(3)\,h & 380.79 & 78.05\,(79)\,\% \\
\ce{^{86}Zr} \cite{negret_nuclear_2015} & 16.5\,(1)\,h & 29.1 & 21.6\,(15)\,\%  \\
 &  & 242.8 & 95.84\,(20)\,\% \\
 &  & 612 & 5.8\,(3)\,\% \\
\ce{^{87}Zr} \cite{johnson_nuclear_2015} & 1.68\,(1)\,h & 1227 & 2.80\,(4)\,\%  \\
\ce{^{88}Zr} \cite{mccutchan_nuclear_2014} & 83.4\,(3)\,d & 392.87 & 97.29\,(0)\,\% \\
\ce{^{89}Nb} \cite{singh_nuclear_2013} & 2.03\,(7)\,h & 920.5 & 1.4\,(3)\,\%  \\
 &  & 1127 & 2.1\,(5)\,\%  \\
 &  & 1259 & 1.2\,(3)\,\%  \\
 &  & 1511.4 & 1.9\,(4)\,\%  \\
 &  & 1627.2 & 3.50\,(7)\,\% \\
 &  & 2572.3 & 2.7\,(6)\,\%  \\
 &  & 2960.1 & 1.8\,(4)\,\%  \\
\ce{^{89m}Nb} \cite{singh_nuclear_2013} & 1.100\,(33)\,h & 588 & 95.57\,(13)\,\% \\
\ce{^{90}Nb} \cite{basu_nuclear_2020} & 14.60\,(5)\,h & 132.716 & 4.13\,(4)\,\%  \\
 &  & 141.178 & 66.8\,(7)\,\%  \\
 &  & 371.307 & 1.80\,(7)\,\%  \\
 &  & 890.64 & 1.80\,(4)\,\%  \\
 &  & 1129.224 & 92.7\,(5)\,\%  \\
 &  & 1270.396 & 1.296\,(25)\,\%  \\
 &  & 1611.76 & 2.38\,(7)\,\%  \\
 &  & 1913.194 & 1.280\,(17)\,\%  \\
 &  & 2186.242 & 17.96\,(17)\,\%  \\
 &  & 2318.959 & 82.0\,(3)\,\%  \\
\ce{^{91m}Nb} \cite{baglin_nuclear_2013} & 60.86\,(22)\,d & 50.1 & 6.5\,(7)\,\%  \\
 &  & 1082.6 & 1.09\,(11)\,\%  \\
 &  & 1204.67 & 2.0\,(3)\,\%  \\
 &  & 1790.6 & 36.3\,(12)\,\%  \\
 &  & 1984.6 & 62.6\,(15)\,\%  \\
\ce{^{90}Mo} \cite{basu_nuclear_2020} & 5.67\,(5)\,h & 122.37 & 64\,(3)\,\% \\
 &  & 162.93 & 6.0\,(6)\,\%  \\
 &  & 203.13 & 6.4\,(6)\,\%  \\
 &  & 257.34 & 77\,(4)\,\%  \\
 &  & 323.2 & 6.3\,(6)\,\%  \\
 &  & 445.37 & 6.0\,(7)\,\%  \\
 &  & 472.2 & 1.42\,(16)\,\%  \\
 &  & 941.5 & 5.5\,(7)\,\%  \\
 &  & 990.2 & 1.02\,(11)\,\%  \\
 &  & 1271.3 & 4.1\,(4)\,\%  \\
 &  & 1387.4 & 1.86\,(24)\,\%  \\
 &  & 1454.6 & 1.9\,(5)\,\%  \\
\ce{^{93m}Mo} \cite{baglin_nuclear_2011} & 6.85\,(7)\,h & 263.049 & 57.4\,(11)\,\%  \\
 &  & 684.693 & 99.9\,(8)\,\%  \\
 &  & 1477.138 & 99.1\,(11)\,\% \Bstrut\\
 \end{longtable}
\endgroup

        

\begingroup
\squeezetable
\begin{longtable}{llll}
\caption{Decay data for cross sections measured on Ti foils}\\
\label{tab:Ti_gamma_data}\\
\hline \hline
Isotope & Half Life & Energy (keV) & $I_\gamma$ (\%) \Tstrut\Bstrut\\ \hline
\endfirsthead
\hline \hline
Isotope & Half Life & Energy (keV) & $I_\gamma$ (\%) \Tstrut\Bstrut\\ \hline
\endhead
\hline \multicolumn{4}{r}{{Continued on next page}} \\ \hline\hline
\endfoot
\hline \hline
\endlastfoot
\ce{^{42}K} \cite{chen_nuclear_2016} & 12.36\,(12)\,h & 1524.6 & 18.08\,(9)\,\%\Tstrut\\
\ce{^{43}K} \cite{singh_nuclear_2015} & 22.3\,(1)\,h & 220.632 & 4.80\,(6)\,\% \\
 &  & 372.76 & 86.8\,(2)\,\% \\
 &  & 396.861 & 11.85\,(8)\,\% \\
 &  & 593.39 & 11.26\,(8)\,\% \\
 &  & 617.49 & 79.2\,(6)\,\% \\
 &  & 1021.698 & 1.96\,(3)\,\% \\
\ce{^{47}Ca} \cite{burrows_nuclear_2007} & 4.536\,(3)\,d & 1297.09 & 67\,(13)\,\% \\
\ce{^{46}Sc} \cite{wu_nuclear_2000} & 83.79\,(4)\,d & 889.277 & 99.984\,(1)\,\% \\
 &  & 1120.545 & 99.987\,(1)\,\% \\
\ce{^{48}Sc} \cite{chen_nuclear_2022} & 43.6704\,(912)\,h & 175.361 & 7.48\,(10)\,\% \\
 &  & 1037.522 & 97.6\,(7)\,\% \\
 &  & 1212.88 & 2.38\,(4)\,\% \\
\ce{^{47}Sc} \cite{burrows_nuclear_2007} & 80.3808\,(144)\,h & 159.381 & 68.3\,(4)\,\% \\
\ce{^{44}Ti} \cite{chen_nuclear_2023} & 60.0\,(11)\,y & 67.8679 & 93\,(2)\,\% \\
 &  & 78.3234 & 96.4\,(17)\,\% \\
\ce{^{48g}V} \cite{chen_nuclear_2022} & 15.9735\,(25)\,d & 944.13 & 7.870\,(7)\,\% \\
 &  & 983.525 & 99.98\,(4)\,\% \\
 &  & 1312.106 & 98.2\,(3)\,\% \\
 &  & 2240.396 & 2.333\,(13)\,\% \Bstrut
\end{longtable}
\endgroup

    \section{\label{app:monitorXstable} Independent and Cumulative Data for Monitor Foil Products}
        \begingroup
\squeezetable
\begin{tiny}
\begin{longtable*}{llllllll}
\caption{Independent and cumulative cross section measurements for \ce{^{nat}Cu}(p,x) reactions. Note that measurements for \ce{^{56g}Co_{(c)}}, \ce{^{58(m+g)}Co_{(c)}}, \ce{^{62g}Zn_{(i)}}, \ce{^{63g}Zn_{(i)}}, and \ce{^{65g}Zn_{(i)}} were derived through variance minimization}\\
\label{tab:cu_xs_data}\\
\endfirsthead
\hline \hline
\endhead
\hline \multicolumn{4}{r}{{Continued on next page}} \\ \hline\hline
\endfoot
\hline \hline
\endlastfoot

\hline \hline

$E_p$\,(MeV)        & 188.565\,(2069) & 173.381\,(2183) & 159.984\,(2301) & 146.453\,(2441) & 131.868\,(2621) & 117.847\,(2835) & 102.805\,(3123) \Tstrut\Bstrut\\
Location            & BNL             & BNL             & BNL             & BNL             & BNL             & BNL             & BNL             \Bstrut\\\hline
\ce{^{51g}Cr_{(c)}} & 11.82\,(19)     & 10.00\,(16)     & 8.483\,(99)     & 7.141\,(81)     & 5.645\,(81)     & 4.829\,(97)     & 2.11\,(12)      \Tstrut\\
\ce{^{52g}Mn_{(c)}} & 5.57\,(11)      & 5.070\,(73)     & 4.507\,(50)     & 3.949\,(43)     & 3.010\,(38)     & 2.252\,(35)     & 2.010\,(46)     \\
\ce{^{54g}Mn_{(c)}} & 16.50\,(27)     & 16.59\,(24)     & 15.71\,(18)     & 14.93\,(16)     & 13.24\,(18)     & 12.33\,(20)     & 8.19\,(19)      \\
\ce{^{56g}Mn_{(c)}} & 2.433\,(83)     & 2.417\,(44)     & 2.203\,(64)     & 2.121\,(33)     & 2.341\,(67)     & 1.791\,(38)     & 1.364\,(87)     \\
\ce{^{59g}Fe_{(c)}} & 1.244\,(30)     & 1.433\,(44)     & 1.366\,(20)     & 1.352\,(20)     & 1.278\,(25)     & 1.295\,(31)     & 1.004\,(42)     \\
\ce{^{55g}Co_{(c)}} & 2.01\,(21)      & 2.04\,(13)      & 2.11\,(19)      & 1.97\,(21)      & 1.96\,(28)      & 1.62\,(26)      & 1.92\,(17)      \\
\ce{^{56g}Co_{(c)}} & 10.30\,(94)     & 13.68\,(87)     & 10.79\,(84)     & 13.82\,(65)     & 13.07\,(59)     & 10.37\,(98)     & 11.21\,(59)     \\
\ce{^{57g}Co_{(c)}} & 39.70\,(67)     & 42.82\,(59)     & 43.66\,(49)     & 44.90\,(47)     & 45.26\,(56)     & 46.56\,(72)     & 43.57\,(95)     \\
\ce{^{58(m+g)}Co_{(c)}} & 43.13\,(83)     & 47.13\,(80)     & 46.48\,(97)     & 48.98\,(79)     & 49.0\,(10)      & 51.3\,(11)      & 52.4\,(12)      \\
\ce{^{60g}Co_{(c)}} & 12.10\,(35)     &                 & 15.04\,(62)     &                 & 13.41\,(75)     &                 & 13.26\,(44)     \\
\ce{^{56g}Ni_{(c)}} & 0.043\,(14)     & 0.0967\,(69)    & 0.1107\,(71)    & 0.136\,(17)     & 0.0848\,(84)    & 0.0602\,(64)    & 0.119\,(55)     \\
\ce{^{57g}Ni_{(c)}} & 1.922\,(52)     & 2.102\,(40)     & 2.108\,(41)     & 2.818\,(63)     & 2.488\,(77)     & 2.375\,(48)     & 1.884\,(56)     \\
\ce{^{61g}Cu_{(c)}} & 30.5\,(11)      & 34.9\,(11)      & 36.1\,(11)      & 41.0\,(14)      & 44.9\,(14)      & 49.7\,(17)      & 57.3\,(20)      \\
\ce{^{64g}Cu_{(i)}} & 28.3\,(25)      & 34.6\,(29)      & 35.9\,(26)      & 42.1\,(31)      & 39.4\,(32)      & 44.9\,(47)      & 51.1\,(39)      \\
\ce{^{62g}Zn_{(i)}} & 2.22\,(10)      & 2.56\,(22)      & 2.678\,(95)     & 3.186\,(98)     & 3.287\,(44)     & 4.05\,(36)      & 4.85\,(13)      \\
\ce{^{63g}Zn_{(i)}} &                 &                 &                 &                 &                 & 7.3\,(15)       &                 \\
\ce{^{65g}Zn_{(i)}} & 1.241\,(34)     & 2.6\,(13)       & 1.84\,(10)      & 2.093\,(91)     & 2.04\,(17)      & 3.20\,(24)      & 2.540\,(78)     \Bstrut\\\hline\hline
$E_p$\,(MeV)        & 91.100\,(1073)  & 79.073\,(1193)  & 72.162\,(1278)  & 66.664\,(1358)  & 62.488\,(1426)  & 59.407\,(1483)  & 56.695\,(1537)  \Tstrut\Bstrut\\
Location            & LANL            & LANL            & LANL            & LANL            & LANL            & LANL            & LANL            \Bstrut\\\hline
\ce{^{51g}Cr_{(c)}} & 2.585\,(65)     & 0.496\,(14)     & 0.482\,(17)     & 0.555\,(22)     & 0.491\,(10)     & 0.360\,(14)     & 0.277\,(15)     \Tstrut\\
\ce{^{52g}Mn_{(c)}} & 1.894\,(51)     & 0.533\,(11)     & 0.1130\,(38)    & 0.02593\,(97)   & 0.01084\,(30)   & 0.00863\,(17)   & 0.003375\,(78)  \\
\ce{^{54g}Mn_{(c)}} & 7.56\,(27)      & 4.097\,(61)     & 4.100\,(51)     & 4.82\,(30)      & 5.086\,(66)     & 4.731\,(58)     & 4.331\,(65)     \\
\ce{^{56g}Mn_{(c)}} & 1.163\,(70)     & 0.82\,(19)      & 0.824\,(56)     & 0.653\,(19)     & 0.288\,(93)     & 0.261\,(23)     & 0.2568\,(93)    \\
\ce{^{59g}Fe_{(c)}} & 0.954\,(34)     & 0.892\,(26)     & 0.889\,(29)     & 0.825\,(29)     & 0.728\,(21)     & 0.624\,(18)     & 0.545\,(18)     \\
\ce{^{55g}Co_{(c)}} & 1.888\,(91)     & 1.106\,(90)     & 0.463\,(52)     & 0.147\,(35)     & 0.069\,(41)     &                 &                 \\
\ce{^{56g}Co_{(c)}} & 10.02\,(24)     & 11.72\,(19)     & 12.83\,(18)     & 10.76\,(13)     & 7.169\,(89)     & 4.139\,(55)     & 2.259\,(36)     \\
\ce{^{57g}Co_{(c)}} & 28.9\,(15)      & 34.4\,(11)      & 34.69\,(92)     & 24.8\,(12)      & 30.00\,(72)     & 40.09\,(82)     & 51.2\,(21)      \\
\ce{^{58(m+g)}Co_{(c)}} & 52.4\,(21)      & 50.0\,(14)      & 46.9\,(12)      & 40.0\,(10)      & 35.14\,(85)     & 31.97\,(59)     & 31.35\,(60)     \\
\ce{^{58m}Co_{(i)}} &                 & 27.4\,(14)      & 24.6\,(14)      &                 &                 &                 & 16.52\,(86)     \\
\ce{^{60g}Co_{(c)}} & 13.29\,(37)     & 12.07\,(80)     & 15.52\,(59)     & 11.53\,(29)     & 11.22\,(36)     & 11.06\,(23)     & 12.26\,(52)     \\
\ce{^{61g}Co_{(c)}} &                 &                 &                 &                 &                 &                 &                 \\
\ce{^{56g}Ni_{(c)}} & 0.0925\,(31)    & 0.0892\,(31)    & 0.0974\,(29)    & 0.0748\,(19)    & 0.0463\,(13)    & 0.02355\,(74)   & 0.01109\,(69)   \\
\ce{^{57g}Ni_{(c)}} & 1.781\,(51)     & 1.305\,(34)     & 1.422\,(31)     & 1.737\,(35)     & 2.142\,(38)     & 2.429\,(46)     & 2.576\,(66)     \\
\ce{^{61g}Cu_{(c)}} & 53.3\,(14)      & 61.9\,(13)      & 71.6\,(11)      & 78.1\,(11)      & 81.4\,(14)      & 81.5\,(15)      & 78.3\,(20)      \\
\ce{^{64g}Cu_{(i)}} & 40.5\,(18)      & 48.2\,(41)      & 51.7\,(43)      & 55.4\,(41)      & 57.0\,(60)      & 60.3\,(80)      & 62.3\,(80)      \\
\ce{^{62g}Zn_{(i)}} & 4.90\,(38)      & 6.11\,(36)      & 6.95\,(38)      & 8.09\,(40)      & 9.56\,(81)      & 10.31\,(81)     & 0.26\,(32)      \\
\ce{^{65g}Zn_{(i)}} & 2.937\,(76)     & 3.441\,(56)     & 4.006\,(55)     & 4.257\,(50)     & 4.439\,(57)     & 4.446\,(58)     & 4.777\,(66)     \Bstrut\\\hline\hline
$E_p$\,(MeV)        & 54.728\,(1580)  & 54.082\,(562)   & 54.078\,(562)   & 52.691\,(1626)  & 52.110\,(578)   & 50.576\,(1675)  & 50.075\,(594)   \Tstrut\Bstrut\\
Location            & LANL            & LBNL            & LBNL        & LANL            & LBNL            & LANL            & LBNL            \Bstrut\\\hline
\ce{^{51g}Cr_{(c)}} & 0.193\,(20)     &                 &                 & 0.132\,(21)     &                 &                 &                 \Tstrut\\
\ce{^{52g}Mn_{(c)}} & 0.0043\,(33)    &                 &                 & 0.0025\,(26)    &                 & 0.0064\,(55)    &                 \\
\ce{^{54g}Mn_{(c)}} & 3.541\,(49)     & 4.078\,(47)     & 4.085\,(26)     & 2.764\,(40)     & 2.547\,(30)     & 1.883\,(32)     & 1.575\,(56)     \\
\ce{^{56g}Mn_{(c)}} & 0.073\,(38)     &                 & 0.006\,(18)     & 0.0424\,(24)    & 0.0854\,(77)    &                 & 0.111\,(13)     \\
\ce{^{59g}Fe_{(c)}} & 0.444\,(16)     & 0.3619\,(70)    & 0.395\,(14)     & 0.342\,(16)     & 0.2769\,(49)    & 0.230\,(11)     & 0.172\,(15)     \\
\ce{^{56g}Co_{(c)}} & 1.179\,(21)     & 1.256\,(43)     & 1.151\,(23)     & 0.5728\,(96)    & 0.573\,(52)     & 0.2784\,(56)    & 0.351\,(54)     \\
\ce{^{57g}Co_{(c)}} & 49.8\,(26)      & 54.47\,(40)     & 56.01\,(28)     & 37.9\,(23)      & 53.07\,(44)     & 40.71\,(85)     & 47.67\,(29)     \\
\ce{^{58(m+g)}Co_{(c)}} & 31.73\,(58)     & 31.55\,(33)     & 32.58\,(47)     & 33.91\,(67)     & 35.76\,(38)     & 37.96\,(82)     & 39.71\,(60)     \\
\ce{^{58m}Co_{(i)}} & 16.67\,(93)     &                 &                 &                 &                 &                 &                 \\
\ce{^{60g}Co_{(c)}} & 10.78\,(30)     & 11.18\,(25)     & 11.04\,(13)     & 10.39\,(18)     & 10.97\,(22)     & 9.57\,(16)      & 10.039\,(75)    \\
\ce{^{61g}Co_{(c)}} &                 &                 & 1.64\,(61)      &                 &                 &                 &                 \\
\ce{^{56g}Ni_{(c)}} & 0.00557\,(81)   & 0.0173\,(72)    &                 &                 &                 &                 &                 \\
\ce{^{57g}Ni_{(c)}} & 2.472\,(69)     & 2.20\,(12)      & 2.164\,(12)     & 2.252\,(53)     & 1.94\,(12)      & 1.882\,(42)     & 1.65\,(15)      \\
\ce{^{60g}Cu_{(c)}} &                 & 27.62\,(54)     & 31.49\,(61)     &                 & 30.2\,(11)      &                 & 27.21\,(38)     \\
\ce{^{61g}Cu_{(c)}} & 84.2\,(14)      & 84.3\,(18)      & 91.81\,(51)     & 86.4\,(15)      & 88.1\,(28)      & 94.6\,(18)      & 94.8\,(22)      \\
\ce{^{64g}Cu_{(i)}} & 63.5\,(57)      &                 &                 & 67.5\,(79)      &                 & 69.0\,(59)      &                 \\
\ce{^{62g}Zn_{(i)}} & 11.44\,(57)     & 10.95\,(36)     & 11.905\,(72)    & 12.15\,(58)     & 11.96\,(38)     & 13.21\,(67)     & 12.28\,(28)     \\
\ce{^{63g}Zn_{(i)}} &                 & 16.8\,(14)      & 18.94\,(28)     &                 & 19.35\,(85)     &                 & 21.65\,(62)     \\
\ce{^{65g}Zn_{(i)}} & 4.826\,(73)     & 4.360\,(42)     & 4.703\,(57)     & 5.116\,(74)     & 4.82\,(12)      & 5.473\,(88)     & 5.000\,(37)     \Bstrut\\\hline\hline
$E_p$\,(MeV)        & 49.529\,(600)   & 47.985\,(613)   & 46.287\,(632)   & 45.814\,(635)   & 43.564\,(660)   & 42.848\,(670)   & 40.769\,(695)   \Tstrut\Bstrut\\
Location            & LBNL        & LBNL            & LBNL        & LBNL            & LBNL            & LBNL        & LBNL            \Bstrut\\\hline
\ce{^{54g}Mn_{(c)}} & 1.392\,(18)     & 0.897\,(56)     & 0.457\,(26)     & 0.586\,(12)     & 0.287\,(26)     & 0.244\,(36)     &                 \Tstrut\\
\ce{^{56g}Mn_{(c)}} & 0.06\,(13)      & 0.09\,(17)      & 0.051\,(51)     &                 &                 & 0.20\,(24)      &                 \\
\ce{^{59g}Fe_{(c)}} & 0.173\,(14)     & 0.0941\,(53)    & 0.101\,(51)     &                 &                 & 0.039\,(31)     &                 \\
\ce{^{56g}Co_{(c)}} & 0.243\,(50)     & 0.209\,(49)     & 0.013\,(28)     & 0.121\,(40)     & 0.119\,(55)     &                 & 0.3142\,(75)    \\
\ce{^{57g}Co_{(c)}} & 46.40\,(27)     & 38.96\,(31)     & 32.31\,(20)     & 28.68\,(20)     & 17.25\,(10)     & 14.725\,(70)    & 5.977\,(50)     \\
\ce{^{58(m+g)}Co_{(c)}} & 41.06\,(51)     & 44.31\,(32)     & 50.21\,(76)     & 50.97\,(40)     & 58.88\,(74)     & 58.25\,(76)     & 62.00\,(86)     \\
\ce{^{60g}Co_{(c)}} & 9.49\,(17)      & 8.991\,(73)     & 8.14\,(33)      & 8.670\,(89)     & 8.51\,(13)      & 7.19\,(19)      & 7.724\,(67)     \\
\ce{^{61g}Co_{(c)}} & 0.49\,(42)      &                 & 1.68\,(28)      &                 &                 & 0.26\,(24)      &                 \\
\ce{^{57g}Ni_{(c)}} & 1.350\,(14)     & 1.14\,(15)      & 0.6966\,(93)    & 0.649\,(34)     & 0.332\,(18)     & 0.2018\,(38)    & 0.088\,(70)     \\
\ce{^{60g}Cu_{(c)}} & 25.41\,(95)     & 22.90\,(63)     & 17.78\,(58)     & 11.2\,(14)      &                 & 7.30\,(52)      &                 \\
\ce{^{61g}Cu_{(c)}} & 102.44\,(50)    & 101.3\,(29)     & 118.53\,(37)    & 112.6\,(23)     & 129.0\,(28)     & 148.40\,(65)    & 152.3\,(23)     \\
\ce{^{62g}Zn_{(i)}} & 13.035\,(75)    & 12.28\,(43)     & 12.977\,(53)    & 11.73\,(43)     & 11.98\,(18)     & 13.000\,(90)    & 11.91\,(25)     \\
\ce{^{63g}Zn_{(i)}} & 23.58\,(33)     & 24.53\,(92)     & 28.68\,(45)     & 24.9\,(17)      & 30.9\,(13)      & 36.42\,(70)     & 39.1\,(23)      \\
\ce{^{65g}Zn_{(i)}} & 5.274\,(71)     & 5.14\,(41)      & 5.69\,(11)      & 5.404\,(31)     & 6.13\,(46)      & 6.140\,(50)     & 6.455\,(52)     \Bstrut\\\hline\hline
$E_p$\,(MeV)        & 39.187\,(719)   & 37.832\,(736)   & 35.231\,(781)   & 34.693\,(788)   & 33.525\,(378)   & 31.846\,(394)   & 30.093\,(411)   \Tstrut\Bstrut\\
Location            & LBNL        & LBNL            & LBNL        & LBNL            & LBNL            & LBNL            & LBNL            \Bstrut\\\hline
\ce{^{54g}Mn_{(c)}} & 0.089\,(33)     & 0.3284\,(52)    & 0.155\,(34)     & 0.319\,(14)     & 0.852\,(85)     & 0.742\,(94)     & 0.542\,(31)     \Tstrut\\
\ce{^{59g}Fe_{(c)}} & 0.010\,(40)     &                 & 0.055\,(98)     &                 &                 &                 &                 \\
\ce{^{56g}Co_{(c)}} &                 & 0.321\,(14)     &                 & 0.309\,(95)     & 0.12\,(73)      & 0.21\,(12)      & 0.25\,(21)      \\
\ce{^{57g}Co_{(c)}} & 2.793\,(28)     & 1.055\,(16)     & 0.238\,(11)     & 0.210\,(17)     & 0.339\,(18)     &                 &                 \\
\ce{^{58(m+g)}Co_{(c)}} & 56.2\,(13)      & 55.44\,(50)     & 44.75\,(72)     & 41.58\,(67)     & 35.45\,(57)     & 22.85\,(25)     & 11.52\,(19)     \\
\ce{^{60g}Co_{(c)}} & 5.82\,(20)      & 5.67\,(10)      & 3.40\,(15)      & 3.487\,(74)     & 3.35\,(51)      & 1.30\,(12)      &                 \\
\ce{^{61g}Co_{(c)}} & 0.30\,(30)      &                 &                 &                 &                 &                 &                 \\
\ce{^{57g}Ni_{(c)}} & 0.0168\,(30)    & 0.081\,(16)     & 0.0014\,(38)    & 0.0401\,(35)    & 0.06\,(15)      & 0.183\,(19)     & 0.182\,(22)     \\
\ce{^{60g}Cu_{(c)}} & 1.92\,(32)      &                 & 0.152\,(79)     &                 &                 &                 &                 \\
\ce{^{61g}Cu_{(c)}} & 177.33\,(82)    & 167.4\,(38)     & 180.82\,(92)    & 161.2\,(32)     & 157.0\,(25)     & 133.5\,(28)     & 101.1\,(18)     \\
\ce{^{62g}Zn_{(i)}} & 13.44\,(11)     & 13.35\,(25)     & 17.15\,(12)     & 17.35\,(49)     & 19.82\,(70)     & 26.51\,(54)     & 37.02\,(80)     \\
\ce{^{63g}Zn_{(i)}} & 46.26\,(67)     & 46.2\,(35)      & 53.88\,(70)     & 50.9\,(31)      & 49.8\,(17)      & 46.9\,(29)      & 47.0\,(22)      \\
\ce{^{65g}Zn_{(i)}} & 6.621\,(90)     & 6.687\,(67)     & 7.387\,(48)     & 7.364\,(38)     & 7.641\,(76)     & 7.97\,(13)      & 8.49\,(12)      \Bstrut\\\hline\hline
$E_p$\,(MeV)        & 30.053\,(881)   & 28.257\,(431)   & 26.340\,(456)   & 24.305\,(485)   & 22.124\,(522)   & 19.764\,(570)   & 17.136\,(637)   \Tstrut\Bstrut\\
Location            & LBNL            & LBNL            & LBNL            & LBNL            & LBNL            & LBNL            & LBNL            \Bstrut\\\hline
\ce{^{54g}Mn_{(c)}} & 0.222\,(28)     & 0.438\,(55)     & 0.560\,(92)     & 0.527\,(10)     & 0.546\,(18)     & 0.419\,(58)     & 0.517\,(33)     \Tstrut\\
\ce{^{56g}Co_{(c)}} & 0.297\,(13)     & 0.32\,(15)      & 0.385\,(18)     & 0.400\,(17)     & 0.32\,(41)      &                 &                 \\
\ce{^{57g}Co_{(c)}} &                 &                 &                 &                 & 0.172\,(12)     &                 &                 \\
\ce{^{58(m+g)}Co_{(c)}} & 12.25\,(28)     & 3.54\,(14)      & 0.422\,(16)     &                 &                 & 0.5058\,(95)    & 0.07\,(12)      \\
\ce{^{58m}Co_{(i)}} &                 &                 &                 &                 &                 & 0.236\,(14)     & 0.0386\,(53)    \\
\ce{^{60g}Co_{(c)}} & 1.064\,(72)     &                 &                 &                 &                 &                 &                 \\
\ce{^{57g}Ni_{(c)}} & 0.138\,(11)     & 0.185\,(15)     & 0.263\,(28)     & 0.285\,(33)     & 0.195\,(27)     & 0.297\,(21)     & 0.403\,(31)     \\
\ce{^{61g}Cu_{(c)}} & 95.9\,(16)      & 62.8\,(14)      & 26.23\,(55)     & 7.03\,(20)      &                 &                 &                 \\
\ce{^{62g}Zn_{(i)}} & 35.21\,(52)     & 51.5\,(10)      & 64.1\,(11)      & 71.44\,(96)     & 67.9\,(11)      & 52.65\,(97)     & 31.37\,(32)     \\
\ce{^{63g}Zn_{(i)}} & 39.8\,(28)      & 36.6\,(17)      & 28.2\,(16)      & 22.21\,(17)     & 27.2\,(11)      & 45.8\,(22)      & 106.3\,(42)     \\
\ce{^{65g}Zn_{(i)}} & 9.15\,(16)      & 9.71\,(29)      & 10.61\,(15)     & 11.79\,(14)     & 14.71\,(11)     & 22.95\,(35)     & 45.62\,(57)     \Bstrut\\\hline\hline
$E_p$\,(MeV)        & \multicolumn{2}{l}{14.166\,(733)} &                 &                 &                 &                 &                 \Tstrut\Bstrut\\
Location            & LBNL            &                 &                 &                 &                 &                 &                 \Bstrut\\\hline
\ce{^{54g}Mn_{(c)}} & 0.572\,(44)     &                 &                 &                 &                 &                 &                 \Tstrut\\
\ce{^{56g}Co_{(c)}} & 0.21\,(26)      &                 &                 &                 &                 &                 &                 \\
\ce{^{57g}Ni_{(c)}} & 0.570\,(34)     &                 &                 &                 &                 &                 &                 \\
\ce{^{62g}Zn_{(i)}} & 2.650\,(84)     &                 &                 &                 &                 &                 &                 \\
\ce{^{63g}Zn_{(i)}} & 256.3\,(75)     &                 &                 &                 &                 &                 &                 \\
\ce{^{65g}Zn_{(i)}} & 117.0\,(13)     &                 &                 &                 &                 &                 &                \Bstrut\\

\end{longtable*}

\end{tiny}
\endgroup
        \begingroup
\squeezetable
\begin{tiny}
\begin{longtable*}{llllllll}
\caption{Independent and cumulative cross section measurements for \ce{^{nat}Nb}(p,x) reactions.}\\
\label{tab:nb_xs_data}\\
\endfirsthead

\hline \hline
\endhead

\hline \multicolumn{4}{r}{{Continued on next page}} \\ \hline\hline
\endfoot

\hline \hline
\endlastfoot
\hline \hline

$E_p$\,(MeV)        & 188.398\,(2070) & 173.201\,(2185) & 159.798\,(2303) & 146.254\,(2443) & 131.650\,(2625) & 117.620\,(2838) & 102.548\,(3129) \Tstrut\Bstrut\\
Location            & BNL             & BNL             & BNL             & BNL             & BNL             & BNL             & BNL             \Bstrut\\ \hline
\ce{^{69g}Ge_{(c)}} &                 &                 &                 & 0.93\,(19)      &                 & 0.73\,(37)      & 0.5\,(13)       \Tstrut\\
\ce{^{69g}As_{(c)}} &                 & 0.31\,(53)      & 0.03\,(11)      & 0.93\,(19)      &                 & 0.72\,(36)      & 0.5\,(13)       \\
\ce{^{71g}As_{(c)}} & 0.74\,(21)      &                 &                 & 0.76\,(21)      &                 &                 &                 \\
\ce{^{76g}Kr_{(c)}} &                 & 1.84\,(99)      & 0.62\,(58)      & 0.08\,(32)      &                 &                 & 0.04\,(29)      \\
\ce{^{77g}Kr_{(c)}} & 38.5\,(19)      & 36.8\,(43)      & 44.\,(14)       & 30.9\,(37)      & 49.0\,(78)      &                 &                 \\
\ce{^{83g}Rb_{(c)}} & 40.6\,(13)      & 38.33\,(54)     & 35.35\,(41)     & 31.77\,(36)     & 28.22\,(35)     & 17.65\,(31)     &                 \\
\ce{^{84g}Rb_{(c)}} & 3.285\,(90)     & 3.088\,(49)     & 2.720\,(42)     & 2.424\,(34)     & 2.144\,(32)     & 1.763\,(42)     & 1.246\,(61)     \\
\ce{^{83g}Sr_{(c)}} & 37.83\,(96)     & 36.43\,(58)     & 33.82\,(41)     & 30.93\,(38)     & 27.50\,(36)     & 17.25\,(39)     & 4.50\,(45)      \\
\ce{^{83g}Y_{(c)}} & 25.3\,(10)      & 22.00\,(44)     & 19.28\,(26)     & 16.98\,(26)     & 15.00\,(22)     & 9.73\,(27)      & 4.48\,(45)      \\
\ce{^{85m}Y_{(c)}} & 58.\,(14)       & 76.8\,(32)      & 82.0\,(13)      & 45.7\,(39)      & 55.0\,(14)      & 45.8\,(12)      &                 \\
\ce{^{86g}Y_{(c)}} & 50.4\,(28)      & 51.3\,(23)      & 52.4\,(23)      & 48.9\,(37)      & 51.5\,(33)      & 44.8\,(30)      & 34.3\,(20)      \\
\ce{^{87g}Y_{(c)}} & 111.2\,(18)     & 115.1\,(15)     & 118.4\,(13)     & 120.8\,(12)     & 124.0\,(15)     & 129.1\,(19)     & 118.7\,(23)     \\
\ce{^{88g}Y_{(c)}} & 25.08\,(66)     & 28.48\,(54)     & 30.76\,(41)     & 28.17\,(35)     & 34.45\,(64)     & 33.8\,(11)      & 28.32\,(63)     \\
\ce{^{90m}Y_{(c)}} & 1.22\,(92)      & 1.1\,(10)       & 1.1\,(10)       & 1.14\,(67)      & 0.96\,(72)      &                 &                 \\
\ce{^{86g}Zr_{(c)}} & 23.0\,(16)      & 24.74\,(86)     & 25.5\,(11)      & 27.4\,(21)      & 26.1\,(20)      & 22.6\,(16)      & 12.51\,(71)     \\
\ce{^{89g}Nb_{(c)}} &                 & 115.\,(27)      & 72.\,(13)       & 70.7\,(36)      & 46.8\,(75)      & 59.8\,(81)      & 155.\,(17)      \\
\ce{^{89m}Nb_{(i)}} &                 &                 &                 & 12.69\,(45)     & 14.00\,(97)     & 13.54\,(91)     & 18.51\,(95)     \\
\ce{^{90g}Nb_{(c)}} & 86.4\,(38)      & 79.0\,(57)      & 102.9\,(44)     & 34.\,(11)       & 124.5\,(36)     & 102.\,(12)      & 102.\,(11)      \\
\ce{^{92m}Nb_{(i)}} & 26.75\,(50)     & 29.96\,(44)     & 31.93\,(40)     & 33.68\,(37)     & 35.58\,(51)     & 38.22\,(74)     & 43.16\,(94)     \\
\ce{^{90g}Mo_{(i)}} & 5.69\,(67)      & 6.3\,(13)       & 6.96\,(96)      & 8.0\,(40)       & 9.41\,(69)      & 11.3\,(35)      & 14.9\,(32)      \\
\ce{^{93m}Mo_{(i)}} & 1.35\,(23)      & 2.16\,(37)      & 0.655\,(69)     & 0.83\,(23)      & 1.39\,(25)      & 0.87\,(12)      & 0.931\,(73)     \Bstrut\\\hline\hline
$E_p$\,(MeV)        & 90.822\,(1075)  & 78.750\,(1196)  & 71.813\,(1283)  & 66.297\,(1363)  & 62.097\,(1433)  & 59.000\,(1491)  & 56.293\,(1546)  \Tstrut\Bstrut\\
Location            & LANL            & LANL            & LANL            & LANL            & LANL            & LANL            & LANL            \Bstrut\\ \hline
\ce{^{83g}Rb_{(c)}} & 6.33\,(15)      & 7.71\,(13)      & 5.353\,(77)     &                 &                 &                 &                 \Tstrut\\
\ce{^{84g}Rb_{(c)}} & 0.638\,(37)     &                 & 0.61\,(43)      &                 &                 &                 &                 \\
\ce{^{83g}Sr_{(c)}} & 4.81\,(20)      & 5.70\,(28)      & 4.58\,(21)      &                 &                 &                 &                 \\
\ce{^{85m}Y_{(c)}} & 29.9\,(88)      & 27.\,(33)       &                 &                 &                 &                 &                 \\
\ce{^{86g}Y_{(c)}} & 31.6\,(14)      & 41.3\,(14)      & 42.5\,(15)      & 32.5\,(13)      & 19.06\,(94)     & 8.88\,(74)      & 3.32\,(67)      \\
\ce{^{87g}Y_{(c)}} & 56.7\,(69)      & 23.4\,(40)      & 29.3\,(41)      & 31.3\,(56)      & 32.3\,(62)      & 40.5\,(68)      & 34.5\,(64)      \\
\ce{^{88g}Y_{(c)}} &                 &                 &                 & 8.51\,(90)      &                 &                 &                 \\
\ce{^{90m}Y_{(c)}} & 0.96\,(12)      & 0.858\,(80)     &                 &                 & 0.744\,(91)     &                 &                 \\
\ce{^{91m}Y_{(i)}} &                 & 38.\,(15)       & 69.\,(33)       &                 & 77.\,(30)       &                 &                 \\
\ce{^{86g}Zr_{(c)}} & 13.61\,(53)     & 20.03\,(61)     & 22.89\,(66)     & 17.29\,(56)     & 9.45\,(40)      & 3.97\,(32)      & 1.33\,(26)      \\
\ce{^{87g}Zr_{(c)}} & 59.2\,(30)      & 31.9\,(21)      &                 & 43.3\,(29)      &                 & 62.0\,(43)      & 54.6\,(42)      \\
\ce{^{89g}Nb_{(c)}} & 115.0\,(45)     & 142.0\,(70)     & 181.9\,(84)     & 203.\,(10)      & 176.\,(10)      & 116.\,(11)      & 93.1\,(56)      \\
\ce{^{90g}Nb_{(c)}} & 168.8\,(51)     & 192.4\,(49)     & 222.0\,(57)     & 257.5\,(59)     & 302.1\,(70)     & 345.5\,(74)     & 402.3\,(97)     \\
\ce{^{92m}Nb_{(i)}} & 38.60\,(93)     & 41.73\,(69)     & 44.21\,(58)     & 45.80\,(58)     & 47.90\,(60)     & 46.93\,(67)     & 48.25\,(71)     \\
\ce{^{90g}Mo_{(i)}} & 15.11\,(85)     & 19.40\,(86)     & 24.0\,(11)      & 31.0\,(12)      & 41.0\,(16)      & 52.4\,(16)      & 67.6\,(22)      \\
\ce{^{93m}Mo_{(i)}} & 0.94\,(15)      & 0.98\,(10)      & 1.03\,(11)      & 1.14\,(27)      & 1.27\,(13)      & 1.27\,(14)      & 1.30\,(15)      \Bstrut\\ \hline\hline
$E_p$\,(MeV)        & 54.309\,(1589)  & 53.674\,(565)   & 52.247\,(1636)  & 51.687\,(581)   & 50.138\,(1685)  & 49.644\,(597)   & 47.529\,(616)   \Tstrut\Bstrut\\
Location            & LANL            & LBNL            & LANL            & LBNL            & LANL            & LBNL            & LBNL            \Bstrut\\ \hline
\ce{^{86g}Y_{(c)}} & 1.4\,(18)       & 1.27\,(84)      &                 &                 &                 &                 &                 \Tstrut\\
\ce{^{87g}Y_{(c)}} & 33.4\,(59)      & 15.1\,(30)      & 29.2\,(52)      & 17.1\,(35)      & 27.5\,(42)      & 18.7\,(52)      & 26.0\,(65)      \\
\ce{^{88g}Y_{(c)}} & 9.\,(79)        & 13.35\,(72)     &                 & 14.84\,(57)     & 10.99\,(65)     & 16.10\,(80)     & 18.39\,(86)     \\
\ce{^{86g}Zr_{(c)}} & 0.50\,(67)      &                 &                 &                 &                 &                 &                 \\
\ce{^{87g}Zr_{(c)}} & 57.6\,(48)      &                 &                 &                 &                 &                 &                 \\
\ce{^{88g}Zr_{(c)}} &                 & 27.71\,(42)     &                 & 33.72\,(39)     &                 & 42.83\,(69)     & 51.68\,(50)     \\
\ce{^{89g}Nb_{(c)}} & 62.1\,(40)      & 28.\,(80)       & 16.4\,(36)      & 10.\,(93)       & 10.3\,(25)      & 4.\,(65)        &                 \\
\ce{^{89m}Nb_{(i)}} &                 & 1.9\,(56)       &                 & 1.4\,(28)       &                 &                 &                 \\
\ce{^{90g}Nb_{(c)}} & 442.\,(11)      & 458.7\,(58)     & 474.\,(13)      & 489.2\,(76)     & 483.\,(14)      & 501.4\,(76)     & 477.9\,(62)     \\
\ce{^{91m}Nb_{(i)}} &                 & 35.0\,(15)      &                 & 36.10\,(90)     &                 & 37.8\,(26)      & 37.17\,(72)     \\
\ce{^{92m}Nb_{(i)}} & 47.82\,(75)     & 52.14\,(99)     & 48.34\,(83)     & 53.3\,(19)      & 48.67\,(97)     & 55.6\,(12)      & 55.7\,(11)      \\
\ce{^{90g}Mo_{(i)}} & 76.8\,(26)      & 81.1\,(33)      & 86.7\,(32)      & 89.4\,(35)      & 90.6\,(34)      & 91.7\,(41)      & 64.1\,(53)      \\
\ce{^{93m}Mo_{(i)}} & 1.32\,(14)      & 1.70\,(23)      & 1.47\,(18)      & 1.713\,(99)     & 1.48\,(21)      & 1.76\,(14)      & 1.74\,(24)      \Bstrut\\\hline\hline
$E_p$\,(MeV)        & 45.348\,(639)   & 43.065\,(665)   & 40.261\,(701)   & 37.284\,(744)   & 34.100\,(797)   & \multicolumn{2}{l}{29.412\,(895)} \Tstrut\Bstrut\\
Location            & LBNL            & LBNL            & LBNL            & LBNL            & LBNL            & LBNL            &                 \Bstrut\\ \hline
\ce{^{87m}Sr_{(i)}} & 0.62\,(52)      & 0.49\,(24)      &                 & 0.2\,(24)       &                 & 0.32\,(24)      &                 \Tstrut\\
\ce{^{87g}Y_{(c)}} & 17.4\,(63)      & 6.4\,(20)       &                 & 0.2\,(36)       &                 &                 &                 \\
\ce{^{88g}Y_{(c)}} & 19.7\,(13)      & 19.7\,(13)      & 16.0\,(14)      & 13.44\,(98)     & 7.98\,(83)      & 1.555\,(97)     &                 \\
\ce{^{88g}Zr_{(c)}} & 63.08\,(60)     & 66.79\,(79)     & 64.4\,(18)      & 53.15\,(62)     & 33.91\,(47)     & 4.841\,(57)     &                 \\
\ce{^{90g}Nb_{(c)}} & 352.\,(35)      & 299.\,(19)      & 129.\,(17)      & 16.9\,(47)      & 1.93\,(51)      & 0.41\,(13)      &                 \\
\ce{^{91m}Nb_{(i)}} & 44.1\,(16)      & 45.83\,(67)     & 56.4\,(13)      & 67.59\,(99)     & 75.6\,(11)      & 74.4\,(10)      &                 \\
\ce{^{92m}Nb_{(i)}} & 57.9\,(20)      & 56.4\,(16)      & 57.5\,(15)      & 57.7\,(20)      & 58.3\,(10)      & 64.9\,(15)      &                 \\
\ce{^{90g}Mo_{(i)}} & 73.\,(12)       & 43.4\,(47)      & 19.4\,(39)      & 3.05\,(86)      & 0.012\,(74)     &                 &                 \\
\ce{^{93m}Mo_{(i)}} & 1.785\,(68)     & 1.68\,(21)      & 1.700\,(82)     & 1.794\,(41)     & 2.034\,(23)     & 2.447\,(34)     &                \Bstrut\\

\end{longtable*}
\end{tiny}
\endgroup
        \begingroup
\squeezetable
\begin{tiny}
\begin{longtable*}{llllllll}
\caption{Independent and cumulative cross section measurements for \ce{^{nat}Ti}(p,x) reactions. Note that measurements for \ce{^{46g}Sc_{(c)}} and  \ce{^{48}V_{(c)}} were derived through variance minimization}\\
\label{tab:ti_xs_data}\\
\endfirsthead

\hline \hline
\endhead

\hline \multicolumn{4}{r}{{Continued on next page}} \\ \hline\hline
\endfoot

\hline \hline
\endlastfoot
\hline \hline

$E_p$\,(MeV)        & 90.732\,(1076)  & 78.652\,(1197)  & 71.707\,(1283) & 66.185\,(1364) & 61.978\,(1434) & 58.875\,(1492) & 56.165\,(1547)       \Tstrut\Bstrut\\
Location            & LANL            & LANL            & LANL           & LANL           & LANL           & LANL           & LANL                 \Bstrut\\\hline
\ce{^{42g}K_{(c)}} & 7.08\,(59)      & 6.89\,(65)      & 7.46\,(86)     & 7.24\,(68)     & 6.61\,(75)     & 5.07\,(67)     & 4.20\,(57)           \Tstrut\\
\ce{^{43g}K_{(c)}} & 1.862\,(69)     & 1.572\,(50)     & 1.430\,(58)    & 1.417\,(64)    & 1.375\,(39)    & 1.476\,(43)    & 1.720\,(41)          \\
\ce{^{47g}Ca_{(c)}} & 0.1075\,(56)    & 0.0985\,(57)    & 0.108\,(12)    &                & 0.0928\,(69)   &                & 0.087\,(27)          \\
\ce{^{46(m+g)}Sc_{(c)}} & 43.17\,(100)    & 45.19\,(71)     & 47.01\,(63)    & 47.49\,(55)    & 48.45\,(54)    & 49.56\,(54)    & 52.17\,(78)          \\
\ce{^{47g}Sc_{(c)}} & 17.85\,(47)     & 18.16\,(38)     & 19.80\,(58)    & 18.64\,(53)    & 17.52\,(34)    & 17.30\,(36)    & 19.16\,(73)          \\
\ce{^{48g}Sc_{(c)}} & 2.263\,(69)     & 2.247\,(59)     & 2.150\,(56)    & 2.071\,(52)    & 1.982\,(50)    & 1.934\,(48)    & 1.863\,(51)          \\
\ce{^{48g}V_{(i)}} & 8.45\,(22)      & 10.30\,(21)     & 10.99\,(18)    & 12.29\,(20)    & 13.31\,(23)    & 13.92\,(18)    & 14.52\,(19)          \Bstrut\\\hline\hline
$E_p$\,(MeV)        & 54.178\,(1591)  & 53.696\,(562)   & 53.545\,(563)  & 52.110\,(1638) & 51.553\,(580)  & 50.000\,(1687) & 49.506\,(595)        \Tstrut\Bstrut\\
Location            & LANL            & LBNL        & LBNL           & LANL           & LBNL           & LANL           & LBNL                 \Bstrut\\\hline
\ce{^{42g}K_{(c)}} & 2.94\,(26)      & 2.292\,(44)     & 2.311\,(21)    & 2.03\,(46)     & 1.69\,(10)     & 1.50\,(47)     & 1.206\,(78)          \Tstrut\\
\ce{^{43g}K_{(c)}} & 1.606\,(48)     & 1.5190\,(91)    & 1.379\,(20)    & 1.592\,(48)    & 1.338\,(17)    & 1.543\,(53)    & 1.245\,(20)          \\
\ce{^{47g}Ca_{(c)}} &                 & 0.053\,(14)     &                &                &                &                &                      \\
\ce{^{46(m+g)}Sc_{(c)}} & 54.01\,(71)     & 55.52\,(33)     & 56.74\,(59)    & 53.99\,(78)    & 58.6\,(10)     & 58.54\,(96)    & 62.2\,(13)           \\
\ce{^{47g}Sc_{(c)}} & 17.76\,(51)     & 19.981\,(97)    & 19.9\,(12)     &                & 18.35\,(65)    &                & 18.22\,(91)          \\
\ce{^{48g}Sc_{(c)}} & 1.836\,(51)     & 1.749\,(19)     & 1.67\,(18)     & 1.781\,(51)    & 1.72\,(31)     & 1.791\,(55)    & 1.73\,(27)           \\
\ce{^{44g}Ti_{(c)}} &                 &                 &                &                &                & 4.5\,(43)      &                      \\
\ce{^{48g}V_{(i)}} & 15.32\,(21)     & 14.10\,(16)     & 15.71\,(59)    & 15.56\,(25)    & 16.53\,(91)    & 17.18\,(45)    & 17.44\,(93)          \Bstrut\\\hline\hline
$E_p$\,(MeV)        & 49.126\,(600)   & 47.387\,(615)   & 45.852\,(633)  & 45.201\,(638)  & 42.914\,(664)  & 42.395\,(672)  & 40.103\,(699)        \Tstrut\Bstrut\\
Location            & LBNL        & LBNL            & LBNL       & LBNL           & LBNL           & LBNL       & LBNL                 \Bstrut\\\hline
\ce{^{42g}K_{(c)}} & 0.864\,(51)     & 0.845\,(63)     & 0.508\,(61)    & 0.633\,(44)    & 0.475\,(42)    & 0.442\,(38)    & 0.320\,(30)          \Tstrut\\
\ce{^{43g}K_{(c)}} & 1.2259\,(98)    & 0.974\,(23)     & 0.8129\,(82)   & 0.6753\,(63)   & 0.4191\,(74)   & 0.3776\,(79)   & 0.1852\,(53)         \\
\ce{^{47g}Ca_{(c)}} &                 &                 & 0.049\,(10)    &                &                &                & 0.000000001631\,(41) \\
\ce{^{46(m+g)}Sc_{(c)}} & 61.38\,(19)     & 63.7\,(11)      & 64.08\,(15)    & 64.7\,(11)     & 62.8\,(11)     & 63.03\,(39)    & 57.1\,(15)           \\
\ce{^{47g}Sc_{(c)}} & 19.623\,(61)    & 18.00\,(67)     & 19.250\,(50)   & 18.38\,(67)    & 18.62\,(54)    & 18.73\,(26)    & 18.99\,(48)          \\
\ce{^{48g}Sc_{(c)}} & 1.691\,(21)     & 1.64\,(22)      & 1.600\,(17)    & 1.59\,(24)     & 1.45\,(20)     & 1.437\,(27)    & 1.28\,(15)           \\
\ce{^{48g}V_{(i)}} & 17.084\,(91)    & 18.74\,(90)     & 19.137\,(85)   & 20.18\,(90)    & 22.27\,(91)    & 22.17\,(12)    & 24.6\,(19)           \Bstrut\\\hline\hline
$E_p$\,(MeV)        & 38.696\,(722)   & 37.116\,(743)   & 34.708\,(786)  & 33.919\,(797)  & 32.975\,(373)  & 31.271\,(388)  & 29.494\,(406)        \Tstrut\Bstrut\\
Location            & LBNL        & LBNL            & LBNL       & LBNL           & LBNL           & LBNL           & LBNL                 \Bstrut\\\hline
\ce{^{42g}K_{(c)}} & 0.371\,(39)     &                 & 0.115\,(17)    &                & 0.105\,(10)    &                &                      \Tstrut\\
\ce{^{43g}K_{(c)}} & 0.1476\,(96)    & 0.0859\,(44)    &                & 0.0515\,(53)   &                &                &                      \\
\ce{^{47g}Ca_{(c)}} & 0.017\,(83)     &                 & 0.0497\,(73)   & 0.01621\,(92)  & 0.01\,(12)     &                &                      \\
\ce{^{46(m+g)}Sc_{(c)}} & 51.80\,(28)     & 45.9\,(13)      & 33.38\,(20)    & 29.3\,(18)     & 24.76\,(44)    & 17.33\,(36)    & 12.45\,(26)          \\
\ce{^{47g}Sc_{(c)}} & 20.03\,(11)     & 20.35\,(57)     & 22.12\,(13)    & 24.5\,(14)     & 22.83\,(76)    & 22.26\,(53)    & 21.33\,(28)          \\
\ce{^{48g}Sc_{(c)}} & 1.169\,(20)     & 1.08\,(18)      & 0.8697\,(97)   & 0.83\,(23)     & 0.69\,(17)     & 0.63\,(17)     & 0.56\,(12)           \\
\ce{^{48g}V_{(i)}} & 25.05\,(15)     & 28.4\,(18)      & 28.40\,(23)    & 30.4\,(28)     & 30.6\,(11)     & 31.6\,(11)     & 34.03\,(97)          \Bstrut\\\hline\hline
$E_p$\,(MeV)        & 29.211\,(895)   & 27.644\,(428)   & 25.692\,(452)  & 23.608\,(483)  & 21.379\,(522)  & 18.925\,(574)  & 16.212\,(647)        \Tstrut\Bstrut\\
Location            & LBNL            & LBNL            & LBNL           & LBNL           & LBNL           & LBNL           & LBNL                 \Bstrut\\\hline
\ce{^{42g}K_{(c)}} & 0.0497\,(39)    &                 &                &                &                &                &                      \Tstrut\\
\ce{^{47g}Ca_{(c)}} & 0.011\,(35)     &                 &                &                &                &                & 0.00002970\,(98)     \\
\ce{^{46(m+g)}Sc_{(c)}} & 12.06\,(19)     & 10.27\,(28)     & 8.89\,(25)     & 7.49\,(23)     & 5.82\,(21)     & 4.50\,(23)     & 3.17\,(27)           \\
\ce{^{47g}Sc_{(c)}} & 20.76\,(20)     & 18.79\,(23)     & 15.08\,(44)    & 11.67\,(34)    & 8.00\,(23)     & 4.20\,(13)     & 1.414\,(47)          \\
\ce{^{48g}Sc_{(c)}} & 0.574\,(38)     & 0.49\,(12)      & 0.36\,(11)     & 0.243\,(89)    & 0.131\,(57)    & 0.039\,(81)    &                      \\
\ce{^{48g}V_{(i)}} & 34.25\,(84)     & 37.0\,(11)      & 42.83\,(62)    & 51.65\,(86)    & 66.74\,(72)    & 104.5\,(11)    & 208.7\,(35)          \Bstrut\\\hline\hline
$E_p$\,(MeV)        & \multicolumn{2}{l}{13.104\,(764)} &                &                &                &                &                      \Tstrut\Bstrut\\
Location            & LBNL            &                 &                &                &                &                &                      \Bstrut\\\hline
\ce{^{46(m+g)}Sc_{(c)}} & 2.12\,(33)      &                 &                &                &                &                &                      \Tstrut\\
\ce{^{47g}Sc_{(c)}} & 0.600\,(77)     &                 &                &                &                &                &                      \\
\ce{^{48g}V_{(i)}} & 375.7\,(53)     &                 &                &                &                &                &                     

\end{longtable*}
\end{tiny}
\endgroup
\bibliography{references}

\end{document}